\documentclass[iop]{emulateapj}%  
\usepackage{amsmath}%  
\usepackage{amsfonts}%  
\usepackage{amssymb}%  
\usepackage{graphicx}%  
\usepackage{epsfig}%
\usepackage[font=small, labelsep=space]{caption}  
\usepackage{natbib}
\usepackage{color, soul}

\newbox\grsign \setbox\grsign=\hbox{$>$} \newdimen\grdimen \grdimen=\ht\grsign
\newbox\simlessbox \newbox\simgreatbox
\setbox\simgreatbox=\hbox{\raise.5ex\hbox{$>$}\llap
     {\lower.5ex\hbox{$\sim$}}}\ht1=\grdimen\dp1=0pt
\setbox\simlessbox=\hbox{\raise.5ex\hbox{$<$}\llap
     {\lower.5ex\hbox{$\sim$}}}\ht2=\grdimen\dp2=0pt

\def\simless{\mathrel{\copy\simlessbox}}
  
\def\JHU{1}
\def\STSCI{2}
\def\MSU{3}  
\def \Siena{4}

\begin{document}  

\slugcomment{\em Accepted for publication in the Astrophysical Journal}
\renewcommand*{\thefootnote}{\fnsymbol{footnote}}

\title{Star Formation Activity in CLASH Brightest Cluster Galaxies\footnotemark[$\star$]}%\footnote{Based on observations obtained at the Southern Astrophysical Research (SOAR) telescope, which is a joint project of the Minist\'{e}rio da Ci\^{e}ncia, Tecnologia, e Inova\c{c}\~{a}o (MCTI) da Rep\'{u}blica Federativa do Brasil, the U.S. National Optical Astronomy Observatory (NOAO), the University of North Carolina at Chapel Hill (UNC), and Michigan State University (MSU).}}%\footnotemark[\Foot]}

\author{Kevin Fogarty\altaffilmark{\JHU}, Marc Postman\altaffilmark{\STSCI}, Thomas Connor\altaffilmark{\MSU}, Megan Donahue\altaffilmark{\MSU}, John Moustakas\altaffilmark{\Siena}}  
  
\altaffiltext{\JHU}{Department of Physics and Astronomy, Johns Hopkins University, 3400 North Charles Street, Baltimore, MD 21218, USA}  
\altaffiltext{\STSCI}{Space Telescope Science Institute, 3700 San Martin Drive, Baltimore, MD 21218, USA}
\altaffiltext{\MSU}{Physics  and  Astronomy  Dept.,  Michigan  State  University, East Lansing, MI 48824, USA}
\altaffiltext{\Siena}{Department of Physics \& Astronomy, Siena College, 515 Loudon Road, Loudonville, NY 12211, USA}
  
\begin{abstract}  
The CLASH X-ray selected sample of 20 galaxy clusters contains ten brightest cluster galaxies (BCGs) that exhibit significant ($>$5 $\sigma$) extinction-corrected star formation rates (SFRs). Star formation activity is inferred from photometric estimates of UV and H$\alpha$+[\ion{N}{2}] emission in knots and filaments detected in CLASH HST ACS and WFC3 observations. UV-derived SFRs in these BCGs span two orders of magnitude, including two with a SFR $\gtrsim$ 100 M$_{\odot}$ yr$^{-1}$. These measurements are supplemented with [\ion{O}{2}], [\ion{O}{3}], and H$\beta$ fluxes measured from spectra obtained with the SOAR telescope. We confirm that photoionization from ongoing star formation powers the line emission nebulae in these BCGs, although in many BCGs there is also evidence of a LINER-like contribution to the line emission. Coupling these data with Chandra X-ray measurements, we infer that the star formation occurs exclusively in low-entropy cluster cores and exhibits a correlation with gas properties related to cooling. We also perform an in-depth study of the starburst history of the BCG in the cluster RXJ1532.9+3021, and create 2D maps of stellar properties on scales down to $\sim$350 pc. These  maps reveal evidence for an ongoing burst occurring in elongated filaments, generally on $\sim$ 0.5-1.0 Gyr timescales, although some filaments are consistent with much younger ($\lesssim$ 100 Myr) burst timescales and may be correlated with recent activity from the AGN. The relationship between BCG SFRs and the surrounding ICM gas properties provide new support for the process of feedback-regulated cooling in galaxy clusters and is consistent with recent theoretical predictions.  
\end{abstract}  
 
%\footnotetext[\Foot]{Based on observations obtained at the Southern Astrophysical Research (SOAR) telescope, which is a joint project of the Minist\'{e}rio da Ci\^{e}ncia, Tecnologia, e Inova\c{c}\~{a}o (MCTI) da Rep\'{u}blica Federativa do Brasil, the U.S. National Optical Astronomy Observatory (NOAO), the University of North Carolina at Chapel Hill (UNC), and Michigan State University (MSU).}


\footnotetext[$\star$]{Based on observations obtained at the Southern Astrophysical Research (SOAR) telescope, which is a joint project of the Minist\'{e}rio da Ci\^{e}ncia, Tecnologia, e Inova\c{c}\~{a}o (MCTI) da Rep\'{u}blica Federativa do Brasil, the U.S. National Optical Astronomy Observatory (NOAO), the University of North Carolina at Chapel Hill (UNC), and Michigan State University (MSU).}
 
\section{Introduction}  

\renewcommand*{\thefootnote}{\arabic{footnote}}
\setcounter{footnote}{0}
  
Brightest cluster galaxies (BCGs) in cool core galaxy clusters exhibit nebular emission features that are thought to be related to the intracluster medium (ICM) in the centers of these objects \citep[e.g.][]{Heckman_1989_CoolingLines, Fabian_1994_CoolingFlows, Crawford_1999_BCS}. Observations of substantial continuum UV and FIR fluxes have been recorded in many of these nominally early-type galaxies as well \citep{Hicks_2010_ClusterUVSFR, Rawle_2012_ClusterIRSFR}. Resolved images of low-redshift (z$\lesssim$0.1) BCGs suggest that while some of this activity is triggered by active galactic nuclei (AGN), most of the emission appears to be powered by recent star formation, located in knots and filaments \citep{Conselice_2001_Perseus, ODea_2004_A1795A2597, McDonald_2009_A1795, ODea_2010_FUVBCG, McDonald_2014_A1795, Tremblay_2012_A2597, Tremblay_2015_BCGUV}. Similar structures have also been observed in the Phoenix cluster ($z=0.596$), which hosts a massive starbursting BCG producing new stars at a rate of nearly 1000 M$_{\odot}$ yr$^{-1}$ \citep{McDonald_2012_Phoenix, McDonald_2013_Phoenix}.  
  
The most plausible candidate for the source of the gas being converted into stars in these galaxies is radiatively cooled ICM plasma \citep{Fabian_1994_CoolingFlows}. In a cool core cluster, the cooling time below a critical radius is less than the Hubble time at the redshift of the cluster, and plasma initially at this radius ought to have cooled. The ICM that manages to cool descends into the gravitational well in order to maintain pressure equilibrium, ultimately condensing into star forming gas inside the virial radius of the BCG \citep{Fabian_1994_CoolingFlows}. However, the cooling inferred from a simple cooling flow model is far more rapid than the star formation rates in these systems \citep[e.g.][]{Heckman_1989_CoolingLines, McNamara_1989_CCSFR}.
  
Activity in BCGs in the form of recent star formation and AGN outbursts are important components of feedback mechanisms proposed to reconcile the tension between the predicted and observed ICM cooling in cool core clusters \citep{McNamara_2007_AGNFeedback, Voit15b}. Specifically, the hot, X-ray emitting ICM of a relaxed galaxy cluster is predicted to radiatively cool more rapidly than is typically observed, and this manifests in star formation rates (SFRs) in BCGs which are roughly an order of magnitude lower than what would be predicted if all the available gas did indeed condense into cold gas \citep{ODea_2008_BCGSF}. However, in the presence of a feedback mechanism, such as energy injection from an AGN, the ICM is partially reheated and prevented from cooling catastrophically. Instead, residual cooling or cooling during an off-mode in the feedback duty cycle can account for the extended star forming structures observed. Studying the residual cooling using observations of BCGs spanning a wide range of activity will allow us to learn about different phases of cooling and feedback, and help us to determine whether a single feedback mechanism accounts for the variety of BCG features we observe.  
  
The high-resolution, multi-band HST observations available from the Cluster Lensing And Supernova survey with Hubble (\cite{Postman_2012_CLASH}; hereafter referred to as CLASH) are ideal for examining the star forming structures in BCGs. In \cite{Donahue_2015_IP}, we examined UV photometry for the entire CLASH sample of 25 galaxy clusters, and found evidence for significant, extended emission attributable to recent star formation in 10 of them. Two BCGs in this sample, RXJ1532.9+3021 and MACS1931.8-2653, stand out due to their strikingly large and luminous UV filaments.  
  
In the present study, we examine BCGs in the subset of CLASH clusters that were X-ray selected. This subsample includes 20 of the 25 CLASH clusters and includes all of the star-forming BCGs as identified by UV features. Using the CLASH HST photometry, along with spectra from the Southern Astrophysical Research (SOAR) telescope and archival \textit{Chandra} data, we investigate the nature of star formation in these BCGs, and provide new constraints on the source of the star formation activity in the structures we observe.  
  
\begin{figure*}[t!]  
\centerline{\includegraphics[width=20cm]{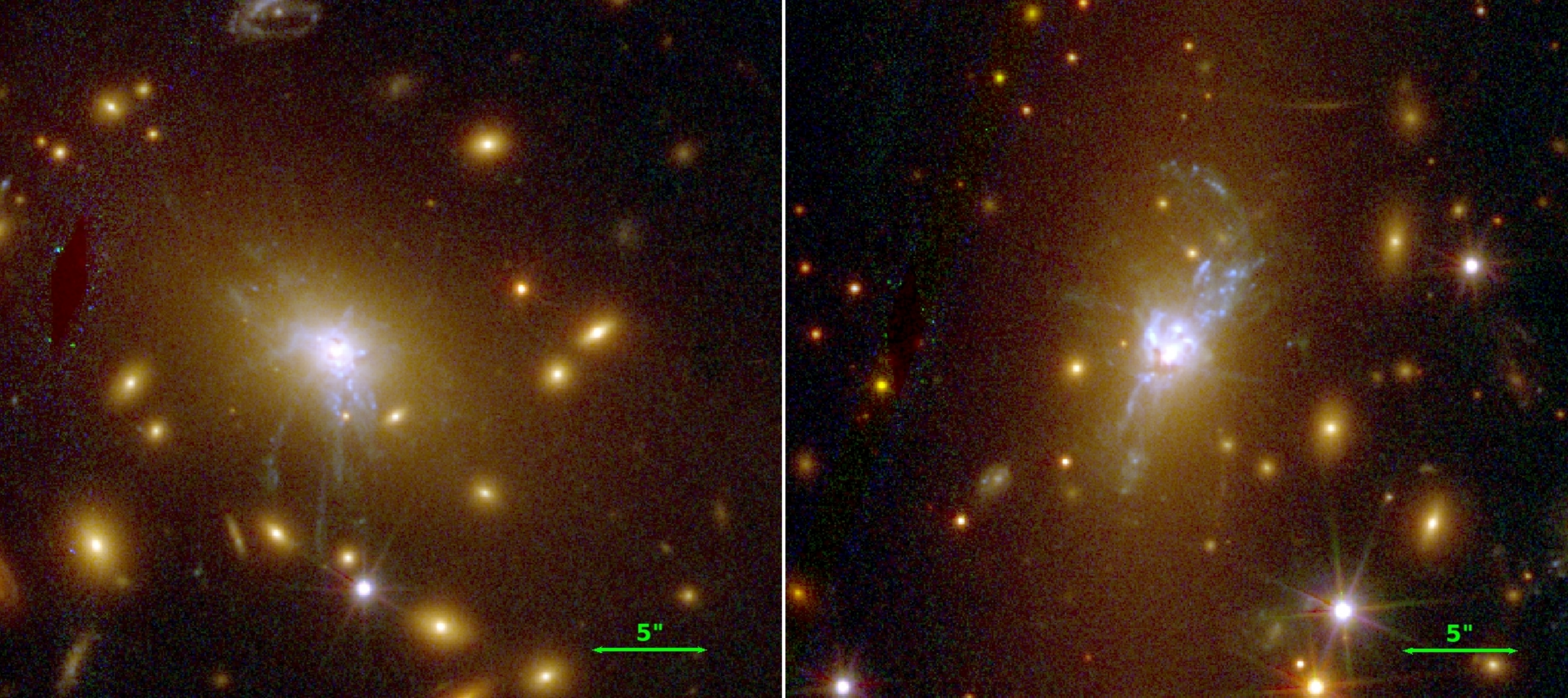}}
\caption[]{Color composite images of the most active star-forming BCGs of RXJ1532.9+3021 (z=0.363) on the left and MACS1931.8-2653 (z=0.352) on the right. The RGB color composites are made using WFC3-IR filters F105W+F110W+F125W+F140W+F160W in red, the ACS filters F606W + F625W + F775W + F814W + F850LP in green, and the ACS filters F435W + F475W in blue.}  
\label{fig:BCG_Colors}  
\end{figure*}

This paper consists of  two parts. First we derive SFRs for all BCGs using HST photometry and characterize the source of nebular line-emission in UV-bright BCGs using a combination of the HST photometry and SOAR-Goodman spectra. Second, we analyse the connection between the star formation and the properties of the ICM. This second part includes a detailed star formation history (SFH) analysis of  the BCG in RXJ1532.9+3021 derived from SED fitting of the CLASH photometry to create maps of stellar population parameters. RXJ1532.9+3021 was chosen for more detailed study because of the spectacular nature of its UV and H$\alpha$ structure (see Figure~\ref{fig:BCG_Colors}).   

We report on the incidence and distribution of reddening-corrected SFRs in our sample, and demonstrate that structures qualitatively comparable with those in the massive outburst in the Phoenix cluster are the sites of BCG activity in these `intermediate' starbursts as well. Structures observed in CLASH clusters also bear similarities to BCGs analysed in \cite{Tremblay_2015_BCGUV}. These SFRs are compared with \textit{Chandra} derived ICM core entropies and predicted cooling rates in the low-entropy core ICM in order to test the hypothesis that star formation is being fed by ICM cooling.  
  
For RXJ1532.9+3021 we are able to compare the properties of BCG filaments with the properties of X-ray cavities in the ICM, which we use to assess recent AGN jet-mode activity. We compare our results to previous analysis of the morphology of the cluster ICM \citep{HlavacekLarrondo_2013_RXJ1532}. The maps produced for this BCG allow us to examine the star formation history in individual knots and filaments, down to $\sim 350$ pc scales. These data allow us to investigate the source of the gas that condenses into the star forming regions and, for the case of RXJ1532.9+3021,  to examine this condensation in significant detail.  
  
Our paper is organized as follows. We describe the observational data in Section 2. In Section 3, we present the analysis of these data. Results are presented in Section 4, and the astrophysical implications are discussed in Section 5. We summarize our main conclusions in Section 6. We adopt the following cosmological parameters throughout this work: $H_o = 70.0$ km s$^{-1}$ Mpc$^{-1}$, $\Omega_m = 0.30$, and $\Omega_{\Lambda} = 0.70$. We assume a \cite{Salpeter_1955_IMF} IMF throughout. 
  
\section{Observations}  
  
\subsection{CLASH HST Observations}  
  
The CLASH program is detailed in \cite{Postman_2012_CLASH}. The 20 X-ray selected clusters were each observed for 15-20 orbits (for a total of at least 20 orbits including archival data) divided among 16 bands of photometry spanning an observer-frame wavelength range of $\sim$2000-17000 $\textrm{\AA}$. We use multi-band mosaics drizzled to a common $0\farcs065$ pixel scale \citep{Koekemoer_2011_Drizzle}. All flux measurements are corrected for foreground reddening using the \citep{Schlegel_1998_Dust} dust maps. A single reddening correction due to dust in the Milky Way was calculated for each BCG in each filter. Additionally, we perform a background subtraction using a combination of iterative 3-sigma clipping on large scales along with a local median flux measurement in an annulus around each BCG in the UV.  
  
Early CLASH WFC3/UVIS observations are affected by non-uniform flat-fielding on scales of hundreds to thousands of pixels.  We accounted for this in observations of BCGs with faint or possibly no significant detection of UV by extracting photometry from multiple identical apertures in WFC3/UVIS images placed on empty patches of sky, and adding the scatter in these apertures to our error budget. For all CLASH observations, we apply a 3\% floor to the total photometric uncertainty to account for all sources of systematic error and absolute flux calibration uncertainty.  
  
\subsection{Chandra X-ray Observations}  
  
All the clusters in the CLASH X-ray sample have archival \textit{Chandra} data. For this work, we use the ICM parameters published in the Archive of Chandra Cluster Entropy Profile Tables (ACCEPT) \citep{Cavagnolo_2009_ACCEPT}. Gas density and temperature profiles were measured in ACCEPT, and the core entropies and cooling times used in the present study are calculated using these profiles. Gas density profiles were measured in concentric annuli 5$"$ wide. Temperature profiles, which were used in combination with the gas density profiles to derive both cooling time and entropy profiles, were measured in concentric annuli containing at least 2500 counts per bin.  
  
\subsection{Spectra}  
   
Optical spectra of the BCGs were obtained for 15 of the 20 CLASH X-ray selected clusters. Objects were observed with the Goodman High Throughput Spectrograph on the SOAR 4.1 meter telescope using either the KOSI 600 grating or the SYZY 400 grating. The KOSI grating's dispersion is approximately 0.65 \AA\, $\textrm{pix}^{-1}$, while the SYZY grating's is approximately 1.0 \AA\, $\textrm{pix}^{-1}$. Their spectral ranges are roughly 2670 \AA\, and 4000 \AA, respectively (see Table 1). Central wavelengths were selected to best include the [\ion{O}{2}] doublet ([\ion{O}{2}]~$\lambda,\lambda3726,29$), the [\ion{O}{3}] doublet (\ion{O}{3}]~$\lambda,\lambda4959,5007$), and H$\beta$ at their redshifted positions. Position angles were chosen to sample observed filamentary structures or other objects of interest. For all observations, the 1.68$''$ long slit was used.\\

Observations were reduced following \cite{Werner_2014_Feedback}. Bias correction and trimming were performed with the \texttt{IRAF}\footnote{IRAF is distributed by the National Optical Astronomy Observatory, which is operated by the Association of Universities for Research in Astronomy (AURA) under a cooperative agreement with the National Science Foundation.}  task \texttt{CCDPROC}. Quartz lamp frames, taken before and after observation images, were used to flat-field the images. Wavelength calibration was performed with FeAr arc lamp exposures, with distortion along the spatial direction corrected for by tracing the position of standard stars. Sensitivity functions were produced from same-night observations of standard stars with the 10.0$''$ long slit that, along with an extinction correction based on \texttt{IRAF}'s extinction file \texttt{ctioextinct.dat} and a correction for airmass, were used to flux calibrate observations. After background subtraction to minimize the contribution of night sky lines, observations were then median combined using \texttt{IMCOMBINE}.\\

%2014MNRAS.439.2291W
   
\begin{table*}[t]
\footnotesize
\caption{\\ SOAR Observations of Brightest Cluster Galaxies}   
\vspace{5mm}  
\centering  
{  
\begin{tabular}{lcrrrrrr}
%\tabletypesize{\scriptsize}
%\tablecaption{SOAR Observations of Brightest Cluster Galaxies}
%\tablewidth{0pt}
%\tablehead{
Cluster &   Obs Date & Exposure Times & Grating  & Range &  PA$^{a}$ & Airmass & Standard Star \\
 & (YYYY-MM-DD) & \multicolumn{1}{c}{(s)} & (l/mm) &  \multicolumn{1}{c}{\AA}  &  \multicolumn{1}{c}{($\ ^\circ$)} &  & \\ \\
\hline  
\hline  \\
%\startdata
Abell 209 					& 2012-09-24 & 1$\times$1200, 1$\times$1800 & 400 & 4514-7555 & 135 & 1.2 & LTT7379\\
Abell 383 					& 2012-10-09 & 2$\times$1200, 1$\times$600 & 400 & 4126-7568 & 0 & 1.3 & LTT1020\\
MACS0329.7$-$0211 & 2012-11-09 & 3$\times$1200 & 400 & 4612-7555 & 125 & 1.1  & LTT7379\\
MACS0429.6$-$0253	& 2012-11-19 & 2$\times$1200, 1$\times$900 & 400 & 4566-7556 & 167 & 1.3  & LTT1020\\
MACS1115.9+0219		& 2013-05-11	& 4$\times$900 & 400 & 4401-8462 & 130 & 1.2  & LTT4364\\ 
MACS1206$-$0847		& 2013-05-11	& 4$\times$900 & 400 & 4398-8457 & 100 & 1.1 & LTT4364\\
MACS1311.0$-$0310	& 2013-05-11	& 1$\times$1500, 1$\times$900 & 400 & 4412-8470 & 40 & 1.5  & LTT4364\\
MACS1423.8+2404		& 2015-02-26	& 3$\times$1200 & 600 & 5049-7724 & 0 & 1.8 & LTT4364\\
MACS1720.3+3536		& 2015-06-13	& 3$\times$1200 & 600 & 4570-7235 & 160 & 2.2  & LTT6248\\
MACS1931.8$-$2653	& 2012-04-17 & 1$\times$1800 & 600 & 4271-6938 & 252 & 1.0 & LTT7379\\
MS2137$-$2353			& 2015-06-13	& 4$\times$1200 & 600 & 4570-7235 & 145 & 1.1  & LTT6248\\
RXJ1347.5$-$1145		& 2012-07-15 & 1$\times$1200, 1$\times$900 & 600 & 4556-7223 & 125 & 1.6 & LTT9491\\
RXJ1532.9+3021   		& 2012-04-17 & 2$\times$1200 & 600 & 4811-7471 & 187 & 2.1 & LTT3864\\
RXJ2129.7+0005			& 2012-07-15 & 4$\times$1200 & 600 & 4560-7228 &  201 & 1.2 & LTT3864\\
RXJ2248.7$-$4431		& 2013-09-08 & 4$\times$1200 & 600 & 4299-6969 &  352 & 1.1 & LTT1020\\
\\
\hline
%\enddata
$^{a}${Position Angle measured east of north.}
\label{tab:CLASH_Soar_obs}
\end{tabular}
} 
\end{table*}
  
\section{Analysis}  
\subsection{Mean UV Luminosities}  
  
We calculate the UV luminosity of CLASH BCGs in order to estimate SFRs using the \cite{Kennicutt_1998_SFR} relation. To do this, we extract the flux from those CLASH filters whose pivot wavelengths fall in the range $1500-2800 \textrm{\AA}$ in the cluster rest frame, corresponding to the wavelength range used by \cite{Kennicutt_1998_SFR} to calibrate the L$_{UV}$-SFR relation.  Since the continuum L$_{UV}$ due to young stars is flat for a \cite{Salpeter_1955_IMF} IMF and continuous star formation, we use fluxes extracted from the relevant UV filters to calculate the average luminosity $\langle L_{\nu,UV}\rangle$. This quantity is one of two SFR proxies we are able to calculate from the HST photometry alone, the other being L$_{H\alpha+[NII]}$, which we discuss in Section 3.2.  
  
For the majority of the BCGs, we extract fluxes from a subset of the CLASH WFC3/UVIS filters, F225W, F275W, F336W, and F390W. For the highest redshift clusters ($z>0.542$), we use ACS filters F435W and sometimes F475W as well. Filter selections are shown in Table 2.  
  
Since we are interested in the UV flux from young stars, we must remove the contribution from the UV-upturn in the quiescent stellar population in the BCGs. The UV-upturn has been well-studied and is due to post AGB stars and extreme blue horizontal branch stars \citep[e.g][]{Brown_2003_UVUpturn, Yi_2008_UVUpturn, Ferguson_1993_UVUpturn, Yi_1998_UVHBStars}. This component of the UV flux mimics low-level star formation and would, if left uncorrected, bias our SFR estimates in UV-faint BCGs.  
  
\begin{figure}[h]  
\centering{\epsfig{file=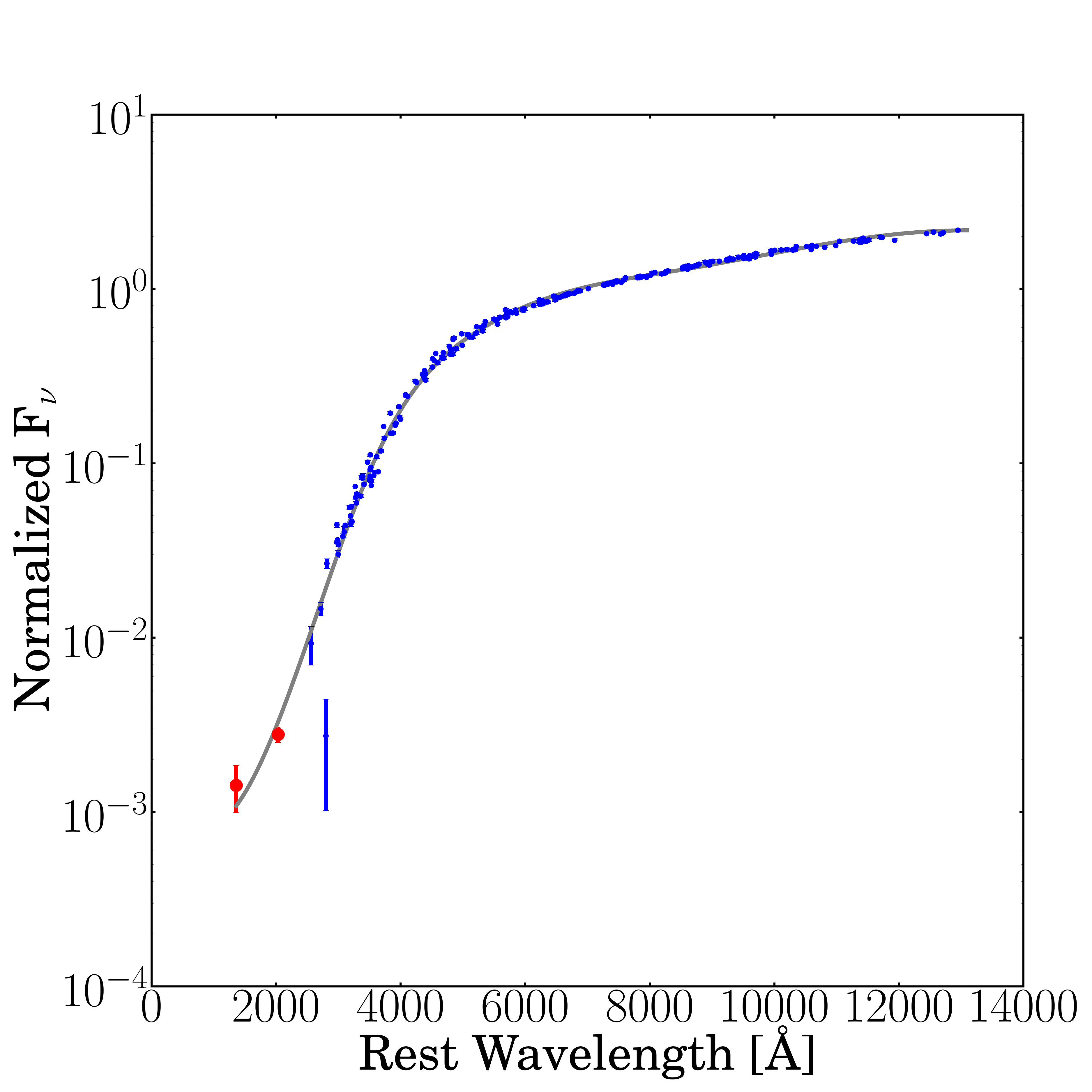,width=8cm,angle=0}}  
\caption{Rest frame SED for quiescent starlight in BCGs. The blue points are ACS/WFC3IR photometry taken from 100 satellite galaxies, and represent the averages of 5 galaxies per CLASH cluster in each of 20 clusters. The red points are derived using the GALEX-2MASS J color calibration from Hicks et al. 2010 for a sample of quiescent BCGs at an average redshift of 0.115. The FUV/NUV rest wavelengths depicted in this figure are for Abell 209 ($z=0.209$), after accounting for the average redshift of the Hicks et al. color calibration. The grey line  represents the spline fit to the data. Fluxes are reported in normalized units of $F_{\nu}$.}  
\label{fig:Template_SED}  
\end{figure}  

A template SED of the UV-upturn population is derived by averaging the normalized ACS-WFC3IR rest-frame photometry of 5 satellite galaxies in each of 20 CLASH clusters (for a total of 100 satellite galaxies). We do not include WFC3/UVIS in the satellite galaxy photometry to avoid complications in estimating the sensitivity to weak UV sources, and instead opt to use the GALEX-2MASS J colors in \cite{Hicks_2010_ClusterUVSFR} to extend the range of our data below 2000 $\textrm{\AA}_{\textrm{rest}}$. These data are fit to a fifth order spline, and the uncertainty in the fit is estimated using a Monte Carlo distribution of spline fits.  
  
We estimate the UV contribution from old stars by subtracting out a model of the underlying early-type galaxy. A 2D non-parametric model of each BCG was fit to the F160W image, which is dominated by the BCG's old stellar population. The model is scaled according to the template SED shown in Figure~\ref{fig:Template_SED} and subtracted from the UV images. The residual UV flux measured after the subtraction is our UV-upturn-corrected estimate due to ongoing star formation activity.   
  
\subsection{H$\alpha$+[\ion{N}{2}] Maps}   
  
The CLASH data allow us to estimate $H\alpha$ emission in the BCGs using a broadband subtraction technique in cases where the emission is strong relative to the stellar continuum. For each BCG, a `line' and `continuum' filter is chosen based on the cluster redshift (see Table 2). One or more satellite early-type galaxies are chosen based on having an IR$-$optical color similar to the BCG, and the mean ratio of the satellite galaxy fluxes between the line and continuum filters is used to scale the continuum filter to the line filter. We subtract the scaled continuum filter image from the line filter image, leaving a residual flux that is primarily due to H$\alpha$+[\ion{N}{2}] emission. The H$\alpha$ nebulae of clusters with significant UV emission are shown in Figure~\ref{fig:Triptychs}.  
  
This method for estimating $H\alpha$+[\ion{N}{2}] is reliable only for line emission with a large equivalent width (EW). $H\alpha$+[\ion{N}{2}] EWs may be approximated by
\begin{equation}
EW = \int{\frac{f_{\lambda} - f_{\textrm{cont}}}{f_{\textrm{cont}}}d\lambda} \approx \textrm{B}_{\textrm{line}}\frac{f_{\textrm{line}}-f_{\textrm{cont}}}{f_{\textrm{cont}}},
\end{equation}
where $f_{\textrm{line}}$ and $f_{\textrm{cont}}$ are the fluxes through the line and continuum filters, and B$_{\textrm{line}}$ is the photometric bandwidth of the line filter. If we assume a 3$\%$ uncertainty for the surface photometry in both the `line' and `continuum' filters, we can reliably recover features with $H\alpha$+[\ion{N}{2}] EWs that are 0.044$\times$B$_{\textrm{line}}$. Because the BCGs are very bright in both the `line' and `continuum' filters, so the dominant source of uncertainty on our photometry will be the absolute flux calibration. For BCGs where we adopt ACS filters (either F775W or F850LP) as the line filters, we can recover $H\alpha$+[\ion{N}{2}] features with EWs $\gtrsim$ 20 $\textrm{\AA}$ in the observer frame. For BCGs where we adopt WFC3IR filters (either F105W or F125W), we can recover features with EWs $\gtrsim$ 40 $\textrm{\AA}$ in the observer frame.   

\pagebreak
\subsection{Other Emission Lines}  
  
Longslit spectra provide coverage of the [\ion{O}{2}], [\ion{O}{3}]~$\lambda,\lambda4959,5007$, and H$\beta$ emission lines. We measure line luminosities by fitting a Gaussian line profile and a continuum to reduced, 1-D longslit spectra using the IRAF task \textbf{splot}. Continuum levels in \textbf{splot} are identified by averaging regions of continuum emission adjacent to emission lines, and continuum-subtracted Gaussian line profiles are fit using the default iterative Levenberg-Marquardt algorithm. An [\ion{O}{2}] luminosity could not be calculated for Abell 383, owing to contamination of the spectrum at $\sim$ 4424 $\textrm{\AA}$. [\ion{O}{3}] $\lambda$ 5007 is unavailable for RXJ1347.5-1145 since the available spectrum does include this line.   
  
[\ion{O}{2}] luminosities are an independent check on the SFR derived from our UV and H$\alpha$ luminosities. While [\ion{O}{2}] luminosities are usually considered to be a less reliable estimator of SFRs due to their dependence on ionization and metallicity, they can provide extra constraining power when used in conjunction with other SFR estimators \citep{Charlot_2001_SFRLines, RosaGonzalez_2001_SFRCalibrations, Kewley_2004_SFROII, Moustakas_2006_SFRIndicators}. We can use H$\beta$ luminosities in a similar fashion. Assuming case B recombination, we can derive an SFR from H$\beta$, using the relation H$\alpha$/H$\beta$ = 2.85  \citep{Veilleux_2002_Spectro}.  
  
[\ion{O}{3}] $\lambda$ 5007 is useful for constructing diagnostic diagrams to separate regions heated by normal stars versus other sources of ionization like AGN and shocks. The classic diagram for distinguishing star forming regions from AGNs is the BPT diagram \citep{Baldwin_1981_BPT, Kewley_2001_BPT, Kauffmann_2003_BPT}. We construct a modified version of the BPT diagram based on a combination of spectral and broadband data. We use an additional diagnostic, when possible, that is nearly insensitive to extinction, and only uses our spectroscopic data. The so-called `blue-line' diagram barely depends on the accuracy of the reddening correction because it is derived solely from equivalent width values. In this work, we use the `blue-line' diagram derived from the [\ion{O}{2}], [\ion{O}{3}], and H$\beta$ lines \citep{Lamareille_2010_BlueLines, Lamareille_2004_BlueLines}. Specifically, this diagram compares the ratios of equivalent widths [\ion{O}{3}] $\lambda$ 5007/H$\beta$ to [\ion{O}{2}]/H$\beta$ and we can measure these ratios directly from the SOAR spectra.     

%\begin{figure}[h]  
%\centerline{\epsfig{file=Reddening_Procedure_RXJ1532.pdf,width=8cm,angle=0}}  
%\caption[]{Illustration of the method for extracting reddening information from UV filters, for RXJ1532.9+3021. The final reddening map is shown in the right. The layout of grid squares used to calculated E(B-V) is shown overlaid on the F275W filter on the left.}  
%\label{fig:Reddening_Proc}  
%\end{figure} 

\begin{figure*}[t!]
\centering
{\epsfig{file=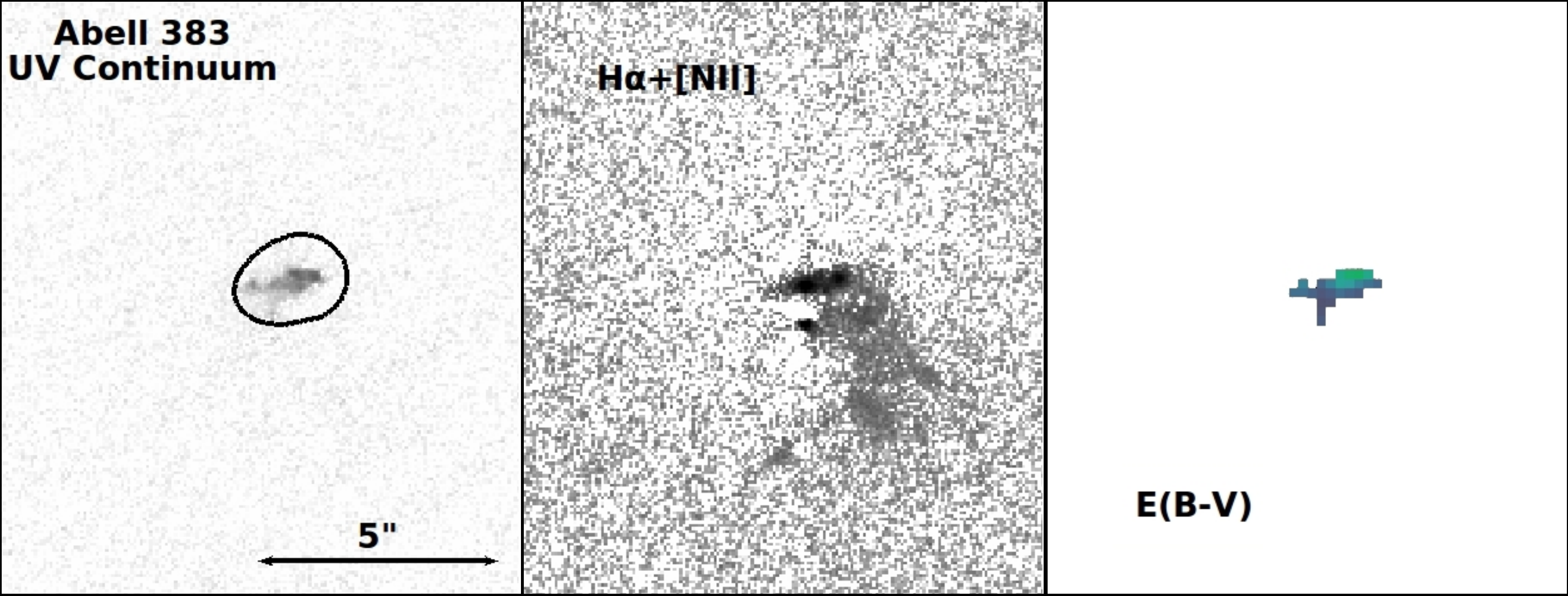,width=14cm,angle=0}
\label{fig:subfigure1}}
{\epsfig{file=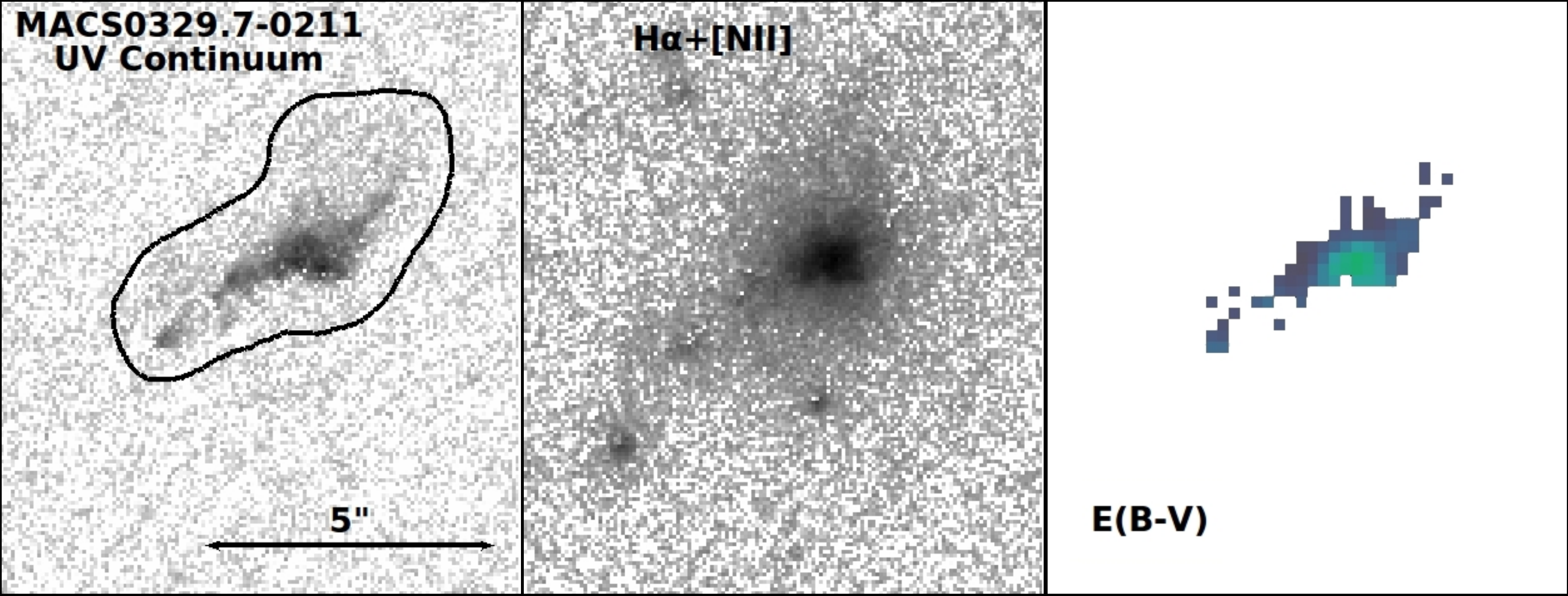,width=14cm,angle=0}
\label{fig:subfigure2}}
{\epsfig{file=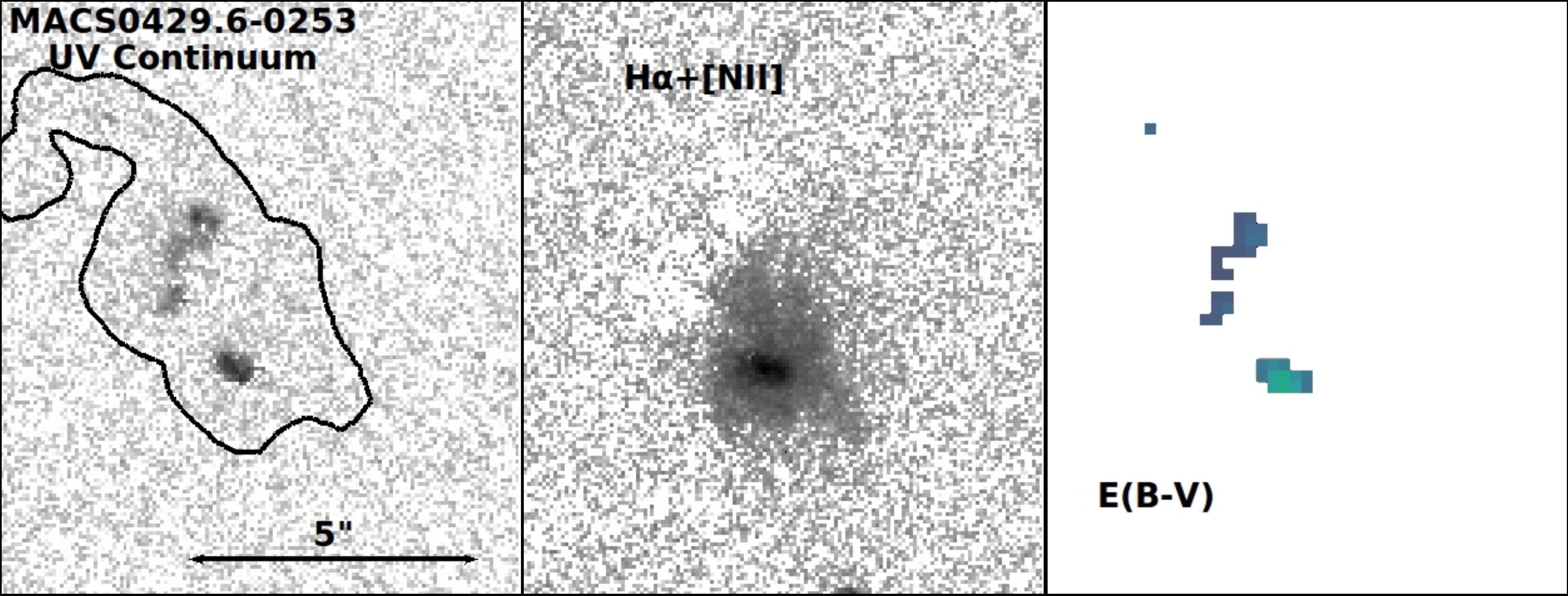,width=14cm,angle=0}
\label{fig:subfigure}}
\caption{The reddening corrected UV luminosity  map for CLASH BCGs are shown in the left panel. The region used to extract L$_{UV}$ for each BCG is outlined in black. Likewise, the H$\alpha$ images are shown in the center panel. The estimated Calzetti reddening map is in the rightmost panel, in units of $E(B-V)$. The uniform color code in the $E(B-V)$ ranges from 0.0 (black) to 0.8 (red), with typical values between 0.1 and 0.5. White regions in the $E(B-V)$ maps have insufficient UV flux to perform the reddening estimate. Galaxies shown are Abell 383, MACS0329, MACS0429, MACS1115, MACS1423, MACS1720, MACS1931, MS2137, RXJ1347, RXJ1532, and RXJ2129, in that order.}
\end{figure*}
\begin{figure*}
\ContinuedFloat
\centering
{\epsfig{file=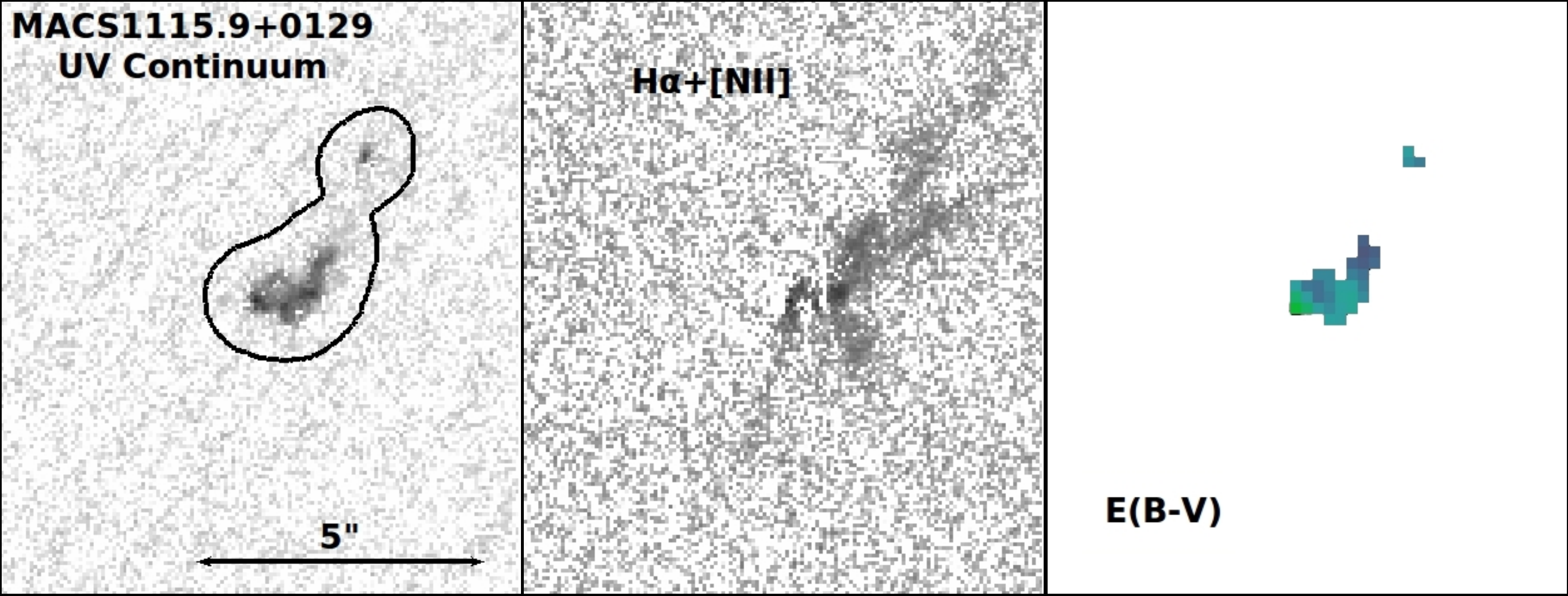,width=14cm,angle=0}
\label{fig:subfigure}}
{\epsfig{file=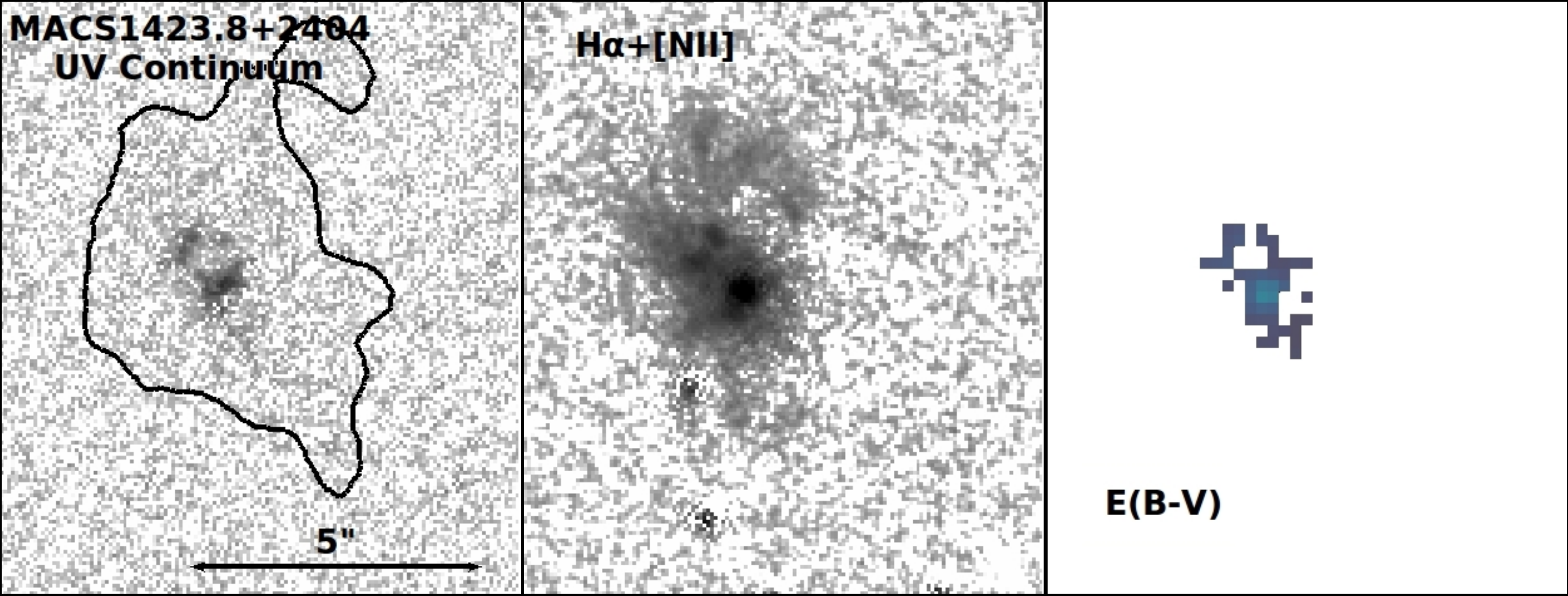,width=14cm,angle=0}
\label{fig:subfigure}}
{\epsfig{file=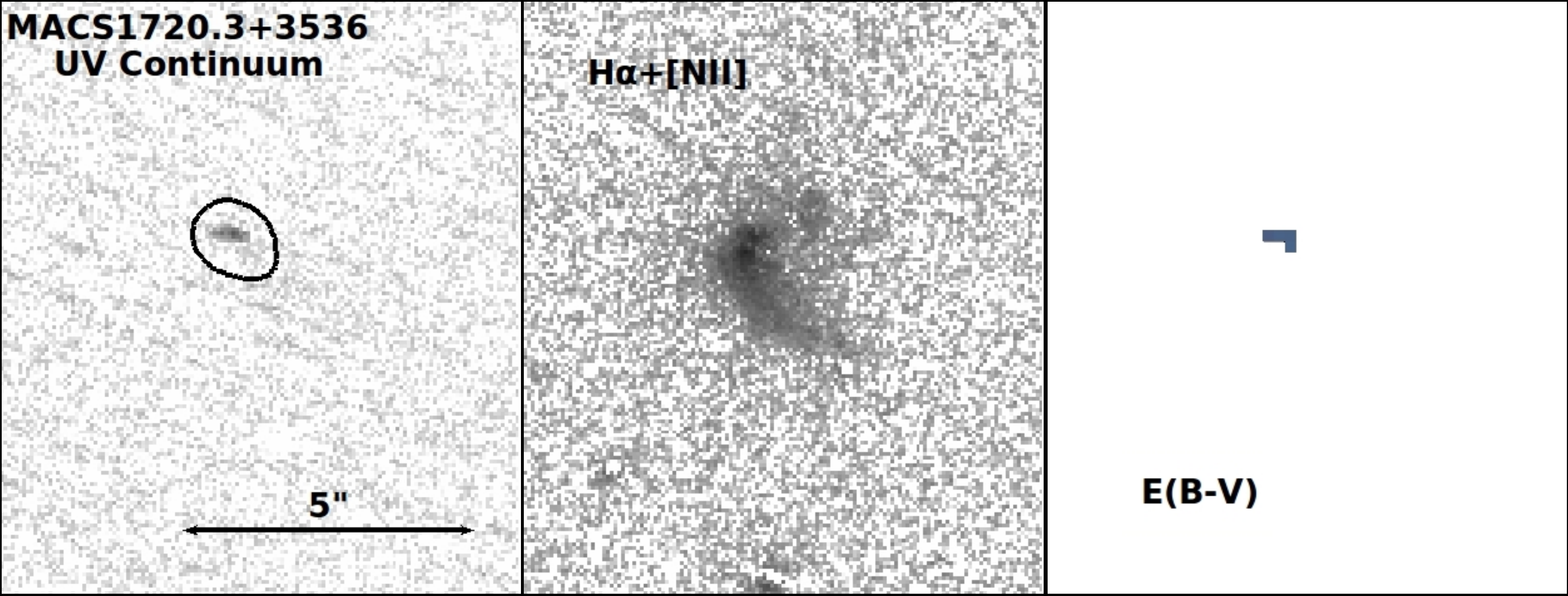,width=14cm,angle=0}
\label{fig:subfigure}}
{\epsfig{file=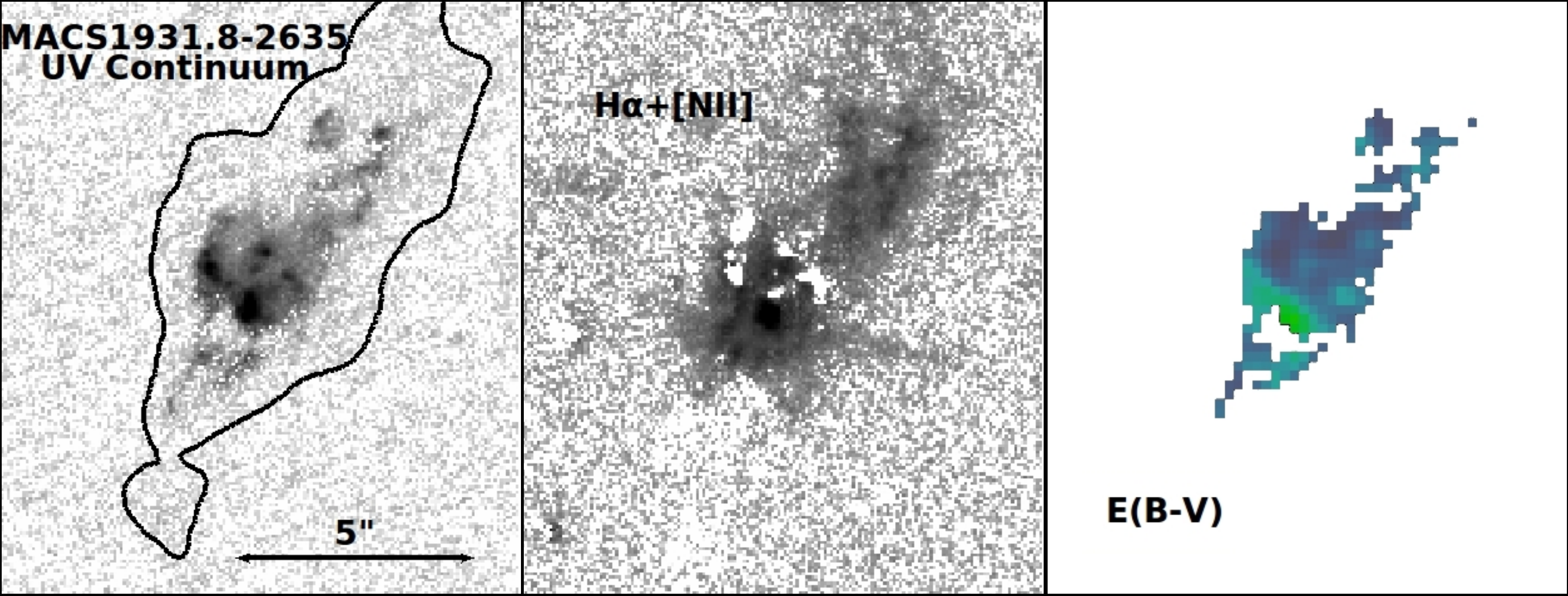,width=14cm,angle=0}
\label{fig:subfigure}}
\caption{\textit{Continued}}
\end{figure*}
\begin{figure*}
\ContinuedFloat
\centering
{\epsfig{file=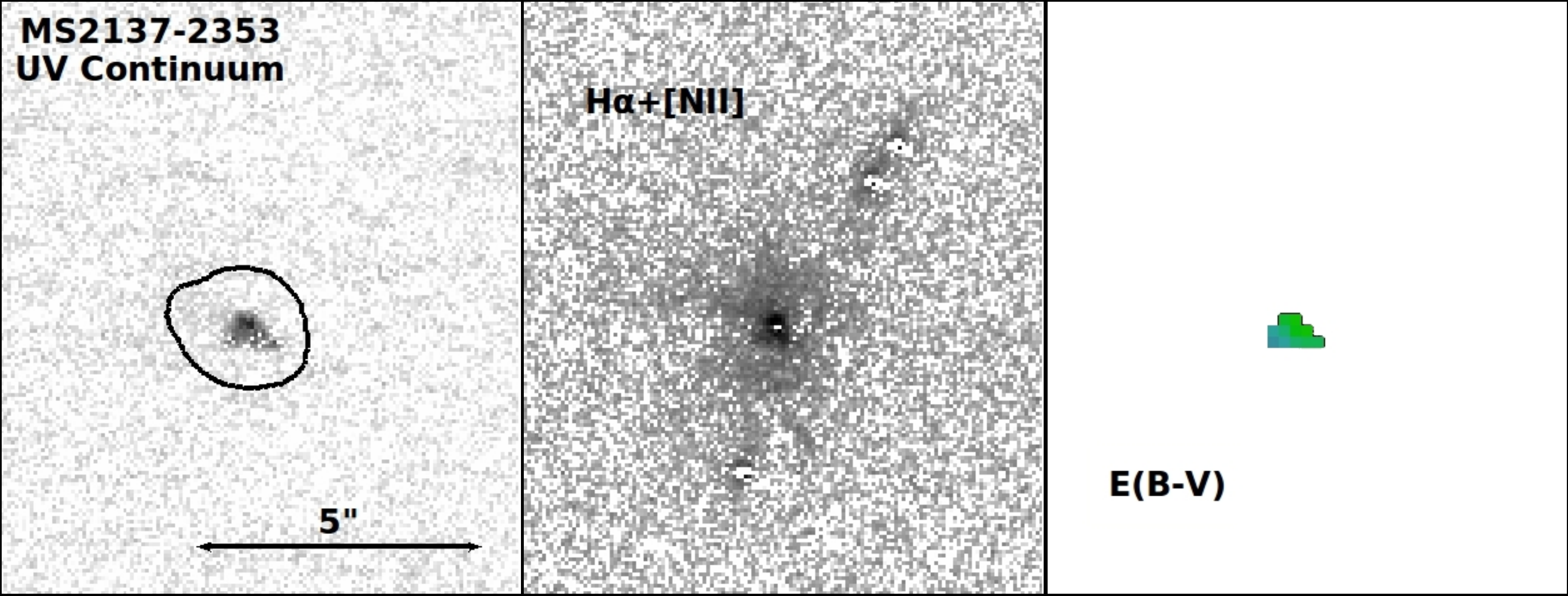,width=14cm,angle=0}
\label{fig:subfigure}}
{\epsfig{file=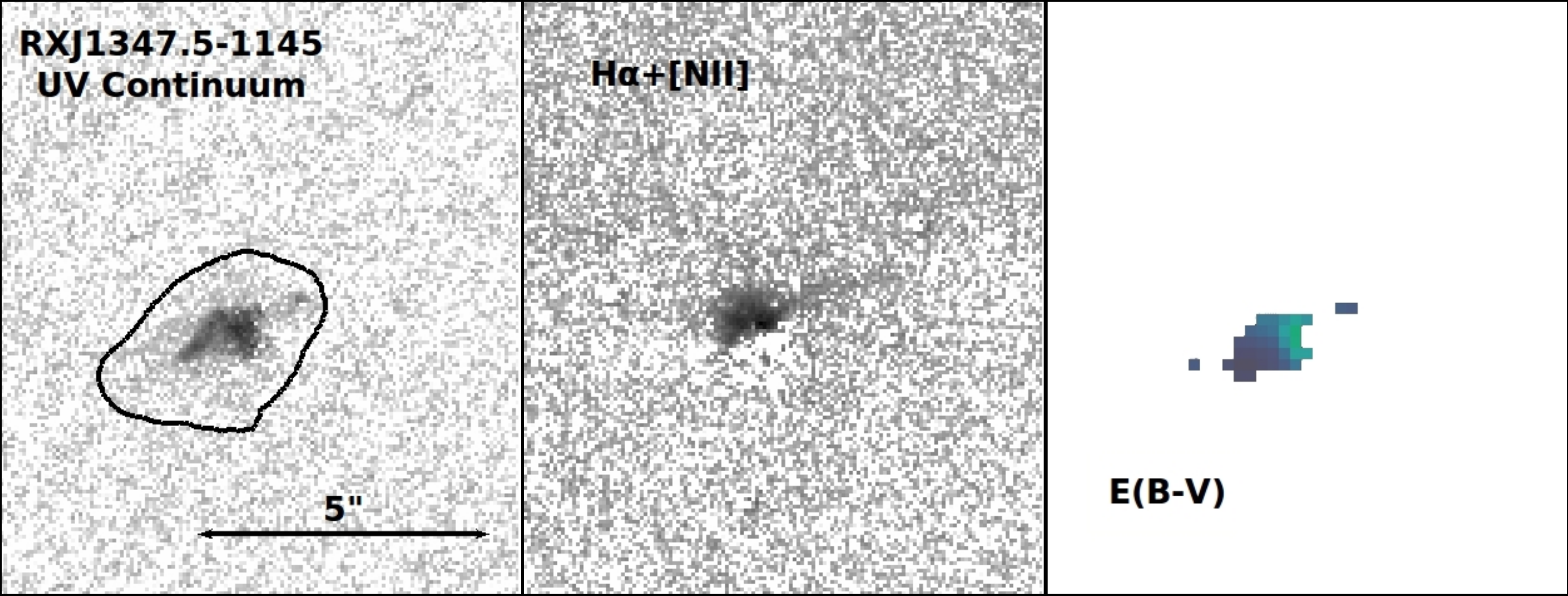,width=14cm,angle=0}
\label{fig:subfigure}}
{\epsfig{file=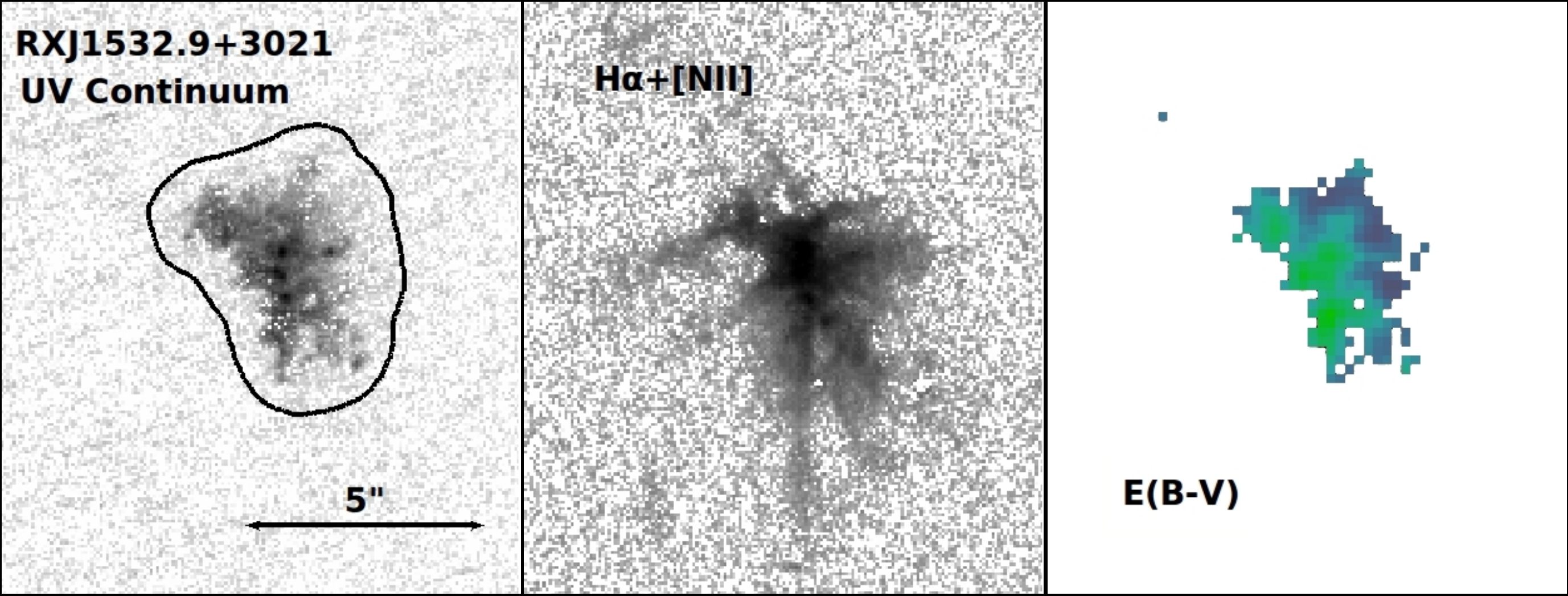,width=14cm,angle=0}
\label{fig:subfigure}}
%\caption{\textit{Continued}}
%\end{figure*}
%\begin{figure*}
%\ContinuedFloat
%\centering
{\epsfig{file=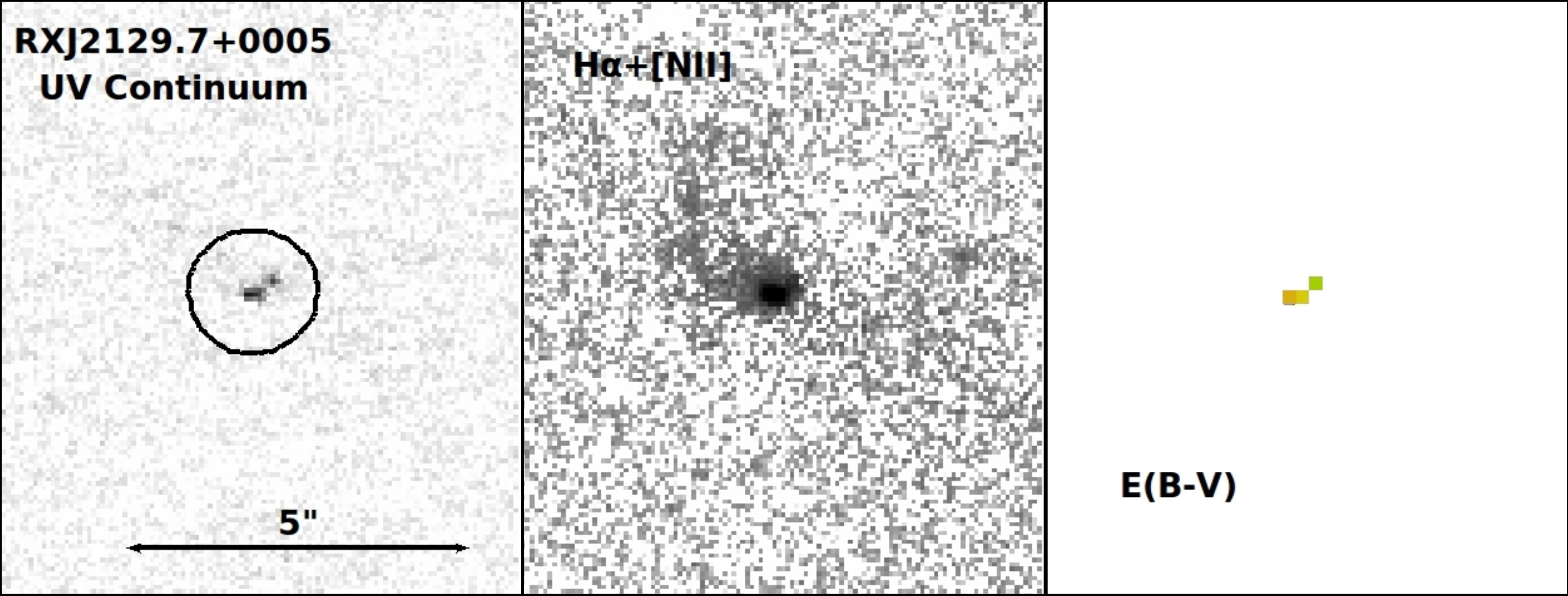,width=14cm,angle=0}
\label{fig:subfigure}}
\caption{\textit{Continued}}
\label{fig:Triptychs}
\end{figure*}

\subsection{Reddening Correction} 
  
HST and SOAR observations (when stated explicitly) are corrected for intrinsic dust reddening in a manner similar to \cite{McDonald_2013_Phoenix}. Assuming continuous star formation, the young-stellar continuum $L_{\nu}\left(\nu\right)$ in the rest wavelengths $1500-2800 \textrm{\AA}$ is flat, and we can thus construct reddening maps for the CLASH BCGs by assuming the gradual slope in flux we observe between pairs of UV images is due to mild dust extinction. In most cases we select the F275W and F336W filter to calculate the reddening. Filter selections for individual BCGs are given in Table 2. Even though this procedure assumes continuous star formation, the effect of SFH is relatively small-- a starburst with a finite burst duration above 10 Myr will have a slope that differs from a model with continuous star formation by $\lesssim 5\%$.   
  
\begin{figure}[h]  
\centerline{\epsfig{file=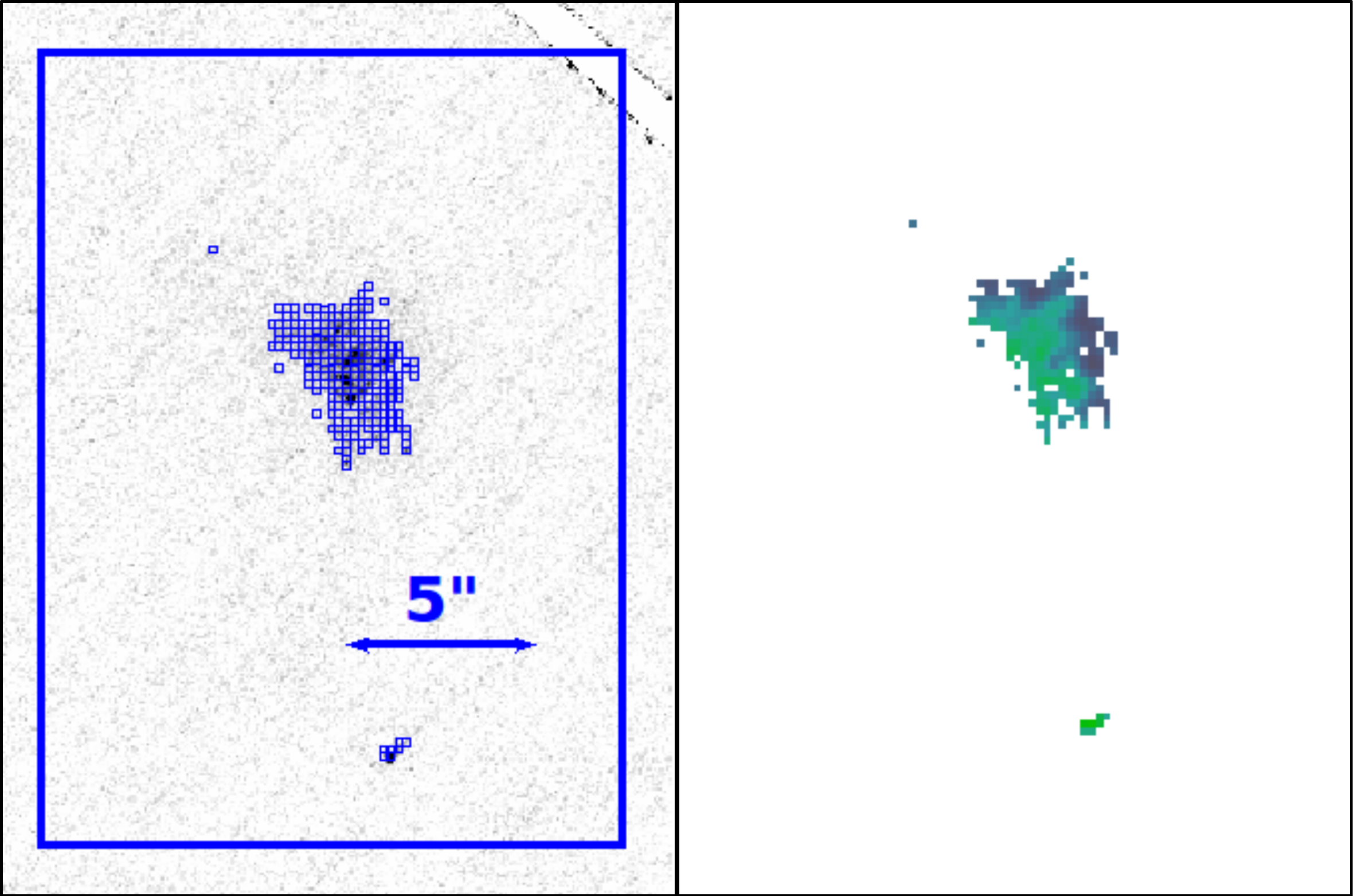,width=8cm,angle=0}}  
\caption[]{Illustration of the method for extracting reddening information from UV filters, for RXJ1532.9+3021. The final reddening map is shown in the right. The layout of grid squares used to calculated E(B-V) is shown overlaid on the F275W filter on the left.}  
\label{fig:Reddening_Proc}  
\end{figure}

We construct spatially resolved reddening estimates by covering the UV emission features in each BCG with grid squares 0.195$"$ on a side, corresponding to 3$\times$3 pixels in the drizzled images. Grid squares of this size are a compromise between sensitivity and spatial resolution, and make our corrections directly comparable to the reddening correction employed in \cite{McDonald_2013_Phoenix}. Squares which do not have a minimum of 5$\sigma$ of UV flux in the two filters used to calculate $E\left(B-V\right)$ are rejected, leaving behind a grid covering just the significant UV emission, like the example given in Figure~\ref{fig:Reddening_Proc}.   
  
We calculate two reddening maps for each BCG, one using a \cite{Calzetti_2000_Extinction} extinction curve, and one using the Milky Way dust curve parameterized in \cite{ODonnell_1994_MWDust}. When using the Milky Way dust model, we avoid the effect of the 2175 $\textrm{\AA}$ bump by both subtracting the best-fit model of the bump in \cite{Fitzpatrick_1986_MWBump} and by choosing UV filters that minimize coverage of the bump. In each grid square, $E\left(B-V\right)$ is calculated by solving for the dust extinction necessary to flatten the slope between the UV-upturn-corrected fluxes in the two filters. The resulting reddening maps are Gaussian smoothed with a 0.195$"$ kernel, in order to blur out the effects of binning.  
  
Out of the 20 clusters examined from the CLASH sample, 11 have sufficient UV flux to estimate intrinsic reddening over at least part of the BCG. Images of the estimated intrinsic reddening maps are shown in the rightmost panel for each of these clusters in Figure~\ref{fig:Triptychs}. We find that, with few exceptions, $E\left(B-V\right) \lesssim 0.5$, which is consistent with the dust content typical of cluster BCGs \citep{Crawford_1999_BCS, McDonald_2011_UV}.   
  
Non-resolved dust extinctions for UV-faint BCGs and faint regions in UV-bright BCGs are calculated using F140W-IRAC colors assuming an underlying early-type stellar population. These bands are suitable for estimating extinction in these galaxies and regions since the F140W band is slightly extinguished at the rest wavelengths of CLASH BCGs while the IRAC bands are essentially reddening free. Furthermore, the F140W-IRAC color at these redshifts is insensitive to the SFH, so the assumed underlying population does not affect the resulting dust estimate. Spitzer/IRAC 3.6$\mu$m and 4.5$\mu$m fluxes are available for all CLASH clusters except Abell 1423 \citep{Moustakas_IP}. We use fluxes measured in 3.0$"$ diameter apertures and apply the aperture corrections used in \cite{Sanders_2007_COSMOS}, selecting either the 3.6 $\mu$m and 4.5 $\mu$m band for each BCG separately in order avoid the polycyclic aromatic hydrocarbon feature at 3.3 $\mu$m rest-frame. 

Detections of H$_{2}$ vibrational modes are also prevalent in the IR between 5-25$\mu$m in starforming BCGs \citep{Donahue_2011_BCGH2}.  We note that the presence of these lines in IRAC filters may affect our estimate of the dust reddening; however, we are mostly relying on this estimate of the reddening in BCGs with little evidence of ongoing star formation where we would not expect there to be wide vibrational H$_{2}$ lines.   
  
We use the spatially resolved extinctions to correct observed fluxes in UV-bright structures and the non-resolved extinctions to correct fluxes outside these structures and in UV-faint BCGs. To correct line luminosities, we adopt the relation $E(B-V)_{*} = 0.44E(B-V)_{\textrm{gas}}$ reported in \cite{Calzetti_2000_Extinction}. While this is the empirically observed relation between the extinction of nebular and stellar emission in starburst galaxies, our choice of extinction model will introduce a systematic uncertainty, since starburst BCGs may differ from the starburst galaxies used to calibrate this relationship.  
  
The reddening correction multiplies our values for L$_{UV}$ in UV luminous BCGs by a factor of $\sim2-5$, which is consistent with what \cite{Donahue_2015_IP} expected, given the typical dust content of an active cool core BCG. For example, \cite{Donahue_2015_IP} reports an unobscured SFR for RXJ1532 of $\sim 40$ M$_{\odot}$ yr$^{-1}$, which is less than the Herschel-estimated value of $\sim 100$ M$_{\odot}$ yr$^{-1}$. The reddening corrected UV SFR for RXJ1532 we find is $97\pm4$ M$_{\odot}$ yr$^{-1}$ (only accounting for statistical uncertainty). For this BCG, the reddening corrected UV SFR yields a rate comparable to the IR estimate. 
  
\begin{table*}[]  
\footnotesize  
\caption{\\ UV and H$\alpha$ Filters for BCG Luminosities}  
\vspace{5mm}  
\centering  
{  
%Name, Redshift, UV filters, E(B-V) filters, UV flux aperture selection, Ha 'cont', Ha 'line'  
\begin{tabular}{lcp{2.5cm}ccc}  
Cluster & z$^{a}$ & \parbox{2.5cm}{\centering UV filters} & E(B-V) filters & H$\alpha$ `continuum' & H$\alpha$ `line' \\  
\hline  
\hline \\  \vspace{1mm}
Abell 209 & 0.209 & \parbox{2.5cm}{\centering F225W, F275W \\ F336W} & --- & F850LP & F775W \\ \vspace{1mm}  
Abell 383 & 0.187 & \parbox{2.5cm}{\centering F225W, F275W} & F225w, F275w & F850LP & F775W \\ \vspace{1mm}  
Abell 611 & 0.288 & \parbox{2.5cm}{\centering F225W, F275W\\ F336W} &  --- & F775W & F850LP \\ \vspace{1mm}  
Abell 1423 & 0.213 & \parbox{2.5cm}{\centering F225W, F275W \\ F336W} & --- & F850LP & F775W \\ \vspace{1mm}  
Abell 2261 & 0.224 & \parbox{2.5cm}{\centering F225W, F275W\\ F336W} & ---  & F625W & F775W \\ \vspace{1mm}  
MACS0329.7$-$0211 & 0.450 & \parbox{2.5cm}{\centering F225W, F275W\\ F336W, F390W} & F275W, F390W & F775W & F850LP \\ \vspace{1mm}  
MACS0429.6$-$0253 & 0.399 & \parbox{2.5cm}{\centering F225W, F275W\\ F336W} & F275W, F336W & F775W & F850LP \\ \vspace{1mm}  
MACS0744.9+3927 & 0.686 & \parbox{2.5cm}{\centering F275W, F336W\\ F390W, F435W} & --- & F140W & F105W \\ \vspace{1mm}  
MACS1115.9+0129 & 0.352 & \parbox{2.5cm}{\centering F225W, F275W\\ F336W} & F275W, F336W & F775W & F850LP \\ \vspace{1mm}  
MACS1206.2$-$0847 & 0.440 & \parbox{2.5cm}{\centering F225W, F275W\\ F336W, F390W} & F275W, F390W & F775W & F850LP \\ \vspace{1mm}  
MACS1423.8+2404 & 0.545 & \parbox{2.5cm}{\centering F225W, F275W\\ F336W, F390W\\F435W} & F275W, F390W & F125W & F105W \\ \vspace{1mm}  
MACS1720.3+3536 & 0.391  & \parbox{2.5cm}{\centering F225W, F275W\\ F336W} & --- & F775W & F850LP \\ \vspace{1mm}  
MACS1931.8$-$2635 & 0.352 & \parbox{2.5cm}{\centering F225W, F275W\\ F336W} & F275W, F336W & F775W & F850LP \\ \vspace{1mm}  
MS2137$-$2353 & 0.313 & \parbox{2.5cm}{\centering F225W, F275W\\ F336W} & F275W, F336W & F775W & F850LP \\ \vspace{1mm}  
RXJ1347.5$-$1145 & 0.451 & \parbox{2.5cm}{\centering F225W, F275W\\ F336W, F390W} & F275W, F390W & F775W & F850LP \\ \vspace{1mm}  
RXJ1532.9+3021 & 0.363$^{b}$ & \parbox{2.5cm}{\centering F225W, F275W\\ F336W} & F275W, F336W & F775W & F850LP \\ \vspace{1mm}  
RXJ2129.7+0005 & 0.235 & \parbox{2.5cm}{\centering F225W, F275W\\ F336W} & F275W, F336W & F850LP & F775W \\ \vspace{1mm}  
RXJ2248.7$-$4431 & 0.348 & \parbox{2.5cm}{\centering F225W, F275W\\ F336W} & --- & F775W & F850LP \\ \vspace{1mm}  
CLJ1226.9+3332 & 0.890 & \parbox{2.5cm}{\centering F336W, F390W\\ F435W, F475W} & --- & F105W & F125W \\ %\vspace{1mm}  
MACS1311-0310 & 0.499 & \parbox{2.5cm}{\centering F225W, F275W\\ F336W, F390W} & --- & F125W & F105W \\ \\
\\
\hline 
\end{tabular}  
\begin{flushleft}  
$^{a}$ Redshifts are the same as those quoted in \cite{Postman_2012_CLASH}, unless otherwise stated. \\
$^{b}$ \citep{Crawford_1999_BCS}
\end{flushleft}  
}  
\end{table*}

\subsection{Broadband Aperture Photometry}  
  
To extract photometry for estimating the mean UV luminosities $\langle L_{\nu, UV}\rangle$ in UV-bright BCGs, we chose regions that contain all of the substantial UV flux with the aid of the ds9\footnote{http://ds9.si.edu} contour tool (see Figure~\ref{fig:Triptychs}).  These regions trace surface brightness contours of 7.14$\times$10$^{24}$ erg s$^{-1}$ Hz$^{-1}$ pix$^{-2}$, corresponding to a star formation rate surface density of $10^{-3}$ M$_{\odot}$ yr$^{-1}$ pix$^{-2}$. For the ten UV-faint BCGs (those lacking sufficient UV flux to make a spatially resolved estimate of reddening), we measured fluxes inside circular apertures that were as large as possible but still excluded satellite galaxies. We opted for this strategy for measuring photometry because it makes it straightforward to capture the flux in the highly irregular morphologies present in the CLASH BCGs. We calculated H$\alpha$+[\ion{N}{2}] luminosities using the same apertures selected for significant UV flux.  
    
Line luminosities measured in our spectra were measured in rectangular apertures that, while similar in area in most cases to the apertures we used to measure broadband luminosities, nonetheless do not cover the same regions of the BCGs that were included in our calculations of L$_{UV}$ and L$_{H\alpha+[NII]}$. Therefore, when comparing line luminosities obtained from SOAR spectroscopy  with broadband luminosities, it was necessary to measure the broadband luminosities in the apertures corresponding to the position and shape of the slit.  Since our reduced spectra are 1-D, we used the mean value of $E(B-V)$ in each rectangular aperture to estimate the extinction correction for both the spectral line luminosities and the UV and H$\alpha$+[\ion{N}{2}] luminosities we compare them to.  

%\pagebreak

\subsection{Stellar Population Properties}  
  
We use {\tt iSEDfit} to calculate the probability distributions for model approximations of the SED in RXJ1532.9+3021, either in single apertures or in individual pixels to create stellar parameter maps. See \cite{Moustakas_2013_iSEDFIT} for details on {\tt iSEDfit}. SEDs are composed of fluxes extracted from identical apertures (or individual pixels), in each of the 16 bands of CLASH HST photometry.  We do not correct for fluxes for intrinsic reddening, since {\tt iSEDfit} allows for local extinction to be treated as a fit parameter.
  
\begin{figure}[h]  
\centering{\epsfig{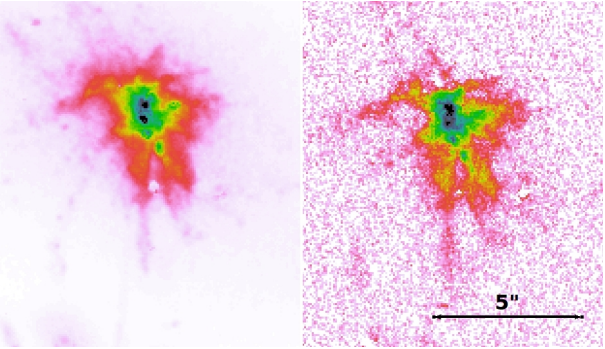}}  
\caption{\textit{Left:} Map of the H$\alpha$+[\ion{N}{2}] luminosity constructed using {\tt iSEDfit}. \textit{Right:} H$\alpha$+[\ion{N}{2}] luminosity map estimated by scaling F850LP and F775W images. We juxtapose the two to demonstrate the morphological similarity between them. The structures in our starburst maps in Figure~\ref{fig:Par_Maps} appear to be dominated by the filamentary structure in the H$\alpha$+[\ion{N}{2}] image.}  
\label{fig:Ha_Map}  
\end{figure}

When creating parameter maps using SEDs fit on individual pixels, we PSF-match each image to the F160W PSF and extract SED for each pixel using the PSF-matched photometry. For approximately normally distributed parameters, we assigned the mean values for the posterior probability density functions (PDF) for model SED parameters for each pixel SED to the locations of pixels in order to create maps of the BCG. The model predicted rest-frame flux for H$\alpha$+[\ion{N}{2}] provides a sanity check on the physical parameter maps for the 16 band SED, since we can compare its morphology to the L$_{H\alpha+[NII]}$ map described in Section 3.2. The two images are shown in Figure~\ref{fig:Ha_Map} with matching coordinates,  where it can be seen that the filamentary features in the L$_{H\alpha+[NII]}$ map correspond to the features derived from the SED. The two images depict the same pair of H$\alpha$ `bulges' in the center of the BCG as well.  
  
\section{Results}  

\subsection{Broadband Luminosities and Star Formation Rates}   
  
\begin{figure}[h]  
\centering{\epsfig{file=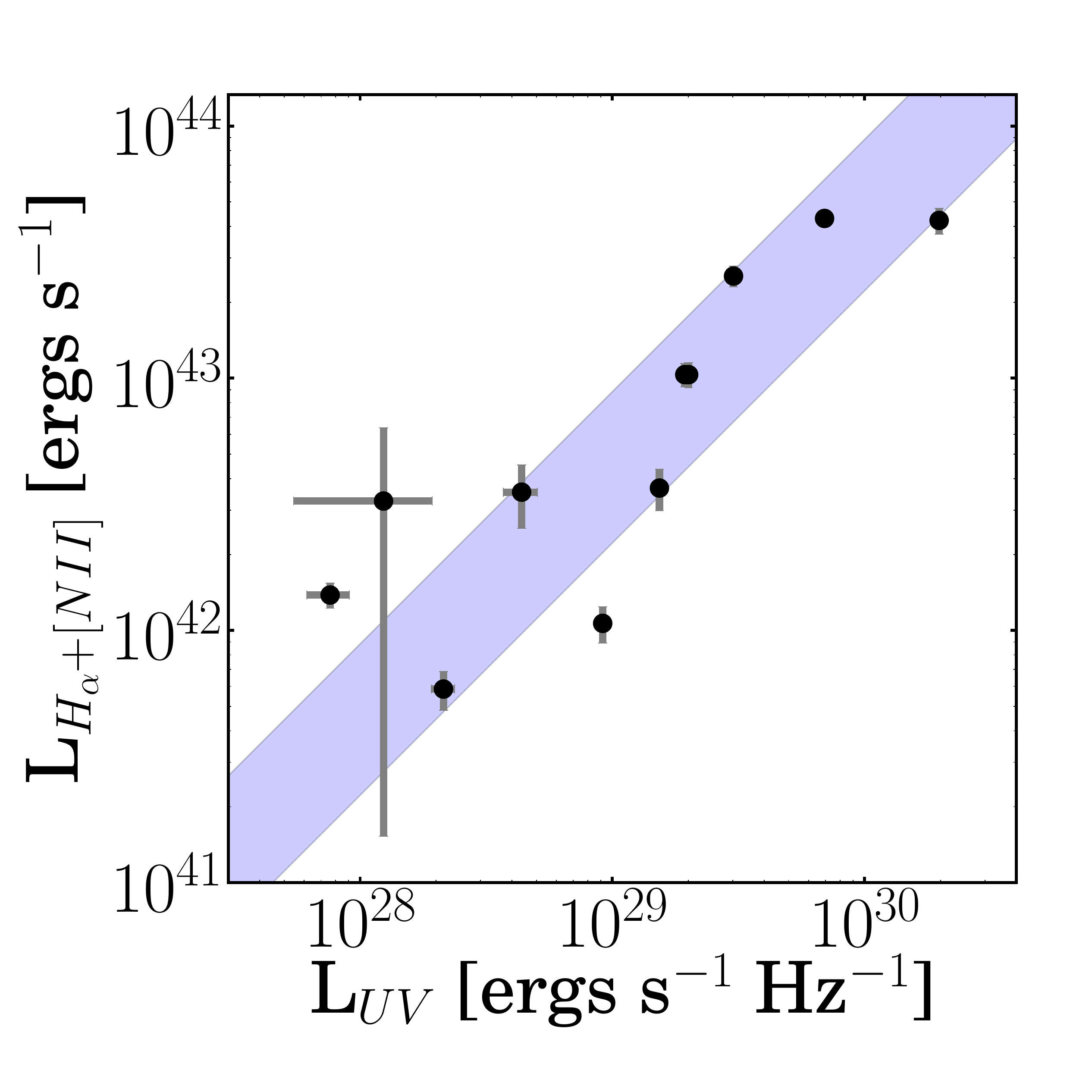,width=8cm,angle=0}}  
\caption{L$_{UV}$ and L$_{H\alpha+[NII]}$ are plotted for the regions tracing UV emission. The luminosities are corrected for dust obeying a Calzetti reddening curve. The light blue band shows the region corresponding to where the two luminosities predict the same SFR according to the Kennicutt SFR calibrations, to within the 0.3 dex scatter in the Kennicutt UV calibration. Absent error bars are too small to see as plotted.}  
\label{fig:UV-Halpha}  
\end{figure}  
  
Mean UV and H$\alpha$+[\ion{N}{2}] luminosities are given in Table 3. We present the reddening corrected $\langle L_{UV}\rangle$ both for a Calzetti reddening law and for Milky Way-type dust. Luminosities are converted to SFRs using the Kennicutt SFR calibrations. There are several sources of potential scatter in this estimate, including contamination of the UV luminosity by AGN activity, and variations in the IMF and SFH of the stellar population (\cite{Kennicutt_1998_SFR} assumes a continuous SFH and a \cite{Salpeter_1955_IMF} IMF). For the L$_{UV}$ based SFR estimates, we have not attempted to correct for these effects. However, the UV features we observe are not likely to be due to AGN activity, which is ruled out by the complicated UV morphology reported in \cite{Donahue_2015_IP}.  

In order to estimate H$\alpha$ based SFRs using \cite{Kennicutt_1998_SFR}, we need to estimate the ratio [\ion{N}{2}]/H$\alpha$. The line ratio of [\ion{N}{2}] to H$\alpha$ can vary between BCGs, and within filamentary structures in BCGs, so whatever choice we adopt will add scatter to our estimate of the H$\alpha$ based SFR \citep{McDonald_2009_A1795, McDonald_2014_A1795, Crawford_1999_BCS}. This ratio is typically 0.5 for optical galaxies \citep{Kennicutt_1992_Lines, Kewley_2001_Starburst, Kewley_2004_SFROII}. However, the ratio [\ion{N}{2}]/H$\alpha$ is often larger than 0.5 in BCGs \citep{Heckman_1989_BCGLines, Crawford_1999_BCS}. For H$\alpha$ luminous BCGs in the X-ray selected sample of clusters in \cite{Crawford_1999_BCS}, the typical [\ion{N}{2}] $\lambda6584$/H$\alpha$ is $1.1\pm0.4$. We adopt this value in order to calculate SFRs using L$_{H\alpha+[NII]}$, bearing in mind that this is a rough approximation with considerable scatter. Nonetheless, we believe that variation in [\ion{N}{2}]/H$\alpha$ is a secondary consideration for the purposes of estimating SFRs, given the scatters in the \cite{Kennicutt_1998_SFR} calibrations between L$_{H\alpha}$ and SFR and between L$_{UV}$ and SFR.

\begin{table*}[t!]  
\footnotesize  
\caption{\\ UV and H$\alpha$ Filters for BCG Luminosities}  
\vspace{5mm}  
\centering  
{  
%Name, Calzetti UV Luminosity in the UV Contour region, Galactic UV Luminosity in the UV Contour region, Ha Contour Region UV, UV  Contour region UV, Ha Contour region Ha, Calzetti UV SFR, Calzetti HA SFR   
\begin{tabular}{lrrrrrrr}  
Cluster & L$_{UV}$ & L$_{UV}$ & L$_{H\alpha+[NII]}$  & UV SFR$^{a}$ & H$\alpha$ SFR$^{a}$ & Area$^{c}$ \\  
&(Cal)$^{a}$ & (MW)$^{b}$  & (Cal)$^{a}$  & & & \\  
& $10^{27}$ ergs s$^{-1}$ Hz$^{-1}$ &  & $10^{41}$ ergs s$^{-1}$ & M$_{\odot}$ yr$^{-1}$ & & kpc$^{2}$ \\  
\hline  
\hline  
Abell 209 & $0.1\pm0.3$ & $0.1\pm0.3$ & -- & $0.01\pm0.04$ & -- & -- \\  
Abell 383 & $21.4\pm2.3$ & $37.2\pm3.3$ & $5.9\pm1.0$ & $3.0\pm0.3$ & $1.9\pm0.6$ & 14.26$\pm$0.02 \\  
Abell 611 & $<0.3^{d}$ & $<0.3$ & -- & $<0.04$ & -- & -- \\  
Abell 1423 &  $0.5\pm0.2$ & $0.6\pm0.3$ & -- & $0.07\pm0.03$ & -- & -- \\  
Abell 2261 & $0.2\pm0.2$ & $0.2\pm0.2$ & -- & $0.02\pm0.02$ & -- & -- \\  
MACS0329.7$-$0211 &  $302.3\pm12.1$ & $349.9\pm12.6$ & $254.0\pm24.1$ & $42\pm2$ & $80\pm21$ & 173.9$\pm$0.2 \\  
MACS0429.6$-$0253 &  $200.1\pm11.2$ & $234.2\pm10.6$ & $103.3\pm11.8$ & $28\pm2$ & $33\pm9$ & $72.2\pm0.2$ \\  
MACS0744.9+3927 &  $4.2\pm0.6$ & $4.6\pm0.7$ & -- & $0.6\pm0.1$ & -- & -- \\  
MACS1115.9+0129 &  $91.6\pm6.5$ & $92.8\pm5.5$ & $10.7\pm1.7$ & $13\pm1$ & $3.4\pm1.0$ & $52.32\pm0.05$ \\  
MACS1206.2$-$0847 &  $20.8\pm4.4$ & $25.4\pm3.9$ & -- & $2.9\pm0.6$ & -- & -- \\  
MACS1423.8+2404 &  $193.7\pm8.2$ & $212.4\pm8.3$ & $103.3\pm11.1$ & $27\pm1$ & $33\pm9$ & $121.2\pm0.2$ \\  
MACS1720.3+3536 &  $7.6\pm1.5$ & $22.5\pm5.7$ & $13.8\pm1.6$ & $1.1\pm0.2$ & $4.4\pm1.2$ & $10.08\pm0.02$ \\  
MACS1931.8$-$2635 &  $1975\pm135$ & $1756\pm79$ & $422.1\pm49.5$ & $280\pm20$ & $130\pm40$ & $331.5\pm0.5$  \\  
MS2137-2353 &  $43.7\pm6.7$ & $22.4\pm2.2$ & $35.3\pm9.9$ & $6.1\pm0.9$ & $11\pm4$ & $15.41\pm0.07$ \\  
RXJ1347.5$-$1145 &  $153.7\pm9.0$ & $175.4\pm8.6$ & $36.7\pm6.9$ & $22\pm1$ & $12\pm4$ & $71.3\pm0.2$ \\  
RXJ1532.9+3021 &  $694.0\pm25.0$ & $765.6\pm24.5$ & $429.9\pm29.1$ & $97\pm4$ & $140\pm40$ & $308.1\pm0.2$ \\  
RXJ2129.7+0005 &  $12.4\pm6.9$ & $9.3\pm3.0$ & $32.5\pm31.1$ & $1.7\pm1.0$ & $17.2\pm16.4$ & $10\pm10$  \\  
RXJ2248.7$-$4431 & $0.8\pm0.4$ & $0.9\pm0.4$ & -- & $0.1\pm0.1$ & -- & -- \\  
CLJ1226.9+3332 & $6.2\pm0.4$ & $6.5\pm0.5$ & -- & $0.9\pm0.1$ & --  & -- \\  
MACS1311.0$-$0310 & $3.9\pm2.9$ & $4.0\pm3.0$ & -- & $0.5\pm0.4$ & -- & -- \\
\\
\hline  
\end{tabular}  
\begin{flushleft}  
$^{a}$ Calzetti model dust was used to calculate the reddening correction. \\   
$^{b}$ Milky Way model dust was used to calculate the reddening correction. \\
$^{c}$ Areas of UV emitting regions as observed with the F336W filter. Uncertainties are calculated by using Monte Carlo draws to sample the distribution of the number of pixels containing positive flux in the regions shown in Figure~\ref{fig:Triptychs}. \\  
$^{d}$ 3$\sigma$ upper limits are shown. \\  
\end{flushleft}  
}  
\end{table*}  
  
The correlation between UV and H$\alpha$+[\ion{N}{2}] luminosities is shown in Figure~\ref{fig:UV-Halpha}. The two luminosities are broadly consistent with the ratio expected from \cite{Kennicutt_1998_SFR}, in that the SFR estimates derived from the UV and H$\alpha$+[\ion{N}{2}] luminosities are consistent with each other. This result is differs from the findings of \cite{McDonald_2010_Ha, McDonald_2011_UV}, since they find on average the UV/H$\alpha$ ratio is slightly lower than that predicted from the Kennicutt relationships. This is most likely because they do not correct for extinction due to dust in the BCG. Indeed, they propose adding a correction of $E\left(B-V\right)=0.2$ to their data, which would make their results consistent with continuous star formation, and this value is typical for the dust extinction we observe in CLASH BCGs.    

\begin{figure}[h]  
\centering{\epsfig{file=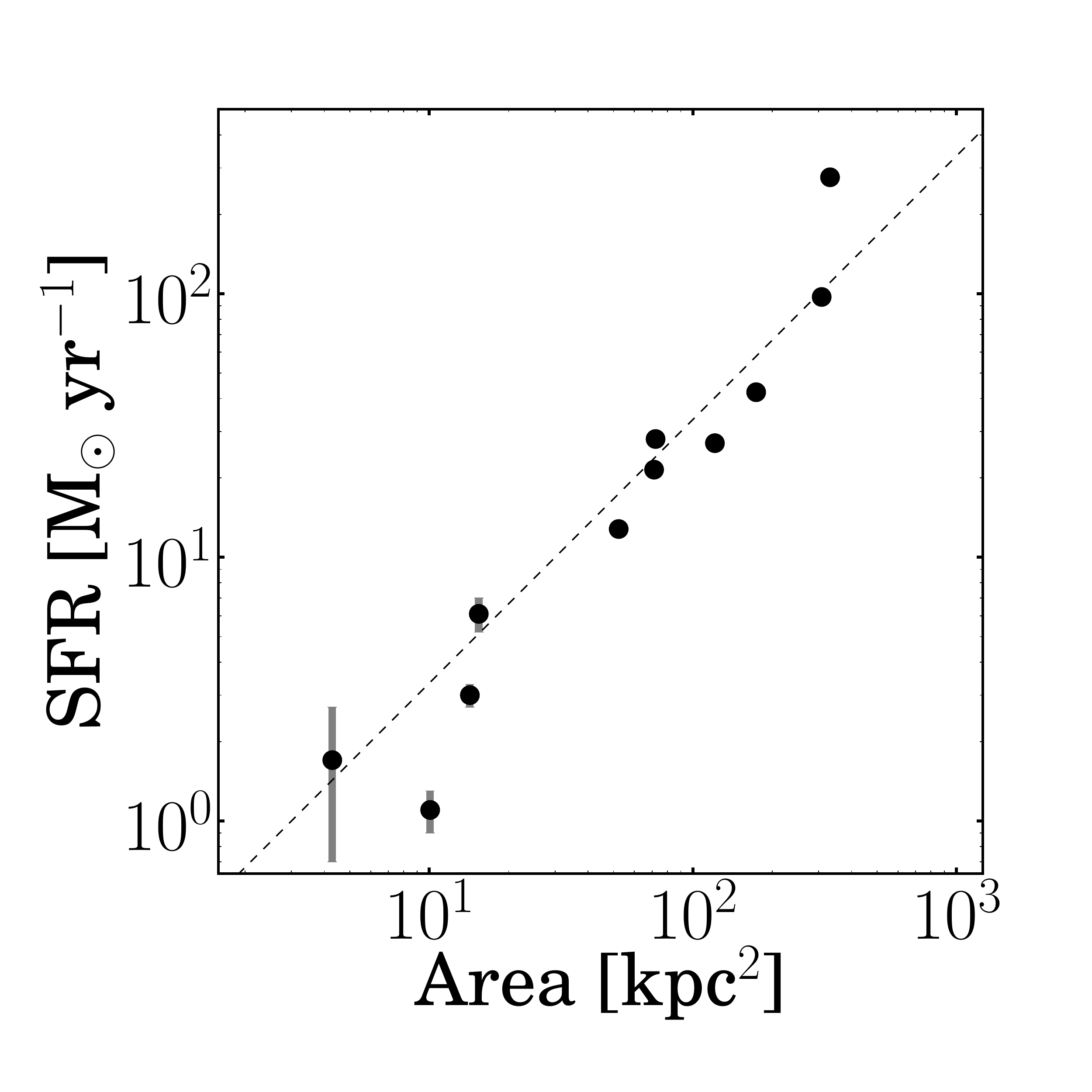,width=8cm,angle=0}}  
\caption{The UV derived SFR is shown as a function of the area of the UV flux emitting regions measured in the F336W filter. The dotted line represents an $\langle\Sigma\textrm{SFR}\rangle$ of $\sim 0.3$ M$_{\odot}$ yr$^{-1}$ kpc$^{-2}$. Absent error bars are too small to be seen as plotted.}  
\label{fig:SFR-Area}  
\end{figure}

UV SFRs are correlated with the areas of the star forming region in CLASH BCGs (Figure~\ref{fig:SFR-Area}). CLASH BCGs have an average SFR surface density ($\langle\Sigma\textrm{SFR}\rangle$) of $\sim 0.3$ M$_{\odot}$ yr$^{-1}$ kpc$^{-2}$, with typical values ranging between $\sim 0.1-0.4$ M$_{\odot}$ yr$^{-1}$ kpc$^{-2}$. Areas of the UV flux emitting regions were measured using the F336W filter. In order to calculate the uncertainty, we sampled the distribution of fluxes in each pixel in the regions shown in Figure~\ref{fig:Triptychs} using a Monte Carlo method, which we used to create a distribution of flux-emitting areas. The exception to this is MACS1931.8$-$2635, which exhibits a $\langle\Sigma\textrm{SFR}\rangle$ of $0.83\pm0.06$ M$_{\odot}$ yr$^{-1}$ kpc$^{-2}$.

\subsection{SOAR Spectra Results}  
  
\begin{figure}[h]  
\centering{\epsfig{file=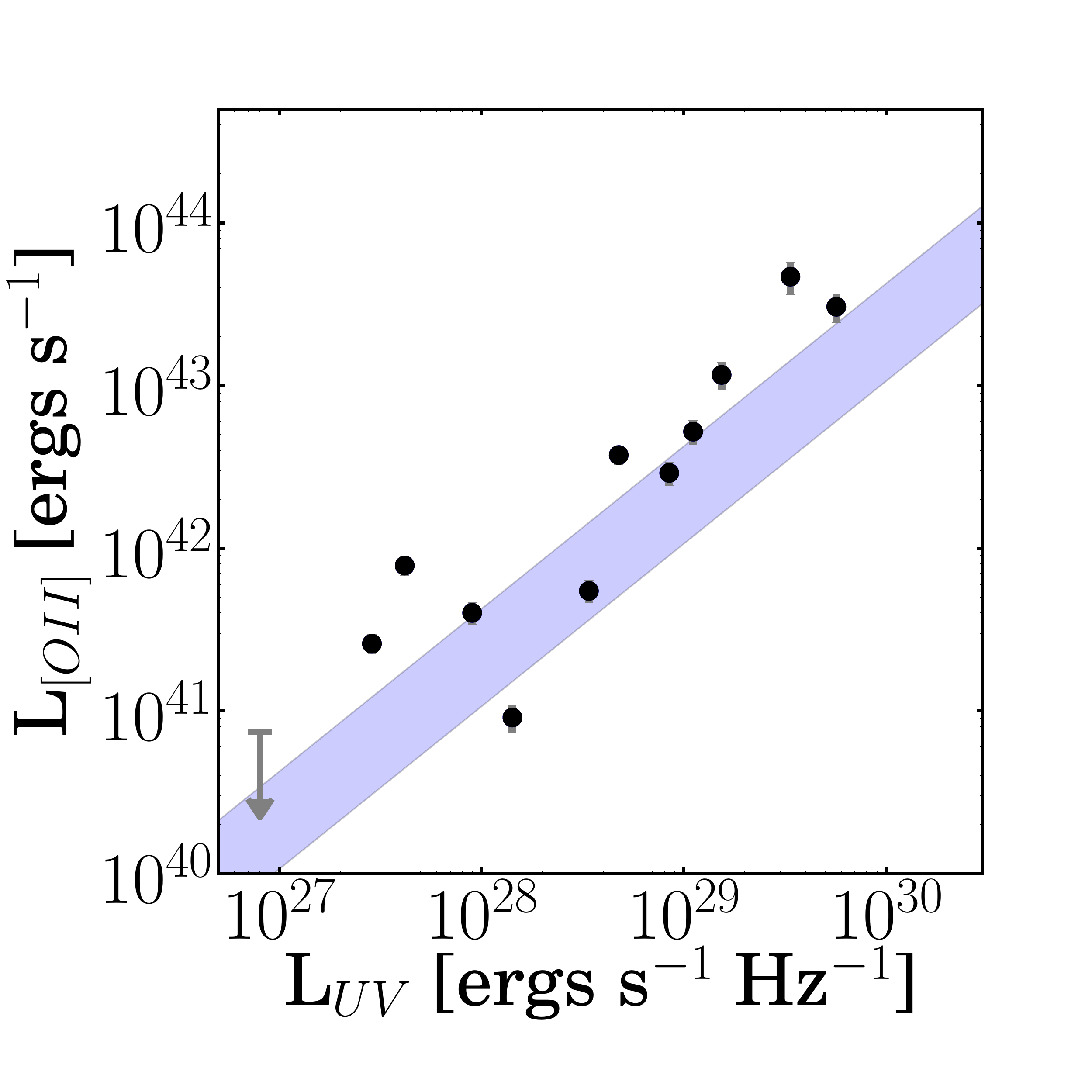,width=8cm,angle=0}}  
\caption{The distribution of BCG UV luminosities, L$_{UV}$, compared to [\ion{O}{2}] luminosities, L$_{[OII]}$, for all BCGs with SOAR coverage of [\ion{O}{2}]. UV luminosities were measured in rectangular apertures that correspond to the spectral slit placements. The light blue band corresponds to where these luminosities predict the same star formation rate to within 0.3 dex, as in Figure~\ref{fig:UV-Halpha}.}
\label{fig:OII-Broad}  
\end{figure}    

\begin{figure}[h]  
\centering{\epsfig{file=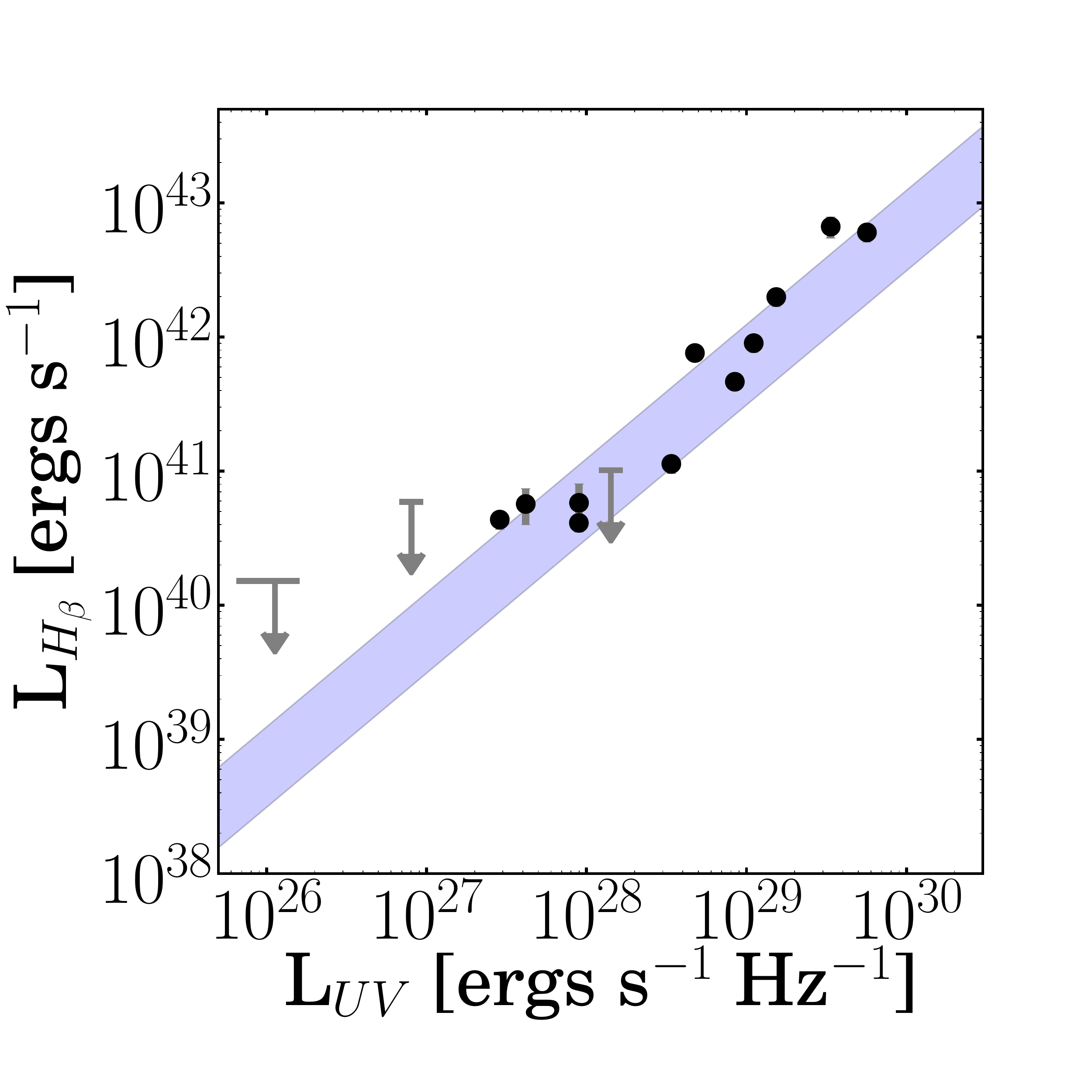,width=8cm,angle=0}}  
\caption{BCG UV and H$\beta$ luminosities for all BCGs with SOAR coverage of H$\beta$. UV luminosities were measured in rectangular apertures that correspond to the spectral slit placements. In order to determine the region in the plot where the two luminosities predict consistent continuous SFRs, we scaled the Kennicutt law relating L$_{H\alpha}$ to the SFR by a factor of 2.85. The light blue band is analogous to thes band depicted in Figures~\ref{fig:UV-Halpha} and~\ref{fig:OII-Broad}. Typical uncertainties on L$_{H\beta}$ are $\sim$10\%, and typical uncertainties on L$_{UV}$ are $\sim$ 5\%.}  
\label{fig:Hbeta-UV}  
\end{figure}

The CLASH BCG UV and [\ion{O}{2}] luminosities, displayed in Figure~\ref{fig:OII-Broad}, scale with each other, but produce divergent SFR estimates. However, UV and H$\beta$ luminosities (Figure~\ref{fig:Hbeta-UV}) have a tight correspondence and produce consistent estimates of the SFR in CLASH BCGs. The agreement between UV and H$\beta$ based SFRs is tighter than the agreement between UV and H$\alpha$+[\ion{N}{2}] derived SFRs, which is to be expected considering the limited precision of SFRs estimated using broadband H$\alpha$+[\ion{N}{2}].

The SFR-L$_{[OII]}$ relation we use is calculated in \cite{Kewley_2004_SFROII} by using a sample mean [\ion{O}{2}]/H$\alpha$ to convert from the \cite{Kennicutt_1998_SFR} SFR-L$_{H\alpha}$ relation to an SFR-L$_{[OII]}$ relation. However, the theoretical value of L$_{[OII]}$/SFR depends on the metallicity of the nebular region (peaking near $Z\sim0.5$ Z$_{\odot}$) as well as ionization parameter (peaking near $q\sim1\times10^{7}$ cm s$^{-1}$) \citep{Kewley_2004_SFROII}. CLASH BCG redenning corrected flux ratios [\ion{O}{3}]/[\ion{O}{2}] are typically $\sim 0.1$, implying an ionization parameter near $q\sim1\times10^{7}$ cm s$^{-1}$ for solar and sub-solar metallicities \citep{Kewley_2002_Ionization}. The combination of these two parameter depedencies may explain the systematic tension between UV and [\ion{O}{2}] SFRs. Furthermore, the offset between UV and [\ion{O}{2}] SFRs we observe in CLASH clusters is consistent with the observation in \cite{Kennicutt_1998_SFR} that L$_{[OII]}$/SFR is typically boosted in starbursts relative to galaxies undergoing continuous star formation by a factor of $\gtrsim 2$.

\begin{figure*}[t!]  
\centering  
\epsfig{file=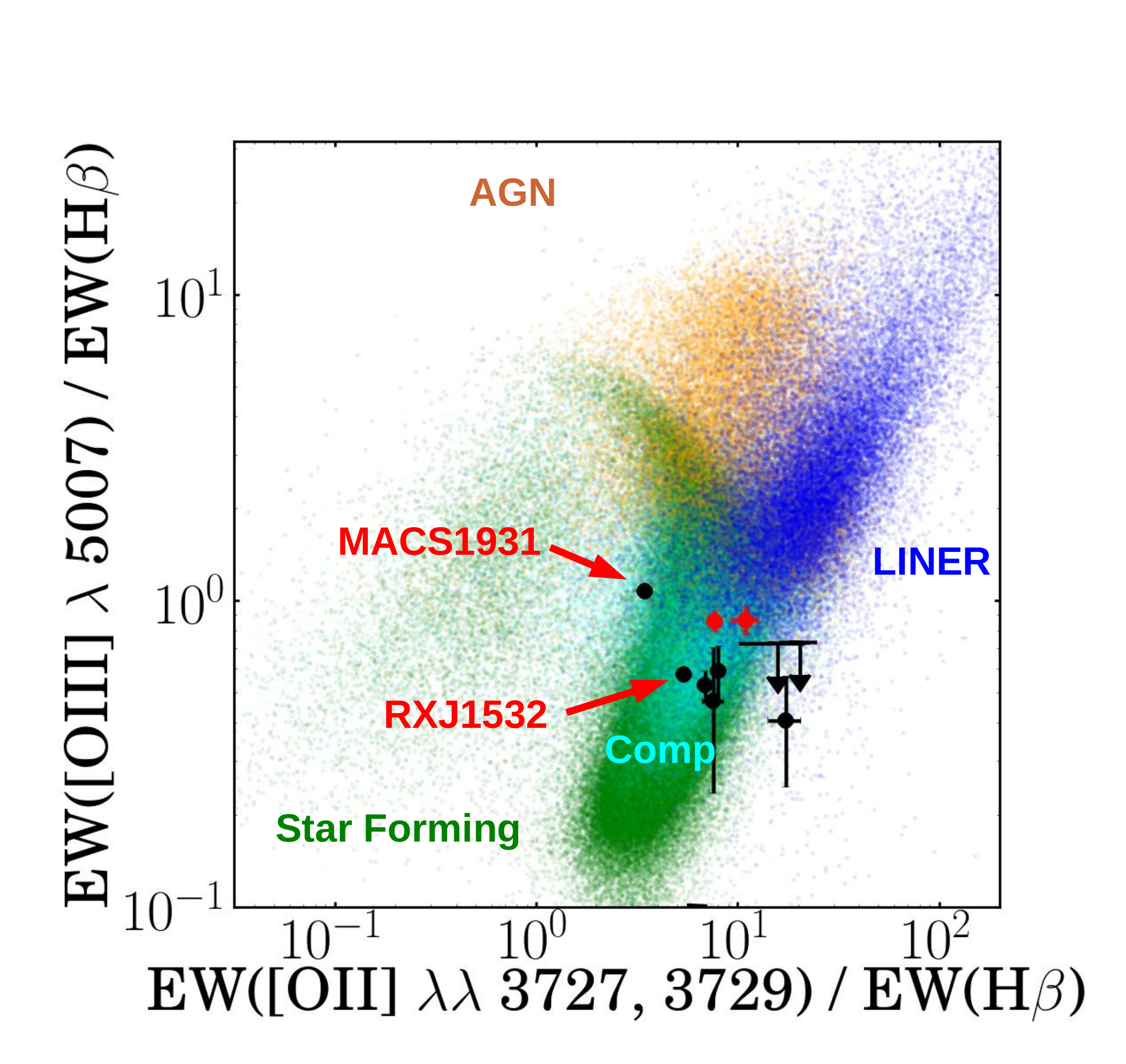,width=8cm,angle=0}  
\label{fig:subfigure1}
%\\  
\epsfig{file=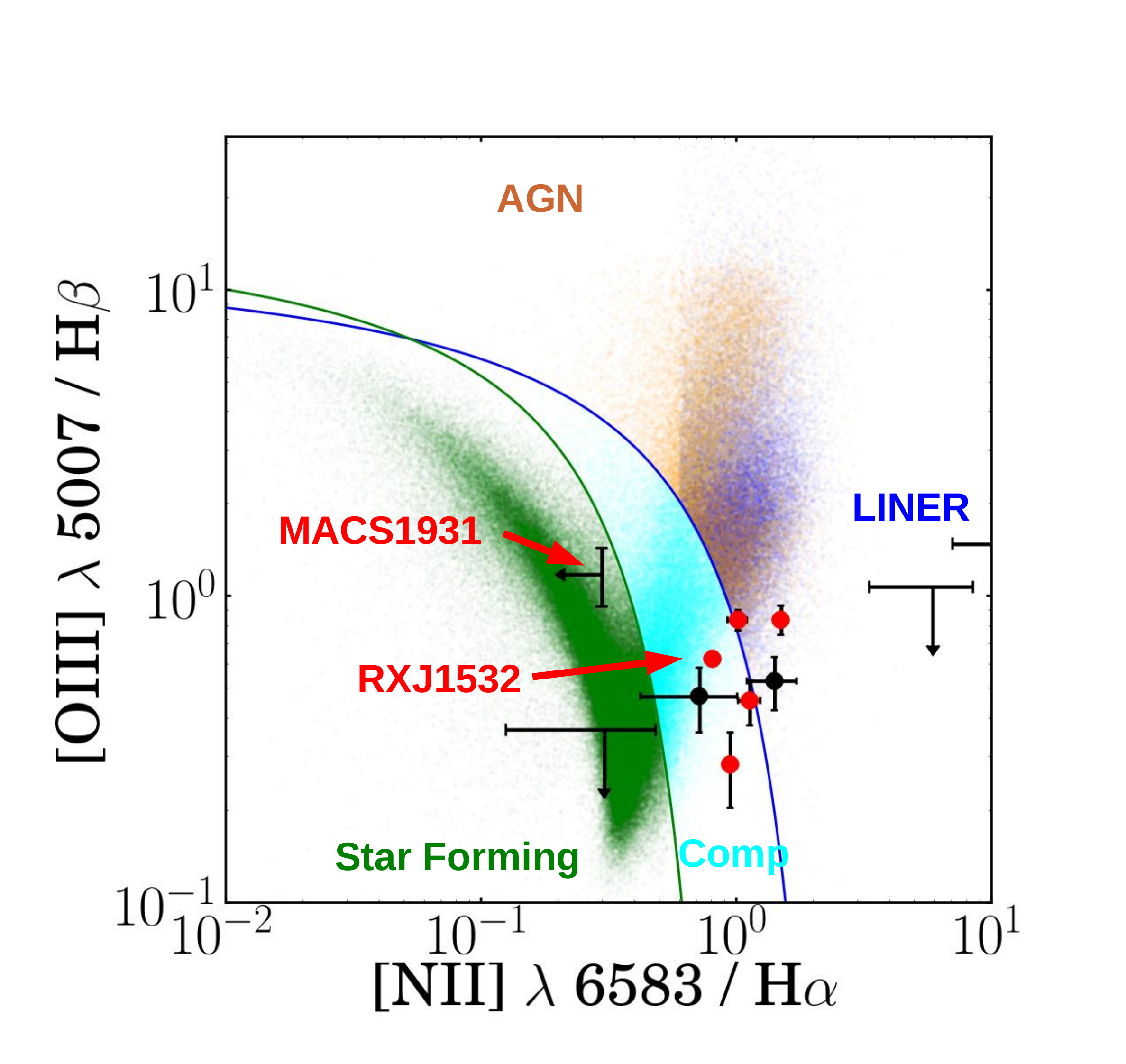,width=8cm,angle=0}  
\label{fig:subfigure2}
\caption{\textit{Left:} The blue line diagnostic diagram for line-emitting CLASH BCGs. Individual BCGs appear labelled on the plot. Galaxies from galSpec are shown color-coded by BPT classification. Green galaxies are starforming galaxies, light blue galaxies are composites, blue galaxies are LINERs and orange galaxies are AGNs. The positions of MACS1931 and RXJ1532 are highlighted, since these two are the largest starbursts in the CLASH BCG sample. Black points are measured from SOAR data, while red point are measured from SDSS data. \textit{Right:} The BPT diagram incorporating CLASH broadband line flux estimates. Red points are measured from SDSS data, while black points are estimated using a combination of SOAR fluxes and HST broadband H$\alpha$+[\ion{N}{2}] flux estimates. For the black points, we defined a proxy for [\ion{N}{2}]/H$\alpha$ on the x-axis of the BPT diagram, $X$, to be $0.75\left(\frac{H\alpha+[NII]}{2.85H\beta}-1\right)$. Region labels match color coding and approximate locations of regions populated by different classifications of galaxy.}    
\label{fig:Diagnostics}  
\end{figure*}

SOAR spectra were also used to constrain the source of the photoionizing emission we observe. We place active CLASH BCGs on the blue-line diagnostic diagram for distinguishing starbursting galaxies from AGN, described in \cite{Lamareille_2004_BlueLines} and \cite{Lamareille_2010_BlueLines}, as well as on the BPT diagram, in Figure~\ref{fig:Diagnostics}. We cannot directly separate H$\alpha$ from [\ion{N}{2}] in our broadband H$\alpha$+[\ion{N}{2}] fluxes, so when available, we use line fluxes from the SDSS Data Release 12\footnote{http://dr12.sdss3.org/} to determine the locations of CLASH BCGs on the BPT diagram \citep{Alam_2015_SDSSDR12}. In order to place the remaining active BCGs on the BPT diagram, instead of comparing [\ion{O}{3}]/H$\beta$ to [\ion{N}{2}]/H$\alpha$, we compare [\ion{O}{3}]/H$\beta$ to $X\equiv 0.75\left(\frac{H\alpha+[NII]}{2.85H\beta}-1\right)$. Our expression for $X$ assumes case B recombination. Regardless of the presence of AGN emission, case B recombination allows us to derive a reasonable estimate of the ratio of H$\alpha$ to H$\beta$, since for systems with hydrogen densities in the range $10^{3}-10^{6}$ cm$^{-3}$, H$\alpha$/H$\beta$ $\sim$ 2.7-3.2 \citep{Netzer_2013_AGN}. Because AGN often produce harder photoionizing spectra than young stellar populations, our assumption will tend to bias $X$ slighter higher than [\ion{N}{2}]/H$\alpha$ in the presence of an AGN. However, such a bias will cause the estimated line ratios to appear more `AGN-like,' so the resulting BPT diagram is a conservative estimate of the contribution of ongoing star formation to the line ratios we observe. The positions of BCGs on the BPT diagram determined using $X$ depend on both the accuracy of our reddening corrections and broadband H$\alpha$+[\ion{N}{2}] estimates. Therefore, their value is primarily as a consistency check of the blue-line diagram.

For the blue-line diagram, we use equivalent widths observed with SOAR for all CLASH BCGs except Abell 383 and MACS1423.8+2404. Since these BCGs have incomplete line flux data from the SOAR spectra (Abell 383 does not have an [\ion{O}{2}] measurement and MACS1423 has an upper limit for [\ion{O}{3}] $\lambda 5007$ estimated from an upper limit of [\ion{O}{3}] $\lambda 4959$) but were observed in Data Release 12, we use SDSS equivalent widths for these lines instead. In general, our SOAR spectra are better suited to observing extended nebular emission in CLASH BCGs because we were able to place the slit to maximize coverage of the nebulae. We overplot our results on the SDSS galSpec\footnote{http://www.sdss.org/dr12/spectro/galaxy\_mpajhu/} galaxy sample \citep{Brinchmann_2004_Galspec, Kauffmann_2003_Galspec, Tremonti_2004_Galspec}. CLASH BCGs tend to lie in a particular region of this diagram, with low [\ion{O}{3}]/H$\beta$ and high [\ion{O}{2}]/H$\beta$ relative to the SDSS dataset. Our results imply that most of the BCGs lie in the composite starforming-LINER region described in \cite{Lamareille_2010_BlueLines}, with the exception of MACS1931.8-2653.  
 
The BPT diagram shows the \cite{Kewley_2001_BPT} line in blue and \cite{Kauffmann_2003_BPT} line in green. CLASH BCGs are distributed in the starforming and composite-starforming regions of the diagram. The CLASH BCGs cluster around log([\ion{O}{3}]/H$\beta$) $\sim -0.3$, which puts them below the BPT discriminating boundary between starforming galaxies and AGN \citep{Kauffmann_2003_BPT}. The exception to this is MACS1931, which is consistent with emission powered predominantly by star formation. We observe an X-ray AGN in the \textit{Chandra} image of MACS1931; however, given the extent of the UV emission region it makes sense to classify the BCG as starbursting. For the most part, the blue-line and BPT indicate consistent sources powering line emission in CLASH BCGs.  
  
Based on these diagnostics, we conclude that the line emission in most of the BCGs is either predominantly due to ongoing star formation, or to a composite starforming-LINER-like source. In particular, the two most UV luminous BCG in our sample, MACS1931 and RXJ1532, are consistent with star formation being the main photoionization mechanism when taking into account both diagnostic diagrams. MACS1720, MS2137 and possibly Abell 383 may be LINERS, although much of their UV and H$\alpha$+[\ion{N}{2}] flux is not in a nuclear emission region. Likewise, the majority of the composite galaxies fall into the composite starforming-LINER classification. This is consistent with previous results finding LINER-like emission in cool-core BCGs \citep{VeronCetty_2000_AGNLines, Edwards_2007_BCGLines}. However, while the emission line diagnostics in these BCGs are LINER-like, they cannot be LINERs since they cannot be powered by a central black hole \citep{Heckman_1989_BCGLines}. Several hypotheses have been proposed for the source of this extended LINER-like emission. Stellar populations may be responsible for this emission, which could be due to photoionization from O-stars and young starbursts, shocks, and old stars \citep{Shields_1992_OLINERs, Olsson_2010_SBLINER, Loubser_2013_BCGs, Gabel_2013_BCGLINER}. Emission lines may also be due to nebular gas being heated by the surrounding medium \citep[e.g.][]{Donahue_1991_Lines, Werner_2013_Heating}, turbulent mixing layers \citep{Begelman_1990_Mixing}, or collisional heating \citep{Sparks_1989_CoolingFlow, Ferland_2009_CollHeating}. 
  
\subsection{Correlation with ICM X-Ray Properties}  
  
\subsubsection{Core Entropies}  
  
\begin{figure}[h]  
\centering{\epsfig{file=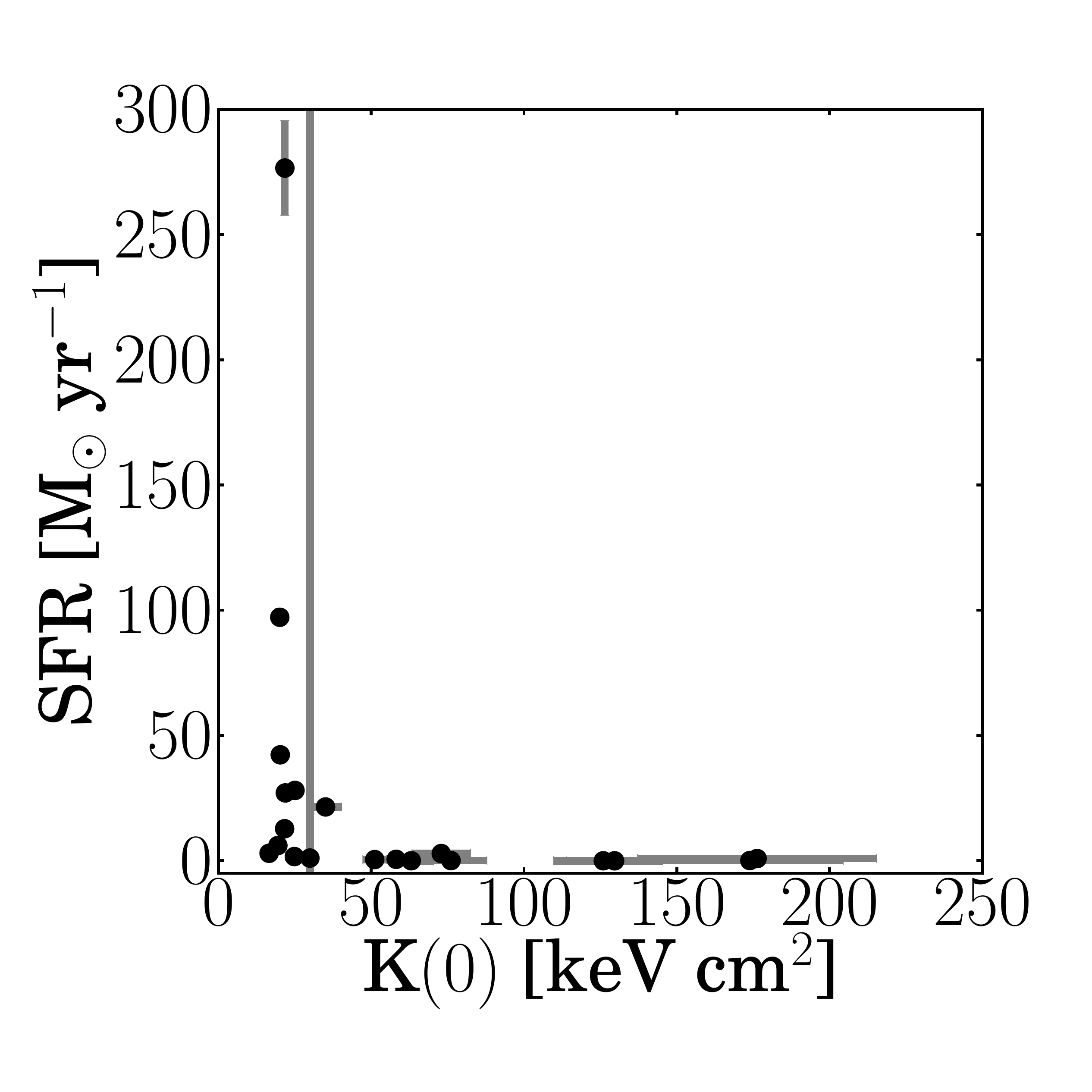,width=8cm,angle=0}}  
\caption{BCG SFRs compared to cluster core entropies. Core entropies below $\sim 30$ KeV cm$^{2}$ are associated with the onset of UV emission in BCGs, and all of the CLASH BCGs with a UV SFR $> 10$ M$_{\odot}$ yr$^{-1}$ reside clusters with a core entropy $\lesssim 30$ KeV cm$^{2}$. The grey vertical line depicts this 30 KeV cm$^{2}$ entropy threshold. Core entropies are taken from the ACCEPT cluster profile archive. For several points, errorbars for the SFR and K$_{0}$ are too small to be depicted on this plot.}  
\label{fig:K0-SFR}  
\end{figure}

SFRs derived from L$_{UV}$ in our sample are correlated with the X-ray properties of the ICM. Figure~\ref{fig:K0-SFR} shows the relationship between CLASH BCG SFRs and ICM core entropies. The core entropy $K_{0}$ used in the present study is defined to be the innermost bin of the entropy profile in ACCEPT, and is a proxy for the existence of a cool core in a galaxy cluster \citep{Hudson_2010_CoolCore}. Entropy as measured by X-rays is defined to be  
\begin{equation}  
K \equiv kT_{x}n_{e}^{-2/3},  
\end{equation}  
where $T_{x}$ is the X-ray temperature in keV, $n_{e}$ electron density in cm$^{-3}$, and k is the Boltzmann constant.  
  
Low values of $K_{0}$ typically accompany activity in BCGs. BCG activity, such as elevated NUV flux relative to the predicted quiescent UV emission, is observed to occur only in clusters where $K_{0}$ is $\lesssim 30$ KeV cm$^{2}$ \citep{Cavagnolo_2008_Entropy, Hoffer_2011_SFREntropy}.  \cite{McDonald_2010_Ha} reported on this phenomenon as well with resolved H$\alpha$ emission maps in low-redshift BCGs. In \cite{Donahue_2015_IP}, a similar entropy threshold was found for the UV-NIR color of CLASH BCGs, indicating the threshold does not change substantially out to z$\sim$0.5.  
  
Here, we demonstrate that a tight correlation exists between reddening corrected SFRs and K$_{0}$. Specifically, all of the BCGs with an SFR $> 10$ M$_{\odot}$ yr$^{-1}$ have a core entropy consistent with a value $\leq 30$ KeV cm$^{2}$ (see Figure~\ref{fig:K0-SFR}). Meanwhile, BCGs that lack significant UV or H$\alpha$+[\ion{N}{2}] luminosities occupy a range of core entropies that extends up to $\sim$ 200 keV cm$^{2}$. Considering the difference between the two observables, the tight correspondence between SFR and core entropy is compelling.
  
\subsubsection{Core $\dot{\textrm{M}}_{g}$ Estimates}  
 
ICM cooling rates inferred by \textit{Chandra} observations of clusters bear little relation to the actual cooling rate in cool-core clusters \citep[e.g][]{McNamara_1989_CCSFR}. However, we may to be able to find a relationship between star formation and a simple proxy for cooling at radii where we hypothesize ICM cooling actually occurs. The quantity we examine, $\dot{\textrm{M}}_{g}\left(r\right)$, is analogous to the predicted cooling rate as a function of radius, and is defined by
\begin{equation}
\begin{split}
%\begin{multline} 
\dot{\textrm{M}}_{g}\left(r\right) &\equiv \frac{4\pi\int^{r}_{0}{\rho_{g}\left(\tilde{r}\right)\tilde{r}^{2}d\tilde{r}}}{t_{cool}\left(r\right)} \\
& = \frac{M_{enc}\left(r\right)}{t_{cool}\left(r\right)},
%\end{multline}
\end{split}
\end{equation}
where $\rho_{g}\left(r\right)$ is the azimuthally average X-ray gas density, and $t_{cool}\left(r\right)$ is the averaged X-ray cooling time at radius $r$. $M_{enc}\left(r\right)$ is the gas mass enclosed in the radius $r$. We use 
\begin{equation*}
t_{cool} = \frac{3}{2}\frac{nkT}{n_{e}n_{H}\Lambda\left(Z,T\right)} 
\end{equation*}
to define the cooling time, and use the assumption in \cite{Cavagnolo_2009_ACCEPT} that $n \approx 2.3 n_{H}$ . The cooling curve $\Lambda\left(Z,T\right)$ is estimated by interpolating the \cite{Sutherland_1993_Cooling} cooling function at solar metallicity.  

We choose to measure $\dot{\textrm{M}}_{g}\left(r\right)$ at $r=35$ kpc ($\dot{\textrm{M}}_{g, r35}$), and at the radius in each cluster where the ratio between the cooling time and free-fall time is $t_{cool}/t_{ff} = 20$ ($\dot{\textrm{M}}_{g, t20}$). We calculate free-fall times by estimating cluster density profiles as the sum of NFW profiles derived from lensing in \cite{Merten_2015_ConcMass} and singular isothermal sphere profiles of BCG stellar density derived from stellar mass estimates in \cite{Burke_2015_BCGMass}. The impact of the stellar mass component on our overall result is not substantial; however, we include it for completeness.

%For convenience, we label our quantities as $\dot{\textrm{M}}_{g}\left(r=35\textrm{kpc}\right)$ and $\dot{\textrm{M}}_{g}\left(t_{cool}/t_{ff}=20\right)$.   

\begin{figure}[h]
\centering
{\epsfig{file=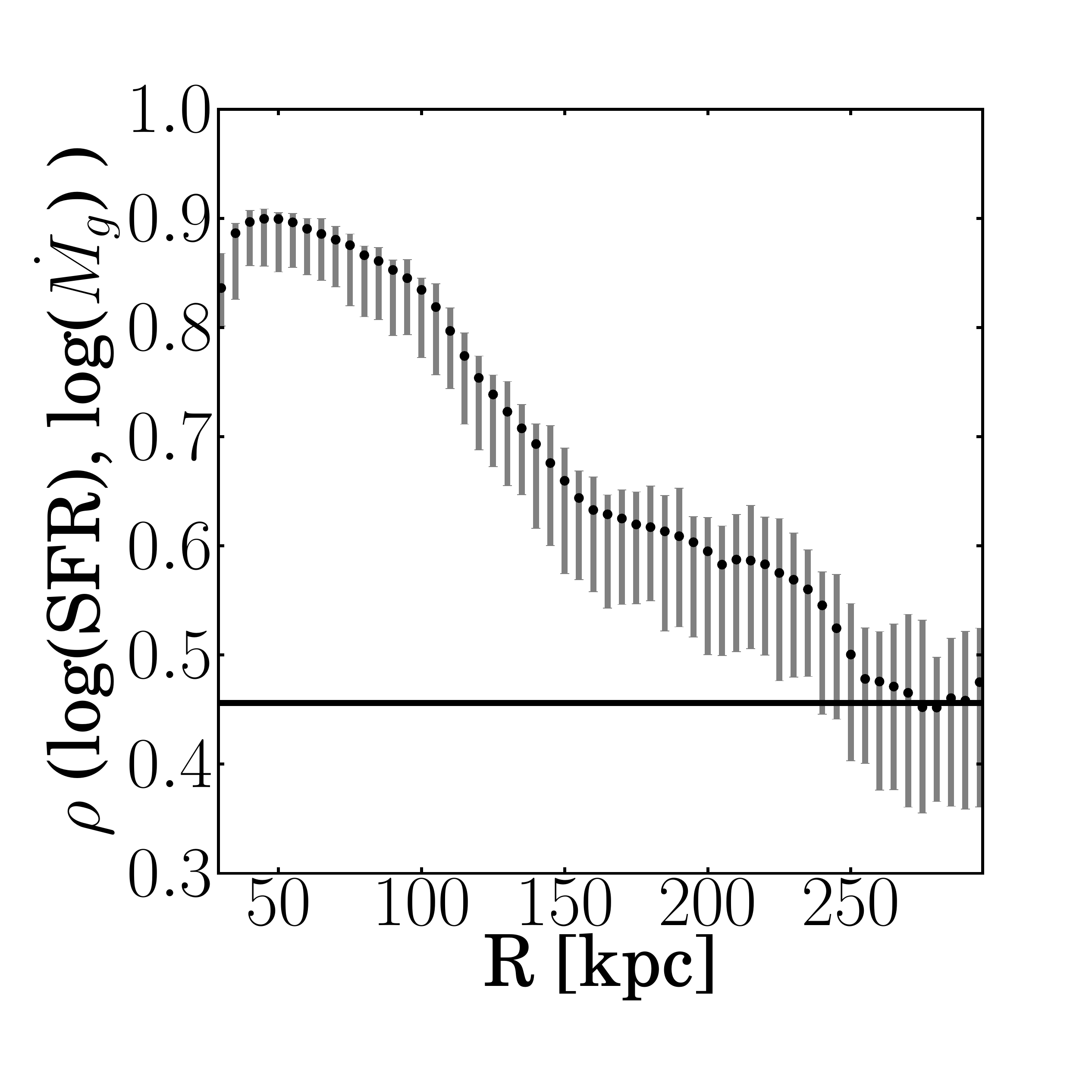,width=8cm,angle=0}} 
\caption{Pearson correlation coefficients are shown between log(SFR) and log($\dot{\textrm{M}}_{g}$) as a function of radius $R$. Green line denote the minimum correlation to rule out the null hypothesis at P$<$0.05. }  
\label{fig:Corr-R}  
\end{figure}

$\dot{\textrm{M}}_{g, r35}$  is useful to measure because 35 kpc is typical of the maximum radius we observe H$\alpha$+[\ion{N}{2}] and UV structures in CLASH clusters, and because this radius maximizes the correlation between UV SFR and $\dot{\textrm{M}}_{g}\left(r\right)$ (see Figure~\ref{fig:Corr-R}). This quantity can be calculated for all the CLASH clusters using ACCEPT data.

The choice of $\dot{\textrm{M}}_{g, t20}$ reflects the finding that BCG activity occurs in clusters with a minimum $t_{cool}/t_{ff}$ between 4 and 20 \citep{Voit15b}. Since BCG activity is associated with potentially cooling ICM gas where $t_{cool}/t_{ff} \leq 20$, $\dot{\textrm{M}}_{g, t20}$ measures the predicted cooling rate in gas that we suspect is directly involved in cooling. Radii where $t_{cool}/t_{ff}=20$ are listed for each cluster in Table 4. Clusters where $t_{cool}/t_{ff}>20$ in the innermost bin of the cooling time profile calculated from ACCEPT data are not included.

\begin{table}[h]  
\footnotesize  
\caption{\\ $t_{cool}/t_{ff}$ Threshold Radii}  
\vspace{5mm}  
\centering  
{  
\begin{tabular}{lr}  
Cluster & Radius [kpc] \\  
\hline  
\hline  
Abell 383 & $50\pm5$ \\
MACS0329.7$-$0211 & $79\pm5$ \\
MACS0429.6$-$0253 & $83\pm9$ \\
MACS0744.9+3927 & $50\pm11$ \\
MACS1115.9+0129 & $108\pm17$ \\
MACS1423.8+2404 & $83\pm8$ \\
MACS1720.3+3536 & $68\pm6$ \\
MACS1931.8$-$2635 & $105\pm14$ \\
MS2137$-$2353 & $83\pm6$ \\
RXJ1347.5$-$1145 & $69\pm18$ \\
RXJ1532.9+3021 & $121\pm12$ \\
RXJ2129.7+0005 & $59\pm8$ \\
MACS1311.0$-$0310 & $83\pm24$ \\
\end{tabular}   
}  
\end{table}  

\begin{figure}[h!]  
\centering  
\epsfig{file=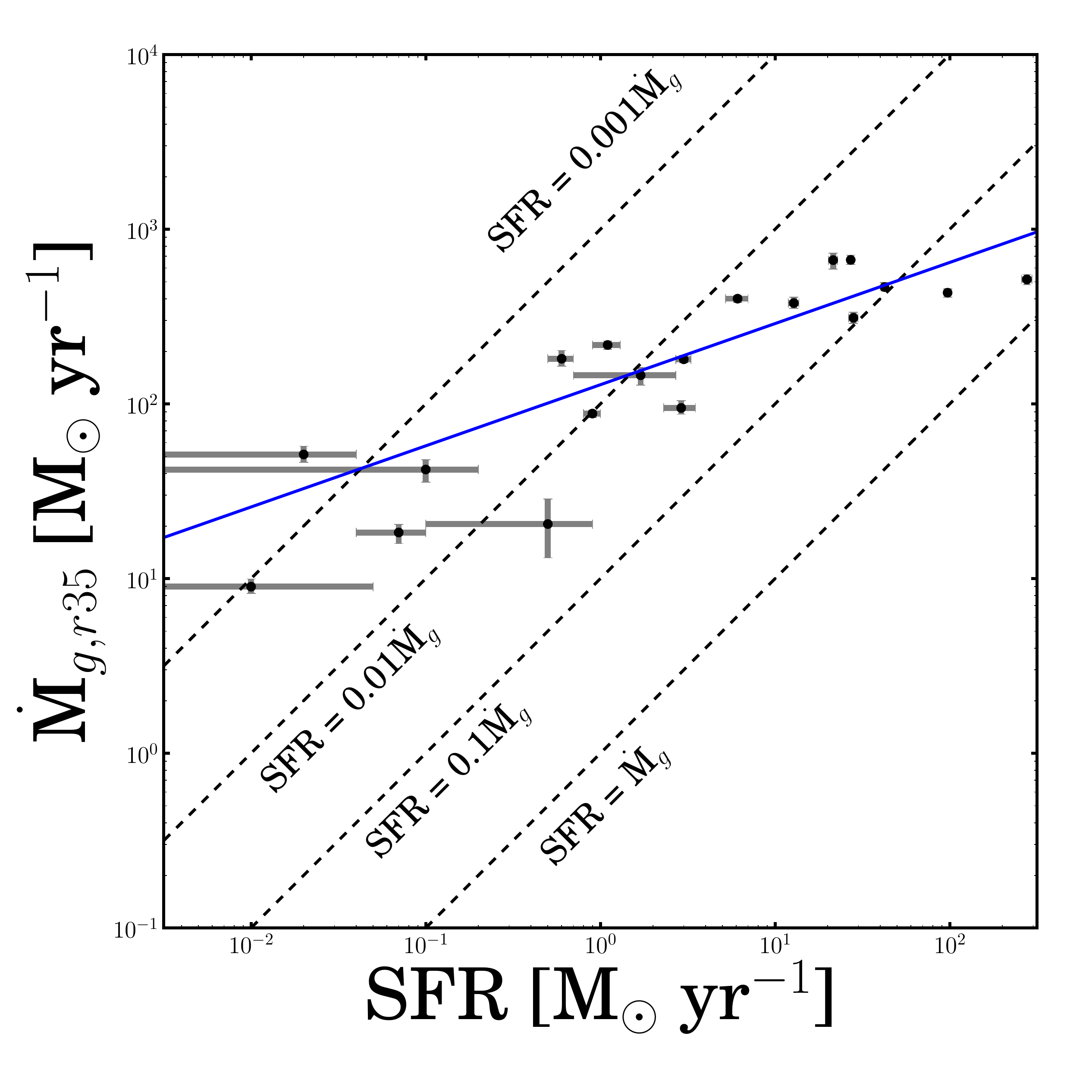,width=8cm,angle=0}  
\label{fig:subfigure1}
\\  
\epsfig{file=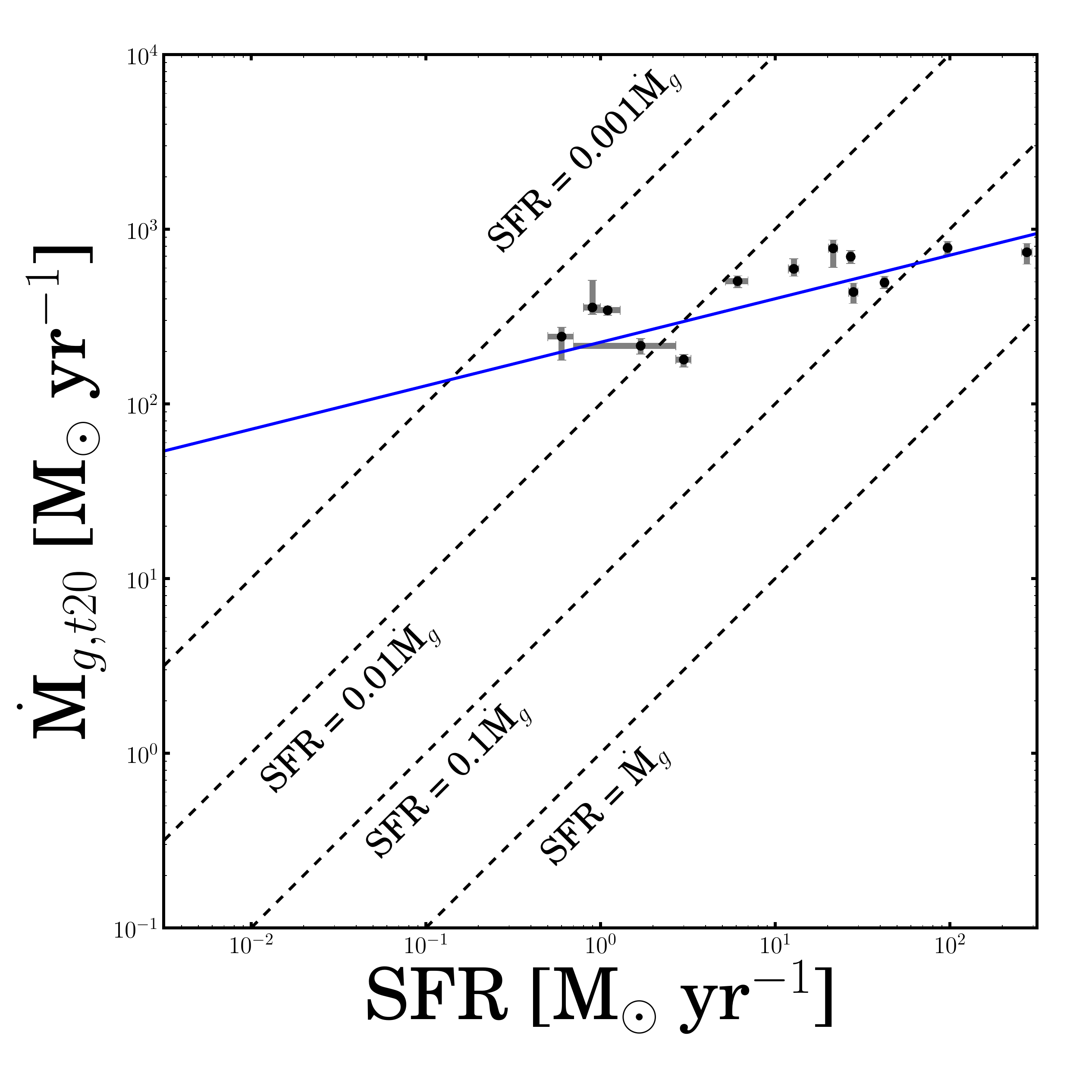,width=8cm,angle=0}  
\label{fig:subfigure2} 
\caption{Relationship between SFR and $\dot{\textrm{M}}_{g}$ measured at 35 kpc (\textit{top}) and radii where t$_{cool}$/t$_{ff}$ = 20 (\textit{bottom}). Solid lines fit the data. Dashed lines indicate (from furthest right) where SFR is 100\% of $\dot{\textrm{M}}_{g}$, 10\%, 1\%, and 0.1\%.}   
\label{fig:Mdot-SFR}  
\end{figure}

We show the relationship between the UV SFR and both $\dot{\textrm{M}}_{g, r35}$ and $\dot{\textrm{M}}_{g, t20}$ in Figure~\ref{fig:Mdot-SFR}. The dashed lines denote, from right to left, where the BCG is forming stars at 100\%, 10\%, 1\%, and 0.1\% of the cooling rate implied by $\dot{\textrm{M}}_{g}$. If we interpret the SFR as a proxy for the actual cooling rate in this system, and interpret $\dot{\textrm{M}}_{g}$ as the `potential' cooling rate  in the absence of feedback, then the larger starbursts are cooling much more efficiently than smaller starbursts, and these lines indicate where the `efficiency' is 100\%, 10\%, 1\%, and 0.1\%.

We fit trend lines to the data for both the SFR-$\dot{\textrm{M}}_{g, r35}$ and SFR-$\dot{\textrm{M}}_{g, t20}$ relations using orthogonal least squares regression. We find that the slope on the trends fit to the two datasets (0.35$\pm$0.05 for SFR-$\dot{\textrm{M}}_{g, r35}$ and 0.27$\pm$0.06 SFR-$\dot{\textrm{M}}_{g, t20}$) are nearly consistent, leading to the conclusion that the two definitions of $\dot{\textrm{M}}_{g}$ measure a similar quantity.

Several limitations impacting our measurements may affect how tightly correlated SFR-$\dot{\textrm{M}}_{g}$ (both $\dot{\textrm{M}}_{g, r35}$ and $\dot{\textrm{M}}_{g, t20}$) appear to be in our data. Gas density profiles in ACCEPT have a limited resolution (between 10$-$30 kpc per bin depending on the CLASH cluster), so values for $M_{enc}$ are typically calculated by interpolating on the central few bins of each profile. Temperature profiles are less well resolved than $\rho_{g}$ profiles, which adds scatter to our estimate of $t_{cool}\left(r\right)$ profiles. Deeper X-ray observations will beat down the systematics in $\dot{\textrm{M}}_{g}$, and a larger sample of cool core clusters will allow us to more precisely constrain the SFR-$\dot{\textrm{M}}_{g}$ relationship and examine the effects of sample selection. With our current data, we establish that a relationship exists between these two quantities, and that this relationship implies that as the BCG SFR increases, there is a steady increase in the ratio of ongoing star formation relative to the predicted cooling time in the reservoir of hot gas. 
  
\subsection{Model Fitting to RXJ1532.9+3021}  
  
We adopt RXJ1532.9+3021 as a case study and use its HST SED to delve into the SFH of the BCG in this galaxy cluster. RXJ1532 exhibits the second highest star formation in our sample and is replete with UV and H$\alpha$ bright filaments and knots. This BCG makes a better case study than the strongest star forming galaxy in our sample, MACS1931.8-2635, because the latter exhibits a strong X-ray AGN which could complicate pixel-scale SED fitting. RXJ1532 also has detailed auxiliary data including an SDSS spectrum covering H$\alpha$ and [\ion{N}{2}] and a deep \textit{Chandra} observation \citep{HlavacekLarrondo_2013_RXJ1532}. RXJ1532 is therefore an excellent prototype for exploring the characteristics of the star forming regions in CLASH BCGs.  
  
\begin{figure*}[t!]  
\centering  
{\epsfig{file=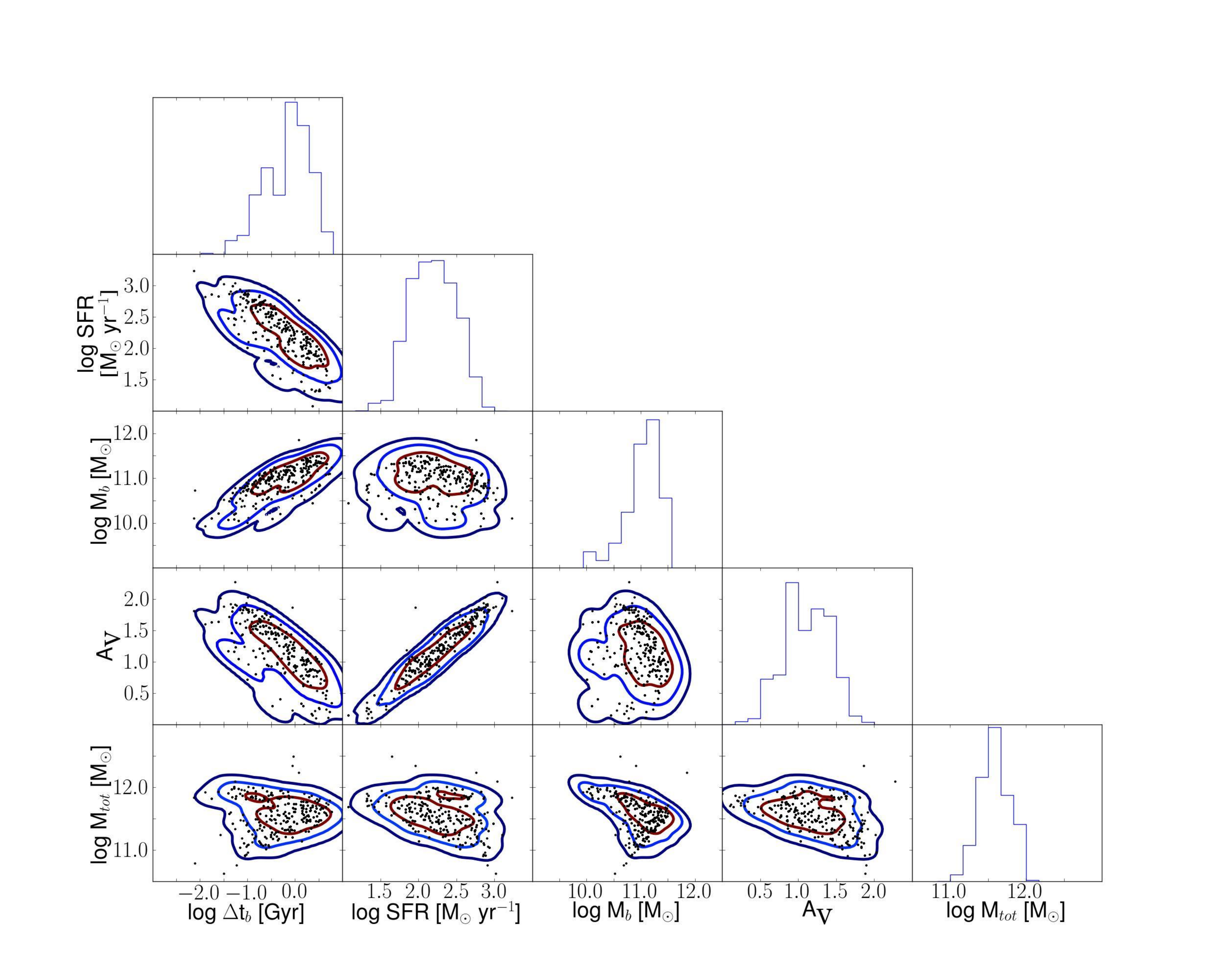,width=16cm,angle=0}}  
\caption{We show the two dimensional posterior probability distributions for $\Delta$ t$_{b}$$\times$SFR, $\Delta$ t$_{b}$$\times$Burst Mass, $\Delta$ t$_{b}$$\times$Total Mass, SFR$\times$Burst Mass, SFR$\times$Total Mass, and Burst Mass$\times$Total Mass. Black points denote the individual Monte Carlo draws used to construct the posterior probability distribution, and contours are the 68$\%$-95$\%$-99.7$\%$ contours calculated from the kernel density estimate smoothed distribution. The marginal distributions for individual parameters are shown by the histograms on the diagonal.}  
\label{fig:PDFs}  
\end{figure*}  
    
We both fit a single-aperture SED from the region used to calculate L$_{UV}$ and constructed stellar parameter maps. The best-fit SFR in the single-aperture SED is 118$^{+215}_{-42}$ M$_{\odot}$ yr$^{-1}$. The average extinction is E(B-V) $= 0.27 \pm 0.07$. Since the posterior probability distribution of the SFR is close to log-normal, the best-fit value we report is the mode of the distribution and the uncertainty we report is the 68.3\% confidence interval. The best-fit starburst is relatively long-lived, with a burst duration log$\Delta t_{b}$ [Gyr] = $-0.16\pm0.47$, and massive, with a total burst mass log M$_{b}$ [M$_{\odot}$] = 11.03 $\pm$ 0.36. The burst parameters in the fit have degeneracies (see Figure~\ref{fig:PDFs}), although the peaks in the probability distribution suggest that we are constraining the parameters.   
  
\begin{figure*}[t]  
\centering  
{\epsfig{file=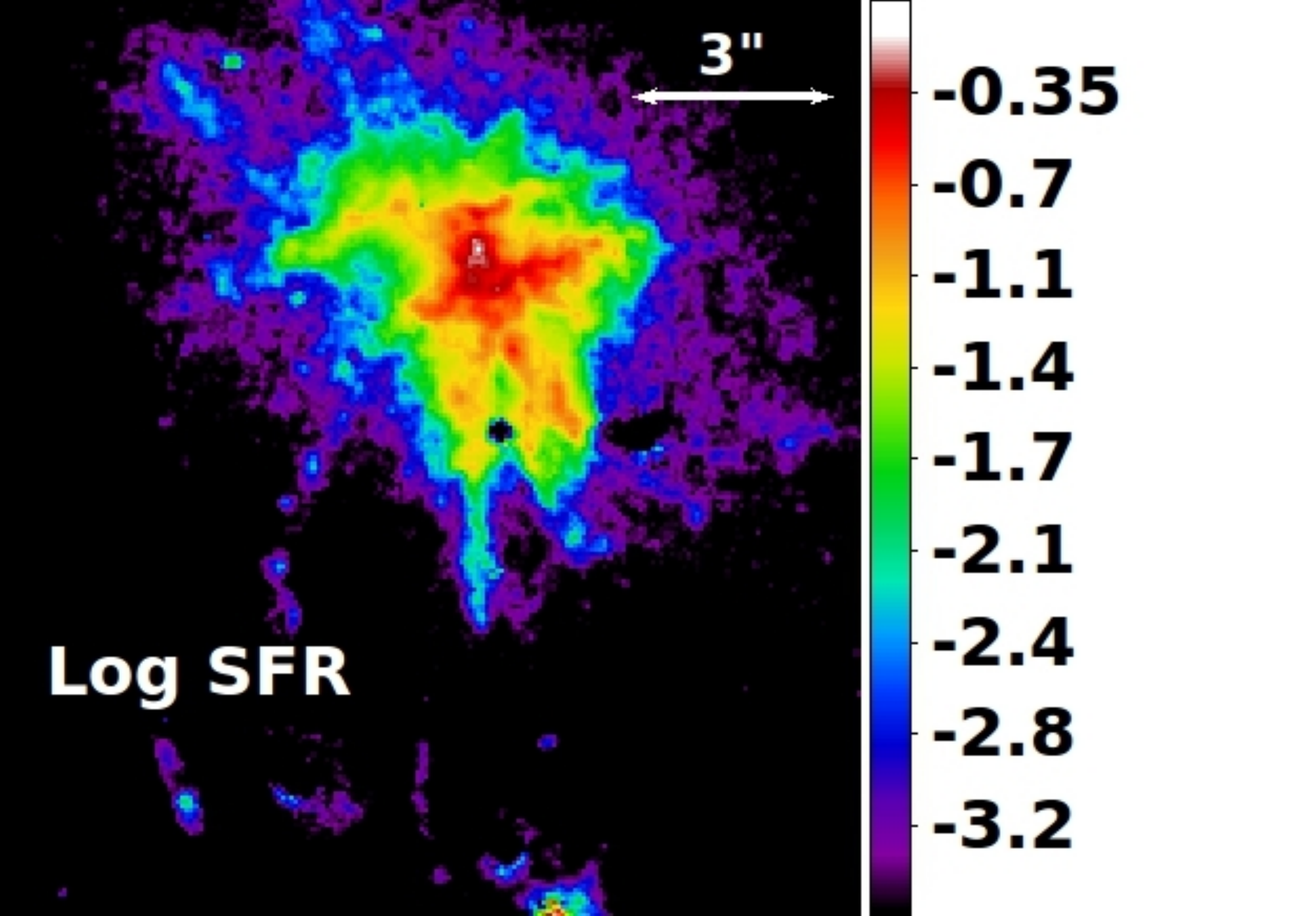,width=8cm,angle=0}  
\label{fig:subfigure1}}   
{\epsfig{file=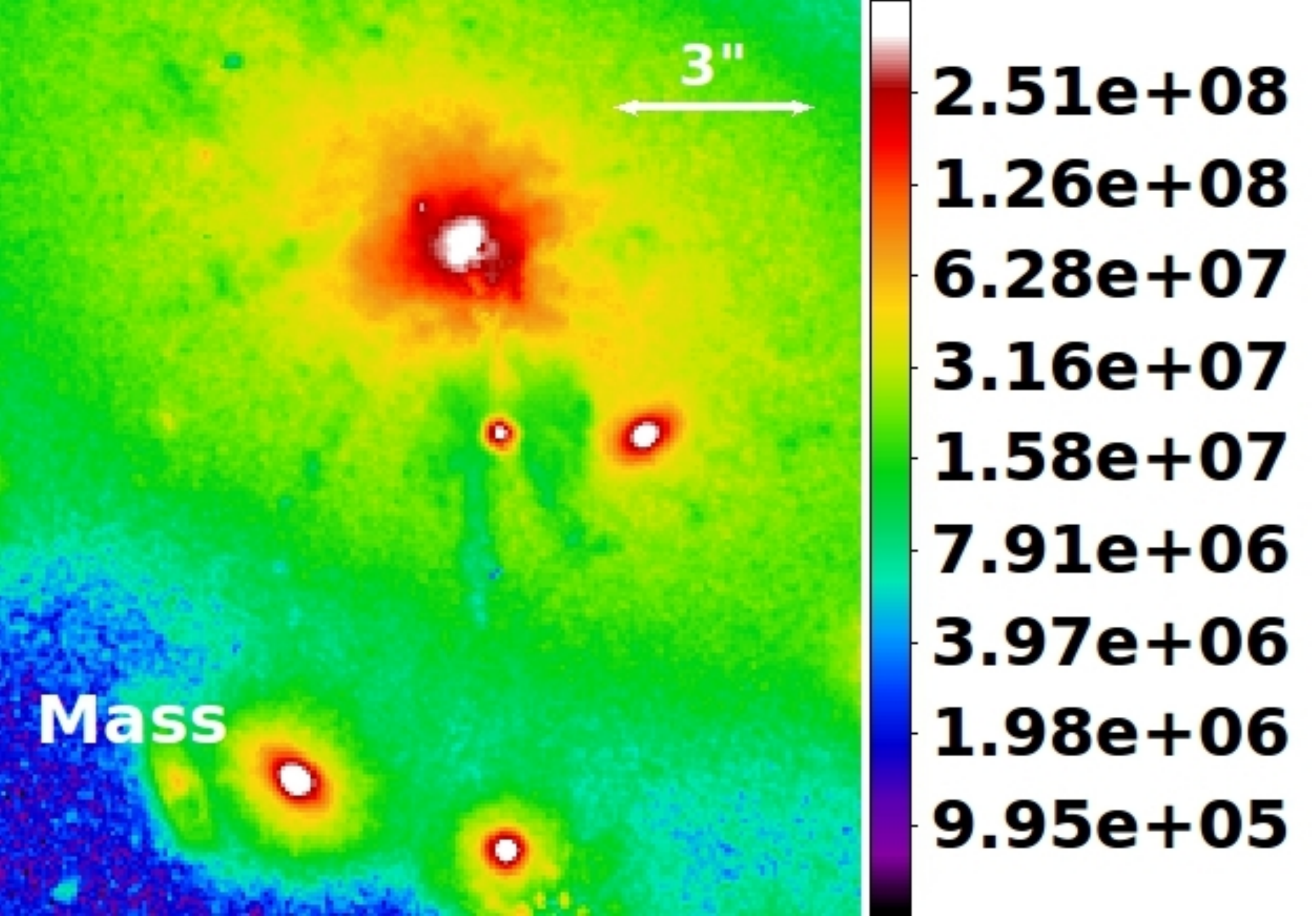,width=8cm,angle=0}  
\label{fig:subfigure3}}  
{\epsfig{file=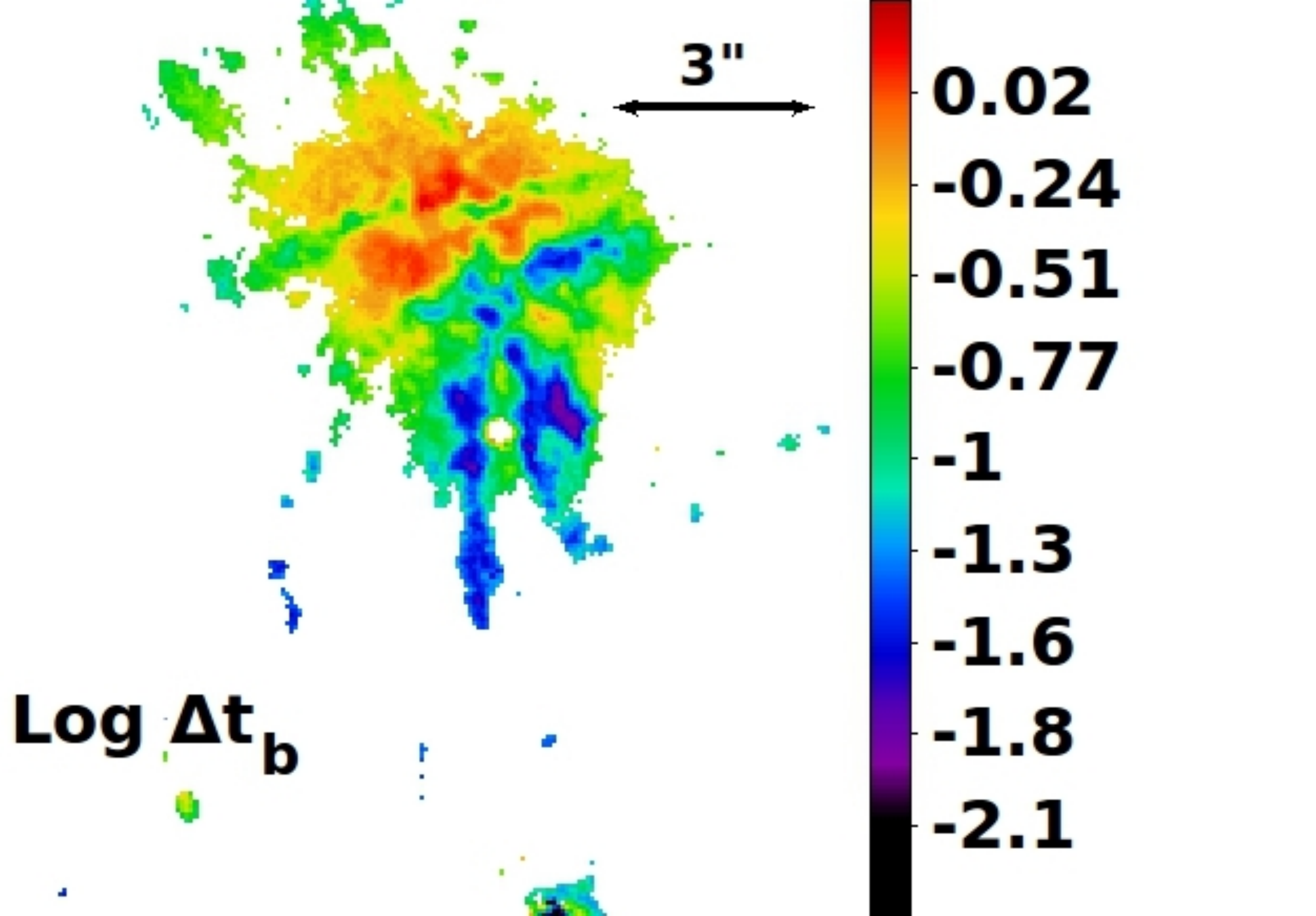,width=8cm,angle=0}  
\label{fig:subfigure4}}
{\epsfig{file=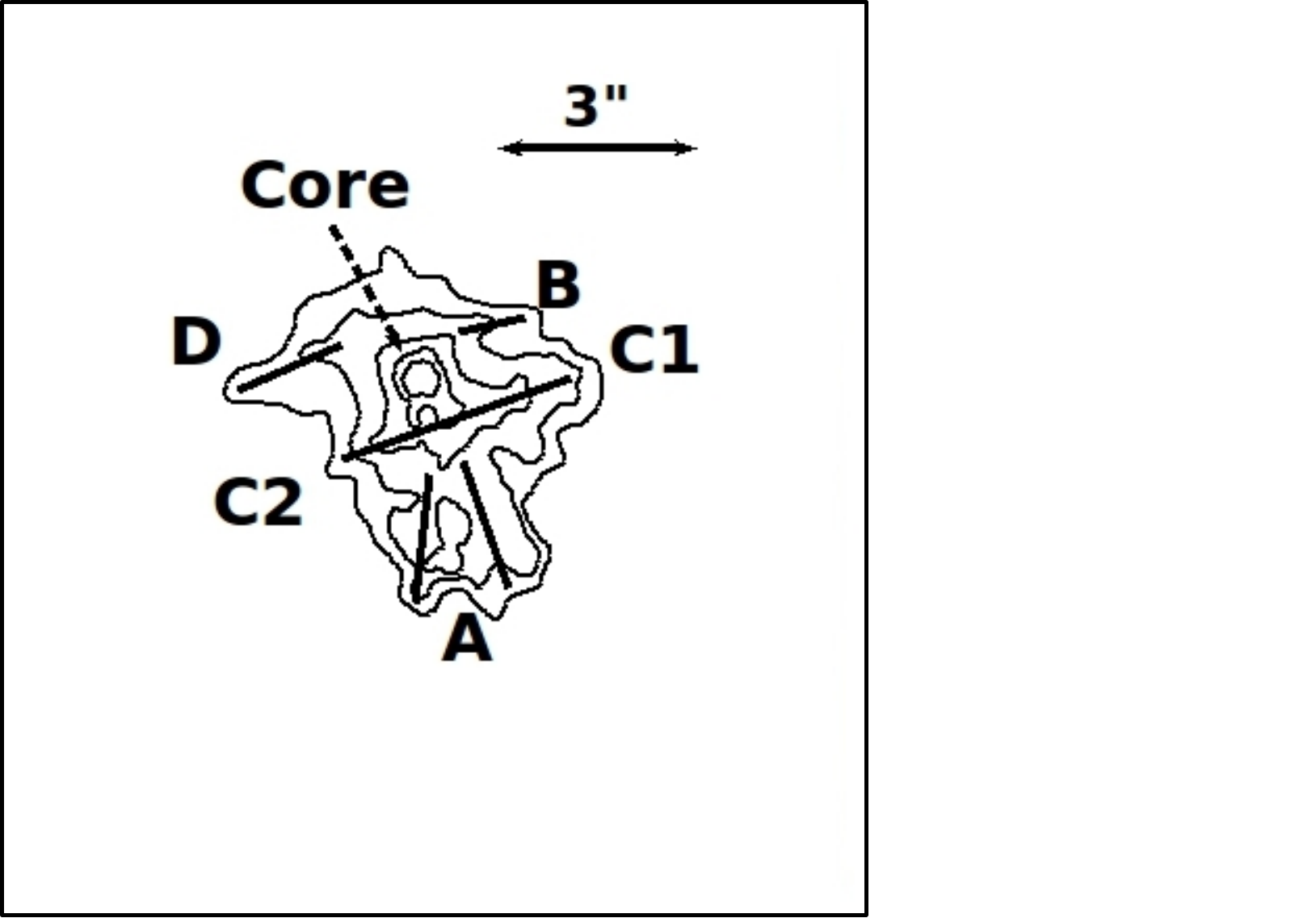,width=8cm,angle=0}
\label{fig:subfigure5}}  
\caption{Two dimensional maps of SFR, log $\Delta$ t$_{b}$, and the total mass surface density for RXJ1532. The bottom right panel provides labels for the morphological features in the parameter maps. The SFR map is in units of M$_{\odot}$ yr$^{-1}$, log $\Delta$ t$_{b}$ in log Gyr, and the total mass in units of M$_{\odot}$ pix$^{-2}$. All values are expectation values. log $\Delta$ t$_{b}$ is masked to only depict regions where the likelihood of a starburst is $\geq 1\sigma$. The pair of filaments to the south of the center of the BCG, and the filament along the NW/SE are noteworthy for being by $\gtrsim$ 1 order of magnitude younger than the average age of the starburst in this system. Possible conical shaped underdensities are roughly along the positions of bright filaments. At the redshift of the cluster, z=0.363, the 3$"$ reference scale in the figures corresponds to about 15.2 kpc.}  
\label{fig:Par_Maps}  
\end{figure*}

We show stellar parameter maps in Figure~\ref{fig:Par_Maps}, including $\Delta t_{b}$, the SFR, and the spatial distribution of the total mass of the BCG. Star formation is concentrated in knots in the center of the BCG, along with a network of six bright filaments and several dimmer filaments. The SFR morphology is consistent with the H$\alpha$ and UV morphology. The sum of the pixel SFR modes in the single-aperture SED region is 119 M$_{\odot}$ yr$^{-1}$, which matches the single-aperture value remarkably well.

%Want DTBURST, SFR, maybe the mass  
  
Two filaments point south (A in Figure~\ref{fig:Par_Maps}), two point northwest (B, C1), one points southeast (C2), and one extends northward before bending to the east (D). The filament pointing southeast and the brighter nothwest pointing filament (C1 and C2) appear to lie along a single axis. The clumps in the core and the filaments in the north account for the bulk of the ongoing star formation. The starbursts in these structures are long-lived, on the order of $10^{8}-10^{9}$ yr, which is consistent with the results of the single aperture SED. The filaments to the south, and the two that lie along a single southeast-northwest axis passing through the center of the BCG, are more than an order of magnitude younger than the rest of the burst system.     
  
The peak and western `bulge' in the total mass corresponds to the central knot morphology in the BCG. Meanwhile, two roughly conical `drop-offs' appear to the east and west, and both of them extend outward from the positions of star-forming filaments. Similar drop-offs are be visible at the positions of the young, southern filaments. These drop-offs may reflect real deficits in the stellar surface density of the BCG, but may also be a consequence of the dust geometry in this system. Since dusty filaments in the BCG screen the elliptical stellar populations behind them, it is possible that the drop-offs are regions where the mass estimate is biased low due to the positions of filaments along the line of sight.  

In the Appendix, we demonstrate that the {\tt iSEDfit} derived values for SFR and $E(B-V)$ agree with the broadband values and show that the SED predicted H$\alpha$+[\ion{N}{2}] line emission feature matches the SDSS spectrum. We also discuss the importance of the H$\alpha$ feature to characterizing the SFH using SED fitting. The particulars of the SFH model (a uniform starburst superimposed on an exponentially decaying SFH) and our choice of parameter space are documented in the Appendix as well.  
  
%\pagebreak
 
\section{Discussion}  
  
Half (10 out of 20) of the X-ray selected sample of CLASH clusters show evidence for significant ($> 5$  $\sigma$) rates of reddening-corrected star formation using both UV and H$\alpha$ indicators. CLASH BCGs occupy regions of line diagnostic diagrams that are typical of composite starforming-LINER galaxies, and in several cases line emission may be primarily powered by star formation. This rate of incidence is substantially higher than previous published rates of incidence of star formation or line emission in X-ray selected cluster BCGs, which are in general closer to 20-30$\%$ \citep[e.g.][]{Crawford_1999_BCS, Edwards_2007_BCGLines}. However, the CLASH X-ray selected sample differs from these populations of galaxy clusters, since it is comprised of high gas temperature (kT$_{x}$ $\geq$ 5 keV) clusters  chosen according to a relaxation criterion based on X-ray morphology.  The CLASH sample of BCGs has an incidence of line emission similar to the incidence of line emission in REXCESS cool core clusters (70$\%$) \citep{Donahue_2010_REXCESS}. Our sample characteristics differ from REXCESS in that it is at higher redshift ($z=0.2-0.7$ compared to $z=0.06-0.18$) and 
% REVISION for Resubmission: end of sentence changed to read:
along with being X-ray selected, CLASH clusters were selected for exhibiting relatively condensed, round X-ray isophotes.

The trends between L$_{UV}$ and L$_{H\alpha+[NII]}$ and between L$_{UV}$ and L$_{H\beta}$ suggest applying the Kennicutt SFR calibrations produces consistent star formation rates. However, SFRs predicted using L$_{[OII]}$ are systematically elevated relative to UV based SFRs. This is in contrast to our findings using L$_{H\alpha+[NII]}$ and L$_{H\beta}$, both of which predict SFRs consistent with L$_{UV}$.  These results may not be unusual for starburst galaxies, although they may also indicate that the [\ion{O}{2}] emission line is being partially powered by an additional source heating the ionized gas.
  
The SFRs in several BCGs are very large. In particular, two galaxies exhibit SFRs $\gtrsim$ 100 M$_{\odot}$ yr$^{-1}$, and an additional five have SFRs $>$ 10 M$_{\odot}$ yr$^{-1}$. The strongest star formers (MACS1931, RXJ1532) are forming stars several times more slowly than the Phoenix cluster, which to date exhibits the largest known SFR of a BCG \citep{McDonald_2013_Phoenix, McDonald_2014_PhoenixGas}. However, MACS1931 is noteworthy because its UV SFR (280 M$_{\odot}$ yr$^{-1}$) is $\sim$ 40\% of the cooling rate in the absence of heating ($\sim$700 M$_{\odot}$ yr$^{-1}$) measured in \cite{Ehlert_2011_MACS1931}. The Phoenix SFR is $\sim$30\% of the cooling rate in the absence of heating measured in \cite{McDonald_2014_PhoenixGas}, so it is plausible that MACS1931 and the Phoenix cluster harbor BCGs undergoing similar feedback events. The presence of an X-ray AGN in each BCG also suggests that the AGN is undergoing a similar evolutionary phase. Furthermore, MACS1931 is forming stars more densely than the rest of the CLASH sample, which suggests it is an outlier relative to other starforming BCGs.
  
\subsection{BCG-ICM Interactions}  
  
Examination of core entropies implies that the extended star forming features in CLASH BCGs are likely due to an interaction between the BCG and the enveloping ICM. Reddening corrected SFRs obey the 30 keV cm$^{2}$ core entropy threshold reported in e.g. \cite{Hoffer_2011_SFREntropy} --  all the strong star formers (SFR $> 10$ M$_{\odot}$ yr$^{-1}$) fall at or below the threshold. From these results we conclude that ongoing star formation in the BCGs is correlated with the thermodynamics of the surrounding ICM. It is plausible that a low ICM core entropy is necessary for the onset of star formation in these BCGs. However, it does not directly trigger star formation, as evidenced by the existence of low-SFR BCGs with core entropies below 30 keV cm$^{2}$.      
  
We also analyze observables related to cooling in the low-entropy ICM surrounding BCGs, in order to better understand the interaction between the low-entropy ICM and BCG starbursts. We define two quantities, $\dot{\textrm{M}}_{g, r35}$ and $\dot{\textrm{M}}_{g, t20}$, which approximate the cooling rate of ICM gas in the vicinity of the BCG. For both definitions of the cooling rate, we observe a similar trend between SFR and $\dot{\textrm{M}}_{g}$. The positive correlations between SFR and both $\dot{\textrm{M}}_{g, r35}$ and $\dot{\textrm{M}}_{g, t20}$ are reasonable since the low-entropy gas near the BCG is a prime candidate for the reservoir of gas that cools to become star forming molecular gas. Since the correlation between SFR and $\dot{\textrm{M}}_{g}\left(r\right)$ drops as $\dot{\textrm{M}}_{g}\left(r\right)$ is measured at larger radii, these findings are consistent with the tension between observed ICM cooling rates those predicted by measuring ICM gas masses with $t_{cool}$ less than a Hubble time.

% REVISION for Re-submission: entire paragraph replaced 
Recent theoretical work has made significant progress in identifying some of the key elements of the AGN feedback processes in clusters of galaxies. Simulations in which cold gas drives the accretion onto an AGN have been done by \cite{Gaspari12, LiBryan14a, LiBryan14b, Li2015}. \cite{Gaspari12} modeled bi-polar AGN jets heating an ICM atmosphere, following up on the work of \cite{McCourt_2012} and \cite{Sharma_2012}, who found that circumgalactic gas in which heating balances cooling globally becomes thermally unstable when $t_{cool}/t_{ff} \simless 10$.  \cite{Gaspari12} found that this criterion also demarcated the transition to a multiphase medium in simulations relying on AGN jet heating fueled by cold gas. \cite{Gaspari_2013, Gaspari_2015} extended this work by focusing on on the sub-pc scale behavior and demonstrated that chaotic cold accretion could significantly increase the accretion rate onto the central black hole, relative to the Bondi rate one would infer from the thermal state of the hot phase alone. \cite{LiBryan14a} examined a mechanism similar to that in \cite{Gaspari12}, implementing bi-polar AGN jet feedback, fueled by cold gas, in an ENZO adaptive mesh simulation.  They found that AGN feedback balanced cooling robustly, and generated episodic behavior on timescales of a Gyr, in alignment with the work of \cite{Gaspari12}. The scale and filamentary nature of the cold filaments in Li \& Bryan simulations resemble those we observe, and the ages of star forming filaments we recover in RXJ1532 are consistent with a Gyr duty cycle.

Star formation in the context of bi-polar AGN feedback triggered by cold gas was first tracked in \cite{Li2015}, and that simulation exhibits episodic outbursts of star formation on Gyr time scales, as well as filamentary structures of cold star-forming gas tens of kpc long, while the minimum threshold of $t_{cool}/t_{ff}$ varies from 5-20 over the course of any single outburst. These simulations reproduce UV morphologies similar to those of cool-core BCGs in CLASH, shown in \cite{Donahue_2015_IP}, and of lower-redshift BCGs studied in \cite{Tremblay_2015_BCGUV}. The earlier simulation work described here inspired the analytic precipitation model framework described in \cite{Voit15a} and \cite{VD15}, which compared the implications of a minimum cooling time to free fall time to the bimodal entropy profiles seen in X-ray observations of clusters from the ACCEPT database \citep{Cavagnolo_2008_Entropy}. \cite{Voit15b} extends this precipitation framework to lower-mass galaxies and derives interesting implications for the connections between galaxy mass, mass of the central black hole, star formation efficiency, and chemical composition.

 Since jet-triggered precipitation ought to have a morphological relationship with the jet and a characteristic timescale set by the AGN duty cycle, we can look for evidence that may support or contradict this prediction by comparing the SED-derived stellar parameter maps in RXJ1532 to X-ray measurements of recent AGN activity. In the single-aperture SED of RXJ1532, the starburst lifetime is log $\Delta t_{b}$ [M$_{\odot}$] = 8.8 $\pm$ 0.5 log$_{10}$yr, so the bulk of the starburst has a lifetime on the order of $\sim$1 Gyr. However, the southern filaments and the northwest-southeast filaments are $10^{7}-10^{8}$ yr old. The \textit{Chandra} X-ray image reveals two well-defined cavities to the east and west of the BCG, and possible evidence of ghost cavities to the north and south \citep{HlavacekLarrondo_2013_RXJ1532}. The cavity refill times, which are the largest estimates of the cavity ages provided by \cite{HlavacekLarrondo_2013_RXJ1532}, are 6.3$\pm$0.7 and 8.2$\pm$0.7 $\times$ $10^{7}$ yr. The cavities appear to be young relative to the timescale of the starburst we recover in our analysis, but match the ages of the young filaments. Our results in RXJ1532 are consistent with an ongoing process of clumps and filaments precipitating out of the ICM when pushed out of equilibrium by jets. Application of the SED fitting techniques developed in this paper to other BCGs in the CLASH sample will determine if this narrative is consistent for all of the star forming BCGs.

Since jet-triggered precipitation ought to have a morphological relationship with the jet and a characteristic timescale set by the AGN duty cycle, we can look for evidence that may support or contradict this prediction by comparing the SED-derived stellar parameter maps in RXJ1532 to X-ray measurements of recent AGN activity. In the single-aperture SED of RXJ1532, the starburst lifetime is log $\Delta t_{b}$ [M$_{\odot}$] = 8.8 $\pm$ 0.5 log$_{10}$yr, so the bulk of the starburst has a lifetime on the order of $\sim$1 Gyr. However, the southern filaments and the northwest-southeast filaments are $10^{7}-10^{8}$ yr old. The \textit{Chandra} X-ray image reveals two well-defined cavities to the east and west of the BCG, and possible evidence of ghost cavities to the north and south \citep{HlavacekLarrondo_2013_RXJ1532}. The cavity refill times, which are the largest estimates of the cavity ages provided by \cite{HlavacekLarrondo_2013_RXJ1532}, are 6.3$\pm$0.7 and 8.2$\pm$0.7 $\times$ $10^{7}$ yr. The cavities appear to be young relative to the timescale of the starburst we recover in our analysis, but match the ages of the young filaments. Our results in RXJ1532 are consistent with an ongoing process of clumps and filaments precipitating out of the ICM when pushed out of equilibrium by jets. Application of the SED fitting techniques developed in this paper to other BCGs in the CLASH sample will determine if this narrative is consistent for all of the star forming BCGs.  
  
\begin{figure}[h]  
\centering{\epsfig{file=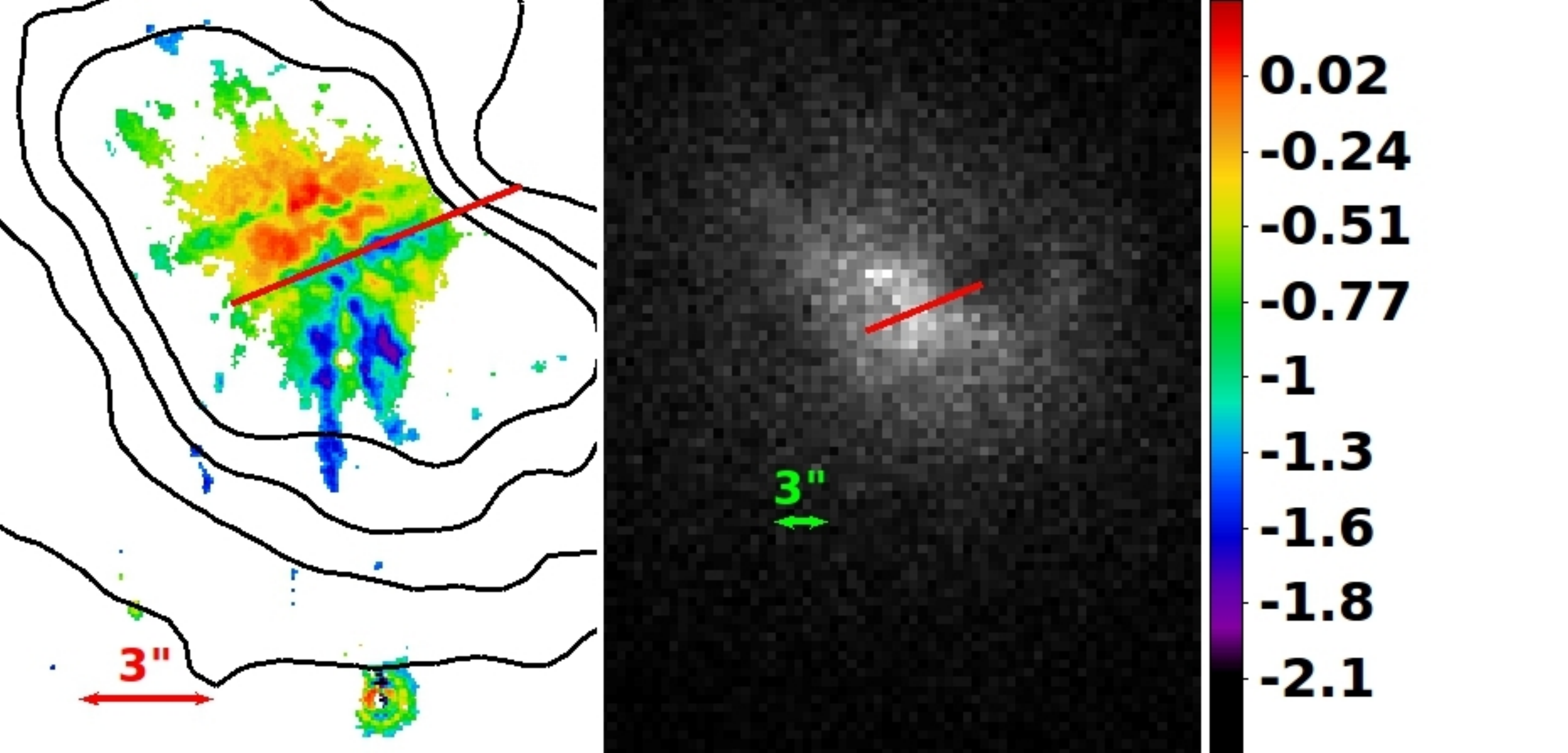,width=8cm,angle=0}}  
\caption{\textit{Left:} The burst log lifetime map for the RXJ1532 BCG, which emphasizes the young filaments, is shown with contours of 0.5-7.0 keV counts from the \textit{Chandra} observation of the cluster shown in black. The red line traces the position of the young filament extending along a northwest-southeast axis. \textit{Right:} The 0.5-7.0 keV \textit{Chandra} image of RXJ1532. The red line is identical to the line shown depicted in the left panel. The red line points directly into the prominent cavity in the northwest of the image, and appears to lie along the axis connecting the northwest cavity to the much fainter southeast cavity, which is evident in the unsharp-masked image presented in \cite{HlavacekLarrondo_2013_RXJ1532}.}  
\label{fig:Cavities}  
\end{figure}

The cavities in RXJ1532 also appear to be aligned with the northwest-southeast oriented young filament, and anti-aligned with the shape of the BCG filament network more generally. This morphological relationship is depicted in Figure~\ref{fig:Cavities}. The young filament traces one of the bright H$\alpha$ filaments as well. The most prominent X-ray cavity corresponds to the brighter (western) end of this filament, suggesting a system with the western edge inclined towards us. Given the available data, we do not rule out a coincidence; however, the corresponding ages and morphologies suggest jet-triggered formation of the young filaments. We hypothesize that the northwest-southeast filament may have been the result of positive feedback triggered by compression from a jet inflating the X-ray cavities.

The narrative we propose for RXJ1532 may be typical for other BCGs in cool core clusters. The SFH of RXJ1532 agrees with a study of line emitting BCGs in the SDSS survey  produced by \cite{Liu_2012_BCGSFR}. While their interpretation of the burst history differs from ours (they assume a stellar population divided into three components-- a recent starburst, young stars, and old stars), they find that the majority of the flux they observe in that sample of BCGs is due to stars forming within $\sim$2.5 Gyr. Furthermore, estimates of molecular gas masses imply that BCG starbursts typically have fuel to last $\sim$1 Gyr \citep{ODea_2008_BCGSF}. While the case of RXJ1532 may be extreme in terms of SFR, its burst history may be the norm for star forming BCGs. The sum of the results described here, taken in addition to the evidence provided by \cite{Donahue_2015_IP}, shows agreement between the resolved star structures we observe in the CLASH BCGs and recent theoretical work on feedback in cool core clusters.  

% REVISION: revised opening sentence:
If gas is condensing out of the ICM and feeding star formation, then 
we expect a general trend between the SFR and our $\dot{M}_g$.
The idea of an AGN jet-driven mechanism for ICM condensation could also account for our finding that as the SFR increases in the BCG, the SFR accounts for a larger fraction of the total cooling implied by $\dot{\textrm{M}}_{g}$. $\dot{\textrm{M}}_{g, r35}$ and $\dot{\textrm{M}}_{g, t20}$ can be interpreted as the cooling rate of gas in these radii in the absence of reheating.  Based on this interpretation, we suspect that larger starbursts occur in BCGs where gas is cooling more efficiently. The correlation between cool core BCG SFRs and `efficiency' can be explained neatly if cooling is localized around AGN jets and cavities, or if in systems with stronger cooling a larger fraction of the gas with $t_{cool}/t_{ff} < 20$ is at $t_{cool}/t_{ff} \lesssim 10$. If larger SFRs occur in BCGs exhibiting AGN feedback in larger areas, a larger fraction of unstable ICM gas inside a given radius may be triggered into condensing by feedback. This scenario is consistent with our finding that SFR is correlated with the area of the star forming regions in BCGs. 
% REVISION: deleted last sentence from above paragraph.

% REVISION: deleted first word in this paragraph:
In the feedback cycle modeled in \cite{Li_2015_SFAGN}, a large amount of gas is cooled quickly with the initial onset of jet feedback (on the order of 10s of Myr), and is slowly consumed by star formation (on the order of a Gyr).  Larger SFRs tend to occur earlier in the feedback cycle, when proportionally more of the ICM is precipitating. If this concern dominates the SFR-$\dot{M}_{\textrm{g}}$ relationship, then we expect to see evidence that larger BCG starbursts are younger on average. In RXJ1532, we see evidence connecting the durations of star forming knots and filaments to the duration of AGN activity (as evidenced by estimates of the ages of X-ray cavities in this system), although a more detailed study of multiple systems will be necessary to establish whether or not there is a relationship between burst age, SFR, and low-entropy ICM gas.

\section{Conclusion}  
  
We have conducted a detailed analysis of the star forming structure in the BCGs in the CLASH X-ray selected sample of galaxy clusters. Using the rich set of CLASH photometry, we estimated the dust reddening in BCGs with significant UV emission and calculated reddening corrected mean UV luminoisities and H$\alpha$+[\ion{N}{2}] luminosities. We compared these measurements to observations of [\ion{O}{2}] and H$\beta$ taken using the Goodman spectrograph. Additionally, we compared the UV derived SFR to X-ray properties calculated using the ACCEPT catalog, including the core entropy, K$_{0}$, and predicted cooling rates for low entropy gas inside $r=35$ kpc ($\dot{\textrm{M}}_{g, r35}$) and inside radii where $t_{cool}/t_{ff}=20$ ($\dot{\textrm{M}}_{g, t20}$). We concluded our analysis by creating a resolved map of the starburst in RXJ1532.9+3021, for which we also have an SDSS spectrum and detailed X-ray data from \cite{HlavacekLarrondo_2013_RXJ1532}.  
  
Using measurements of [\ion{O}{2}], [\ion{O}{3}], and H$\beta$ lines in conjunction with broadband H$\alpha$+[\ion{N}{2}] estimates, we constructed diagnostic diagrams for the CLASH BCGs in order to constrain the line-emission power source in these galaxies. Line emission in CLASH BCGs are powered by a combination of star formation and a LINER-like source (possibly the signature of hot, young stars or interaction between the ICM and nebular gas), while the biggest starburst (MACS1931) has a line emission spectrum dominated by ongoing star formation.  
  
CLASH SFRs span a range of magnitudes up to $\gtrsim 100$ M$_{\odot}$ yr$^{-1}$, and significant, extended star formation occurs in 10 out of 20 BCGs in our sample. Based on comparisons with K$_{0}$ and $\dot{\textrm{M}}_{g}$, we establish a link between the star formation in the BCG and the state of the surrounding ICM. All of the star forming BCGs with an SFR $> 10$ M$_{\odot}$ yr$^{-1}$ are consistent with a $\sim$ 30 KeV cm$^{2}$ entropy threshold, and a trend exists between SFR and $\dot{\textrm{M}}_{g}$. These findings imply SFR is fueled by a reservoir of low entropy gas.   
  
SED analysis of RXJ1532.9+3021 reveals a long-lived starburst, with a log lifetime of 8.8$\pm$0.5 log$_{10}$yr, and a total star formation rate of $118^{+215}_{-42}$ M$_{\odot}$ yr$^{-1}$, which is consistent with our estimates from UV and H$\alpha$ luminosities. The overall burst timescale is much longer than the AGN on-cycle as inferred by the ages of AGN cavities in the ICM of this cluster, although several of the individual filaments are consistent with the $\sim 60-80$ Myr cavity refill times. These results are consistent with recent jet-triggered filaments super-imposed on an older long-lived starburst, which may have been the result of jets from previous AGN on-cycles. The burst history in RXJ1532 is also consistent with another study of stellar populations by \cite{Liu_2012_BCGSFR} conducted on SDSS BCGs, so we hypothesize that in upcoming work we will find similar evidence for sporadic starbursts corresponding to episodes of AGN activity.

The SFH of knots and filaments in RXJ1532 suggest a jet-induced precipitation scenario such as \cite{Li_2015_SFAGN} is responsible for converting the ICM into cold, starforming gas. If true for all CLASH BCGs, this mechanism would explain the relationship between the thermodynamic state of the ICM surrounding CLASH BCGs and the SFRs in the BCGs. The increasing `efficiency' of BCG SFRs relative to the cooling rates implied by $\dot{\textrm{M}}_{g}$ as a function of SFR is plausibly explained by this scenario, as well.

\appendix
%\section{Appendix}
This appendix includes the details of the parameter space chosen to fit the photometry of RXJ1532.9+3021 to a distribution of model SEDs. We describe the parameterization of the SFH, along with the parameter space we defined. Finally, we describe the consistency tests performed on fitting the CLASH SED.   
  
For the 16 band SEDs we construct using all the available bands of CLASH photometry, we used the \cite{Salpeter_1955_IMF} IMF and \cite{Bruzual_2003_BC03} SSP. The SFH we fit consists of a uniform starburst imposed on a background population with an exponentially decaying SFR, thus the SFH is modeled by  
\begin{align}  
\psi_{e}\left(t\right) &= \frac{M_{\textrm{early}}}{\tau}e^{-t/\tau} \\  
\psi_{b}\left(t\right) &=   
\begin{cases}  
F_{b} M_{\textrm{early}} \frac{1-e^{-\left(t_{\textrm{age}}-\Delta t_{b}\right)/\tau}}{\Delta t_{b}} & t_{\textrm{age}} - t \leq \Delta t_{b} \\  
0 & t_{\textrm{age}} - t > \Delta t_{b} \\  
\end{cases}  \\  
\psi_{net}\left(t\right) &= \psi_{e}\left(t\right)+\psi_{b}\left(t\right),  
\end{align}  
which is a variant of the SFH described in \cite{Moustakas_2013_iSEDFIT}. $\psi_{e}$ is the SFH for the background early-type population of stars, which is parameterized by the time constant $\tau$.  $\psi_{b}$ is the burst SFH, and it is parameterized by the burst lifetime $\Delta t_{b}$ and the fractional burst amplitude $F_{b}$. Our SFH consists of one burst for simplicity.  
  
We allowed the age of the galaxy to vary between 6 and 9.5 Gyr, and we allowed $\tau$ to vary between $\frac{1}{5}$ and $\frac{1}{20}$ the age of the BCG, thereby ensuring that the background population corresponds to a quiescent, early-type galaxy. We allowed the metallicity of the stellar population to vary between 0.04 Z$_{\odot}$ and 1.6 Z$_{\odot}$, and the dust attenuation $A_{V}$ to vary between 0 and 5 mag. In order to sample a wide range of possible burst histories, we sampled burst parameters logarithmically, selecting $\Delta t_{b}$ in the range -3 $\leq$ log$\Delta t_{b}$ [Gyr] $\leq$ 0.8, and $F_{b}$ in the range -2 $\leq$ log$F_{b}$ $\leq$ 1.0. This parameter space is summarized in Table A1. We drew 10$^{4}$ models from this parameter space for each SED we fit to.
  
\begin{table}[h]
\renewcommand\thetable{A1}
\label{A1}
%\tablenum{A1}  
\footnotesize  
\caption{\\ SED Fitting Parameter Choices}  
\vspace{5mm}  
\centering  
{  
%Name, Calzetti UV Luminosity in the UV Contour region, Galactic UV Luminosity in the UV Contour region, Ha Contour Region UV, UV  Contour region UV, Ha Contour region Ha, Calzetti UV SFR, Calzetti HA SFR   
\begin{tabular}{lcc}  
Parameter & Range & Units \\  
\hline  
\hline  
SSP & BC03$^{a}$ & \\  
IMF & Salpeter$^{b}$ & \\
Model Draws & 10$^{4}$ & \\   
t$_{\textrm{age}}$ & [6, 9.5] & Gyr \\  
$\tau$ & [0.05, 0.2] & t$_{\textrm{age}}$ \\  
Metallicity & [0.04, 1.6] &  Z$_{\odot}$ \\  
A$_{\textrm{V}}$ & [0.0, 5.0] & mag\\  
log [\ion{O}{3}]/H$\beta$ & [-0.5, 0.5] & dex\\  
log $\Delta$ t$_{b}$  &  [-3.0, 0.8] &  log Gyr \\  
log $F_{b}$ & [-2.0, 1.0] & dex\\  
\end{tabular}  
\begin{flushleft}  
$^{a}$ \cite{Bruzual_2003_BC03} \\  
$^{b}$ \cite{Salpeter_1955_IMF}  
\end{flushleft}  
}  
\end{table}  
  
\subsection{SED Fitting Consistency Checks}  
  
The equivalent width of [\ion{N}{2}]+H$\alpha$, EW([\ion{N}{2}]+H$\alpha$), is exquisitely sensitive to the SFH of a galaxy, since it is a measure of the ratio of an SFR indicator to the red continuum \citep{Kennicutt_1998_SFR, Leitherer_2004_AgeDating}. Since this value was measured directly by SDSS, and our model SEDs predict line strengths, we can compare the best fit EW([\ion{N}{2}]+H$\alpha$) for the CLASH SED with the SDSS measured value. This in turn indicates how reliable our estimate of the burst duration is.  
  
\begin{figure}[h]
\figurenum{A1}  
\centering{\epsfig{file=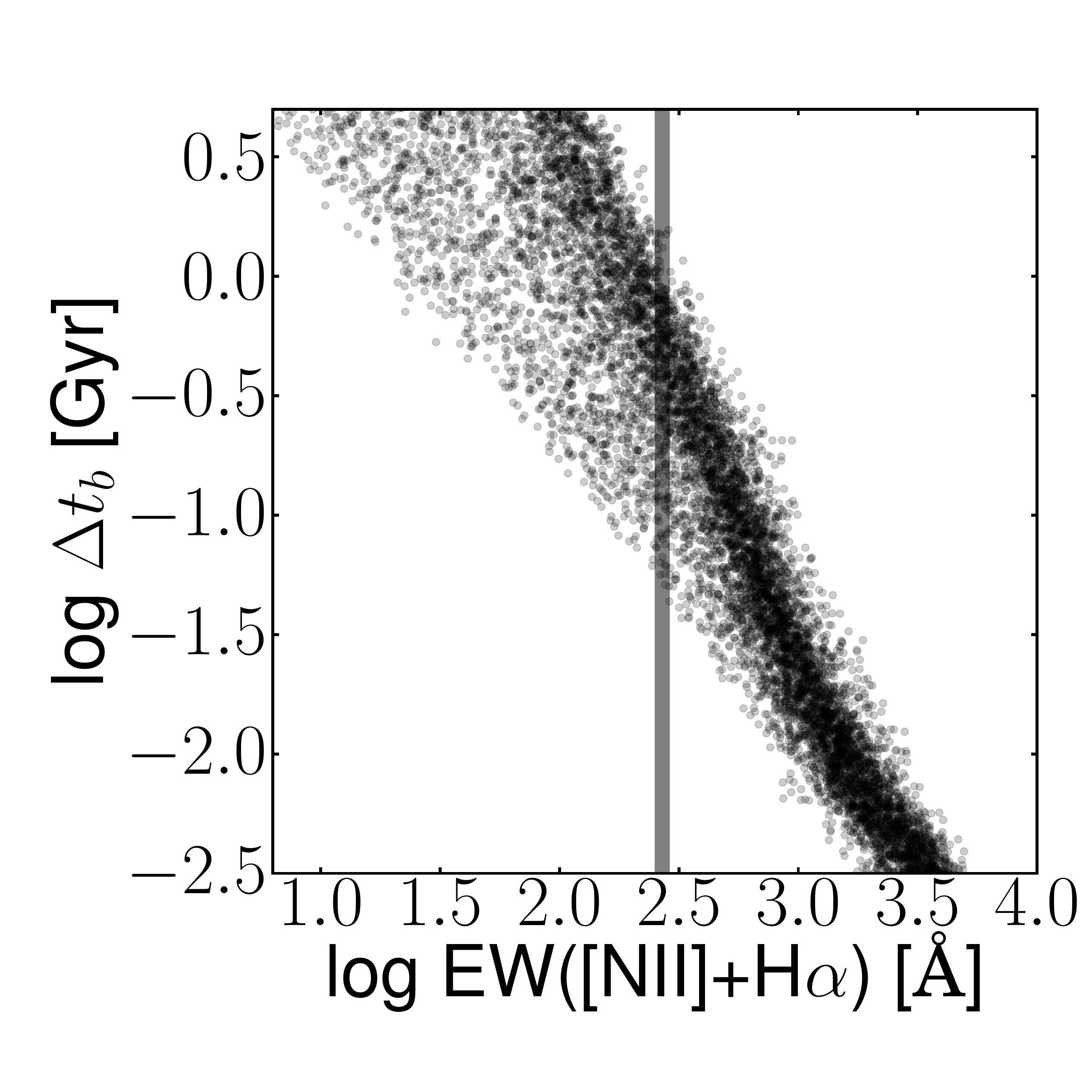,width=8cm,angle=0}}  
\caption{We plot the distribution in $\Delta$ t$_{b}$ and EW([\ion{N}{2}]+H$\alpha$ for all the models gridded in the parameter space summarized in Table A1. Each point corresponds to the values of these two parameters for each individual model, and the vertical line denotes  269.3 $\textrm{\AA}_{rest}$, the SDSS measured EW. This plot reveals a densely populated trend tracing out a curve of decreasing burst lifetimes as a function of EW, with a wider, more sparsely populated envelope.}  
\label{fig:DTB_EW}    
\end{figure}  
  
 We perform this comparison by fitting the SED of fluxes extracted in a 3$''$ diameter aperture centered on the coordinates of the SDSS fiber used to take spectra of RXJ1532.  Our predicted EW([\ion{N}{2}]+H$\alpha$) = 269$\pm$120 $\textrm{\AA}_{rest}$, which matches the SDSS measurement of 269.3$\pm$2.4 $\textrm{\AA}_{rest}$. The probability distribution we recover also shows that the burst duration ($\Delta t_{b}$) is log$\Delta t_{b}$ = $8.8\pm0.5$ [yr]. and burst mass (M$_{b}$) is log M$_{b}$  = 10.83 $\pm$ 0.35 [M$_{\odot}$]. The overlap between the SDSS fiber and the observable UV structure in RXJ1532 is substantial, so it is not surprising that we recover the same burst history and burst mass for this SED as the SED fit in the Results.  
  
As shown in Figure~\ref{fig:DTB_EW}, $\Delta t_{b}$ is highly dependent upon the equivalent width. The sensitivity of the equivalent width to the burst history makes it straightforward to constrain a relatively narrow range of burst durations for the model SED.  The agreement between the spectral equivalent width and our SED fit value is important because it shows that the H$\alpha$+[\ion{N}{2}] feature is detected strongly enough in the CLASH photometry that a meaningful constraint on the burst history can be made with it.   
  
We used {\tt iSEDfit} to fit only the three bands of WFC3/UVISIS photometry that were used to estimate L$_{UV}$, and show that the results agree with our broadband estimates. We fit photometry extracted from the region used to calculate L$_{UV}$ and assumed fit parameter constraints that are consisent with \cite{Kennicutt_1998_SFR}. For this step, we assumed the metallicity in the stellar population is 1 Z$_{\odot}$, that the observed age of the BCG, $t_{\textrm{age}}$, is at least 6 Gyr, and that the decay timescale for the SFH, $\tau$, is 100 Gyr in order to force a continuous SFR model. We also assume a \cite{Salpeter_1955_IMF} IMF, \cite{Bruzual_2003_BC03} SSP, and a \cite{Calzetti_2000_Extinction} reddening law.  
  
Using this method, we calculated an SFR of 99$\pm$24 M$_{\odot}$ yr$^{-1}$, and an average reddening of E(B-V) $= 0.26 \pm 0.04$. As before, the SFR PDF is log-normal and we report the mode of the distribution. This result is consistent with the UV photometry results. The agreement implies that while taking a single-aperture SED fit washes out the correlation seen between the spatial distribution of the dust and the SFR, this effect is not a strong source of systematic variation in our fits.   
   
%\clearpage  
%%%%%%%%%%%%%%%%%%%%%%%%%%%%%%%%%%%  
%Bibliography  
%%%%%%%%%%%%%%%%%%%%%%%%%%%%%%%%%%%  
%\bibliographystyle{apj}  
\bibliography{CLASH_Refs}  
\end{document}